 \documentclass[prc,amsmath,amfonts,showpacs,eqsecnum,floatfix,twocolumn]{revtex4}

 \usepackage{axodraw}   
 \usepackage{graphicx}  
 \usepackage{pstcol}    
 \usepackage{bm}    
 \usepackage{tabularx}    
\begin{document}
\hyphenation{Rijken}
\hyphenation{Nijmegen}
 
\title{
    Extended-soft-core Baryon-Baryon Model ESC08\\   
    II. Hyperon-Nucleon Interactions \\
% {\bf (version: best14june)} 
}
\author{M.M.\ Nagels}
\affiliation{ Institute of Mathematics, Astrophysics, and Particle Physics \\
 University of Nijmegen, Nijmegen, The Netherlands}               
\author{Th.A.\ Rijken}
\affiliation{ Institute of Mathematics, Astrophysics, and Particle Physics \\
 University of Nijmegen, Nijmegen, The Netherlands}               
\affiliation{Nishina Center for Accelerator-Based Science, Institute for Physical \\
 and Chemical Research (RIKEN), Wako, Saitama, 351-0198, Japan }
\author{Y.\ Yamamoto}
\affiliation{Nishina Center for Accelerator-Based Science, Institute for Physical \\
 and Chemical Research (RIKEN), Wako, Saitama, 351-0198, Japan }

\pacs{13.75.Cs, 12.39.Pn, 21.30.+y}

\date{version of: \today}
 
\begin{abstract}                                       
The YN results are presented from a new version of the Extended-soft-core (ESC) 
potential model for Baryon-baryon (BB) scattering. The potentials 
consist of local- and non-local-potentials due to (i) One-boson-exchanges 
(OBE), which are the members of nonets of pseudoscalar-, vector-, scalar-, and
axial-vector mesons, (ii) Pomeron and Odderon exchanges, 
(iii) Two pseudoscalar exchange (PS-PS),
and (iv) Meson-Pair-exchange (MPE). Both the OBE- and Pair-vertices are regulated 
by gaussian form factors producing potentials with a soft behavior near the origin.
The assignment of the cut-off masses for the BBM-vertices is dependent on the 
$SU(3)$-classification of the exchanged mesons for OBE, and a similar scheme 
for MPE.        
In addition to these standard ingredients of the ESC-models also the possible 
short range repulsion due to the quark Pauli-principle in the BB-channels
is included in the analysis,
for the first time in a systematic way, in this paper.

The present version of the ESC-model, called ESC08, describes 
nucleon-nucleon (NN) and hyperon-nucleon (YN)  as well as the S=-2 
hyperon-hyperon/nucleon (YY) in a unified way using broken 
$SU(3)$-symmetry. Major novel ingredients with respect to the former version ESC04
are the inclusion of 
(i) short-range gaussian odderon-potentials corresponding to the odd number gluon-exchange,
(ii) exceptional short range repulsion in specific YN and YY channels
due to Pauli-forbidden six-quark cluster $(0s)^6$-configurations. 
Further new elements are
(i) the extension of the $J^{PC}=1^{++}$ axial-vector meson coupling, 
(ii) the inclusion of the $J^{PC}=1^{+-}$ axial-vector mesons, and (iii) 
a completion
of the $1/M$-corrections for the meson-pair-exchange (MPE) potentials.
Like in the ESC04-model, 
the  octet and singlet coupling constants and $F/(F+D)$-ratio's of the model are
conform the predictions of the quark-antiquark pair-creation (QPC) model with
dominance of the $^3P_0$-mechanism.
This not only for the OBE-couplings but also for the MPE-couplings and $F/(F+D)$-ratio's.
 
Broken $SU(3)$-symmetry serves  to connect the $NN$,  the $YN$ and the $YY$
channels.  The fitting of $NN$ dominates the determination of the 
couplings and the cut-off masses. Only a few parameters are
strongly influenced by the $YN$ data, and by the constraints for the
$YY$-interactions following from G-matrix analyses of hypernuclei.
In particular, the meson-baryon coupling constants are calculated via $SU(3)$
using the coupling constants of the $NN$-analysis as input.
In contrast to ESC04, we do not consider medium strong flavor-symmetry-breaking (FSB) 
of the coupling constants.
The charge-symmetry-breaking (CSB) in the $\Lambda p$ and $\Lambda n$ channels, which is 
an $SU(2)$ isospin breaking, is included in the OBE-, TME-, and MPE-potentials.

For the ESC08-model we performed a simultaneous fit to the combined NN and
YN scattering data, supplied with constraints on the YN  and YY interaction
originating from the G-matrix information on hypernuclei.
In addition to the usual set of 35 YN-data and 3 $\Sigma^+p$ cross-sections 
from a recent KEK-experiment E289, 
we added 
 11 elastic and inelastic $\Lambda p$ 
and 3 elastic $\Sigma^-p$ 
cross-sections at higher energy. 
We obtained within this simultaneous fit $\chi^2/NN_{data}= 1.081$ and 
$\chi^{2}/YN_{data} = 1.08$. In particular, we
were able to fit the precise experimental datum $r_{R}=0.468 \pm 0.010$ for
the inelastic $\Sigma^-p$ capture ratio at rest rather well.      

Besides the results for the fit to the scattering data, which
defines the model largely, also the application to hypernuclear
systems, using the G-matrix method, is rather important in establishing 
the ESC-model. Different versions of e.g. the ESC08-model give different 
results for hypernuclei. The reported G-matrix calculations are performed for
$YN$ ($\Lambda N$, $\Sigma N$, $\Xi N$) pairs in nuclear matter  and also
for some hypernuclei. The obtained
well depths ($U_\Lambda$, $U_\Sigma$, $U_\Xi$) reveal distinct features of the
ESC-model.

The inclusion of a  quark core Pauli-repulsion can make the $\Sigma$-nucleus
interaction sufficiently repulsive, as seems to be required by the available 
experimental evidence. Furthermore, the ESC08-model gives small spin-orbit 
splittings in $\Lambda$-hypernuclei, which is also  indicated by experiment.

%\vspace{2mm}
%\noindent {\bf esc08c.yn10.tex: VERSION NOV. 2014}

 \end{abstract}
 \pacs{13.75.Cs, 12.39.Pn, 21.30.+y}

\maketitle

\twocolumngrid
\section{Introduction}
\label{sec:1}

%---------------------------------------------------------------------------------
%---------------------------------------------------------------------------------
%General things on the ESC-model:
This is the second in a series of papers \cite{NRY12a,NRY12b,NRY12c}
, henceforth referred to I, II, and III respectively, 
on the results of the Extended-soft-core (ESC) model for 
low and intermediate energy baryon-baryon interactions using the ESC08-model. 
The first results on the BB-channels and applications to hypernuclei were given 
in the review \cite{PTP185}.
With the ESC04-models \cite{Rij04a,Rij04b,Rij04c}, it was shown 
that a very successful description of the presently available 
baryon-baryon scattering data
could be achieved within the ESC-approach to the nuclear force problem.
Also, such a description was obtained with meson-baryon coupling parameters
which can be understood rather nicely within the context of the $^3P_0$-
quark-pair creation mechanism \cite{Mic69,LeY73}. This latter mechanism 
has been shown to be dominant in the framework of lattice QCD \cite{Isg85}.
The simultaneous and unified treatment of the $N\!N$ and $Y\!N$ channels in ESC04, 
using broken $SU(3)$-flavor, has given already a rather successful 
potential model for the low and intermediate energy baryon-baryon scattering data. 
Furthermore, the basic ingredients of the model are physically motivated by
the quark-model (QM) and QCD.

\noindent The G-matrix calculations showed that basic features of hypernuclear
data are also reproduced rather well, improving several weak points of 
the soft-core OBE-models \cite{NRS78,MRS89,RSY99}. However, there remained the 
problem that the meson-exchange models seem to be unable to give a        
positive well depth $U_\Sigma$. A second problem posed the very small 
spin-orbit splittings in $\Lambda$-hypernuclei \cite{Hiy00,HT06}.
In this paper we extend and refine the ESC-model in order to provide
improvements and answers to these issues.

\noindent First, we list the new ingredients of the here presented version
ESC08c, which are more or less in line with the ESC-approach as presented so far.
In this category, the following additions to the ESC04-model are made for the 
present ESC08-model:\\
\noindent (i) For the axial-vector mesons with $J^{PC}=1^{++}$, the A-mesons, next to the 
$\gamma_5\gamma_\mu$-coupling also the derivative $\gamma_5 k_\mu$-coupling is exploited.\\
\noindent (ii) The axial-vector mesons with $J^{PC}=1^{+-}$, the B-mesons, are
included as well. 
The latter have potentials of the same type as the pseudo-scalar
mesons, but have an opposite sign. We notice that now the set of the exchanged
quantum numbers for OBE-potentials is identical to that for MPE-potentials.\\
\noindent (iii) For the meson-exchange we have included the Brown-Downs-Iddings
anti-symmetric spin-orbit potentials from pseudoscalar-, vector-, scalar-, and axial-
meson exchange \cite{BDI70}. \\
\noindent (iv) We have completed the $1/M$-corrections for meson-pair-exchange (MPE),
in particular for the $J^{PC}=1^{++}$- and $J^{PC}=1^{+-}$-axial pairs. 
This also leads to new important contributions to the 
anti-symmetric-spin-orbit interaction \cite{ALS07}.\\
\noindent (v)  For the diffractive contribution we have next to the Pomeron-exchange
\cite{note.diffr} added the Odderon-exchange \cite{Odd03}.              
Whereas in QCD the Pomeron can be associated with colorless even number (2,4, ...) of 
gluon-exchanges, the Odderon is associated with the colorless odd number (3,5, ...)
of gluon-exchanges. At low energies the Pomeron has $J^{PC}=0^{++}$, but the
Odderon has $J^{PC}=1^{--}$.

\noindent Secondly, we have opened the possibility to incorporate possible effects 
of a 'structural' or channel-dependent repulsion due to Pauli-blocking. 
This repulsion originates from a
'forbidden-state' in the SU(6)$_{fs}$ Quark-Cluster-Model (QCM) \cite{Oka00,Fuj07}.
This is the analog of a well known effect in $\alpha\alpha$-scattering discovered
in the sixties \cite{Tam65}.
This 'forbidden-state' is the [51]-irrep and this irrep occurs with a large weight in the
two $J^{P}= 1/2^{+}$-baryon states in the SU(3)$_f$-irreps $\{10\}$ and $\{8_s\}$. 
These irreps are prominent in the $\Sigma^+p(^3S_1)$- respectively the 
$\Sigma N(^1S_0)$-states. These are precisely the states where according to e.g. the 
G-matrix calculations the ESC-models possibly lack some repulsion. This repulsion seems to
be indicated by experiment \cite{Dab99,Nou02}. The [51]-irrep also occurs in the other NN-,  
YN-, and YY-channels, but with roughly equal weights, see \cite{Oka00},
apart from a few S=-2 channels, e.g. $\Xi N(I=1,S=0)$.

\noindent  We account for the 'exceptional-repulsion' in a phenomenological way 
by enhancing the "pure" Pomeron-coupling. So the effective Pomeron-repulsion consists 
of the pure Pomeron-exchange contribution augmented with a fraction of Pauli-blocking repulsion,
which varies for the different BB-channels. (The other typical quark-cluster
effects like e.g. one-gluon-exchange (OGE) annex quark-interchange is in ESC-models
taken care of by meson exchange.) In this work we try to determine the strength 
of this Pauli-blocking effect in BB-channels. The fit to NN determines the sum of both
the pure Pomeron-repulsion and the Pauli-blocking repulsion. The fit to YN determines
the fraction of Pauli-blocking in it.

The ESC08-model realizes a fusion between the soft-core meson-exchange 
potentials and QCM-aspects of the baryon-baryon interactions and can be called 
a 'hybrid' ESC-model.
The soft-core meson-exchange model has been described in detail in previous
papers, \cite{Rij04a,Rij04b,Rij04c}. Therefore, we may refer here to those papers
for a description of (a) the physical background, (b) the employed formalism,
(c) the description of the potentials, either in details or in references to 
papers where further information may be obtained. 
In this paper we will derive (i) the new OBE-potentials employed here for the
first time in the context of the ESC-model, (ii) the Odderon-potentials, and 
(iii) a derivation of the short-range phenomenology connected to the quark
Pauli principle within the context of the SU(3)$_f$-formalism as used in the
Nijmegen potentials. Next to these items, we will also give the new $1/M$-corrections
for the axial-meson-pair-exchange potentials, where we restrict ourselves to the 
spin-spin and tensor contributions. The YN symmetric and anti-symmetric
spin-orbit potentials will be described in another paper.

In \cite{Rij04a,Rij04b} a detailed description of the basic features of the 
ESC-models has been given and motivated. Many of these were already present 
in the Nijmegen soft-core \cite{MRS89} and hard-core \cite{NRS77} OBE-models.
We refer the reader to these references for the description and discussion of
the items such as: (broken) SU(3)-flavor, charge-symmetry-breaking (CSB) in YN,
meson-mixing in the pseudo-scalar-, vector-, scalar- meson SU(3)-nonets, the
role of the quark-antiquark pair-creation $^3P_0$-model for BBM- and BBMP-couplings.
Also, in e.g. \cite{Rij04b} one finds a recapitulation of the goals of our
continued investigation of the baryon-baryon systems.

%---------------------------------------------------------------------------------
%gaussian form factors:
In the soft-core Nijmegen OBE- and ESC-models the form factors are taken
to be of the gaussian-type. In the (non-relativistic) QM's a gaussian behavior 
of the form factors for ground-state baryons is most natural. The two-particle
branchpoints, corresponding to e.g. $\pi\pi, \pi\rho,\ K\rho$-etc.,
are in the ESC-models accounted for by the MPE-potentials. Gaussian residue
functions are used in regge-pole models for two-particle reactions at high-energy
and low momentum-transfers.

%---------------------------------------------------------------------------------
%Medium-strong SU(3)-breaking:
As pointed out in \cite{Rij04a,Rij04b}
$SU(3)$-symmetry and the QPC-model give strong constraints on the coupling 
parameters. The $^{3}P_{0}$-model also offers the possibility to introduce
a scheme for hypercharge breaking a la Gell-Mann-Okubo for the BBM-couplings.
In order to keep some more flexibility in distinguishing the $N\!N$- and
the $Y\!N(S=-1)$-channels, such a medium-strong breaking was explored in
the NSC97 \cite{RSY99} and ESC04 \cite{Rij04b}. In the present study we 
do not apply such a breaking.  The results show that a scheme of
SU(3) symmetric couplings with only mass breaking can give an excellent description 
of all BB interactions.

%%---------------------------------------------------------------------------------
The content of this paper is as follows. In section II we review very briefly
the scattering formalism, the Lippmann-Schwinger equation for the T- and V-matrices.
Similarly, in section III 
the $NN$ and $S=-1$ $Y\!N$-channels on the isospin and particle basis, and
the use of the multi-channel Schr\"{o}dinger equation is mentioned.
The potentials in momentum and configuration space are defined by
referring to the description given in \cite{Rij04a}. 
Also SU(3)-breaking is reviewed briefly.
In section IV on the OBE-potentials,  the additions for ESC08 in comparison with 
the ESC04-model are described. Here, we give the new potentials in momentum and
configuration space.
In section V the SU(3) structure of the MPE-potentials is given and
the additions in comparison with the ESC04-model are listed.      
The latter are the axial $J^{PC}= 1^{+-}$ $(\pi\omega)$-pair potentials,
which is the content of Appexdix~\ref{app:MPE.ax.2nd}.
In section VI the short-range phenomenology is discussed. We derive the
incorporation of the 'exceptional' Pauli-repulsion, which shows up 'exceptionally' 
large in the SU(3)-irreps $\{10\}$ and $\{8_s\}$.\\
In section VII the simultaneous $NN \oplus YN \oplus YY$ fitting procedure is reviewed.
In section VIII the results for the coupling constants and $F/(F+D)$-ratios
for OBE and MPE are given. 
They are discussed and compared with the predictions of the 
QPC-model. Here, also the values of the $BBM$-couplings are displayed
for pseudo-scalar, vector, scalar, and axial-vector mesons.\\
In section IX the $YN$-results for ESC08c from the combined 
$NN \oplus YN \oplus YY$-fit  
are discussed.
In section X we discuss the fit to the $Y\!N$ scattering data.          
In section X, the hypernuclear properties of ESC08 are studied through
the G-matrix calculations for $YN$ ($\Lambda N$, $\Sigma N$, $\Xi N$)
and their partial-wave contributions. Here, the implications of possible
three-body effects for the nuclear saturation and baryon well-depths 
are discussed. Also, the $\Lambda \Lambda$ interactions in ESC08 are 
demonstrated to be consistent with the observed data of $^{\ 6}_{\Lambda\Lambda}$He.
In section XI we finish by a final discussion, draw some conclusions, and an outlook.
In Appendix~\ref{app:MPE.SU3} we display the full SU(3) contents of the 
 MPE-couplings, and in Appendix~\ref{app:MPE.ax.2nd} for completeness the 
 $J^{PC}=1^{+-}$ axial-pair potentials are given.
Finally, in Appendix~\ref{app:H} the antisymmetric spin-orbit potentials are 
derived explicitly for strange meson-exchange K, K$^*$, $\kappa$, and K$_1$.
%------------------------------------------------------------------------

%-------------------------------------------------------------------------------
\section{Scattering formalism, the  Lippmann-Schwinger Equation, Potentials} 
\label{sec:LS}
In this paper we treat the nucleon-nucleon (NN) and hyperon-nucleon (YN)
reactions with strangeness $S=0,-1$. 
Since in general there are both 'direct' and 'exchange' potentials, the ordering of 
the baryons in the incoming and outgoing states needs special attention. For 
keeping this ordering clear, we consider for definiteness 
the hyperon-nucleon reactions 
\begin{equation}
  Y (p_{1},s_{1})+N(p_{2},s_{2})      \rightarrow 
  Y'(p_{1}',s_{1}')+N'(p_{2}',s_{2}')\ .
\label{eq:LS.1}\end{equation}
Like in \cite{MRS89}, whose conventions we will
follow in this paper, we will also refer to $Y$ and $Y'$ as
particles $1$ and $3$ and to $N$ and $N'$ as particles
$2$ and $4$.  The four momentum of particle $i$ is $p_{i}=(E_{i},{\bf k}_{i})$
where
$E_{i}=\sqrt{{\bf k}_{i}^{2}+M_{i}^{2}}$ and $M_{i}$ is the mass/
The transition amplitude matrix $M$ is related to the $S$-matrix via
\begin{equation}
  \langle f|S|i \rangle= \langle f|i \rangle -
  i(2\pi)^{4}\delta^{4}(P_{f}-P_{i})
  \langle f|M|i \rangle\ ,
\label{eq:LS.2}\end{equation}
where $P_{i}=p_{1}+p_{2}$
and   $P_{f}=p_{1}'+p_{2}'$
represent the total four momentum for the initial state $|i\rangle$
and the final state $|f\rangle$. The latter refer to the
two-particle states, which we normalize in the following way
 
\begin{eqnarray}
&&  \langle{\bf p}_{1}',{\bf p}_{2}'|{\bf p}_{1},{\bf p}_{2}\rangle
  = (2\pi)^{3}2E({\bf k}_{1})
  \delta^{3}({\bf p}_{1}'-{\bf p}_{1})\cdot \nonumber\\
  && \hspace{1cm}\times(2\pi)^{3}2E({\bf k}_{2})
  \delta^{3}({\bf p}_{2}'-{\bf p}_{2})\ .
\label{eq:LS.3}\end{eqnarray}

We follow section II of \cite{MRS89} in detail. The
transformation to the non-relativistic normalization of the
two-particle states leads to states with
\begin{eqnarray}
&& ({\bf p}_{1}',s_{1}';{\bf p}_{2}',s_{2}'|{\bf p}_{1},s_{1};
 {\bf p}_{2},s_{2} ) =
  (2\pi)^{6}\delta^{3}({\bf p}_{1}'-{\bf p}_{1})\cdot \nonumber\\
  && \hspace{1cm} \times   \delta^{3}({\bf p}_{2}'-{\bf p}_{2})\
  \delta_{s_{1}',s_{1}} \delta_{s_{2}',s_{1}}\ .
\label{eq:LS.4}\end{eqnarray}
For these states we define the $T$-matrix by
\begin{equation}
  (f|T|i)=\{4M_{34}(E_{3}+E_{4})\}^{-\frac{1}{2}} \langle f|M|i \rangle
          \{4M_{12}(E_{1}+E_{2})\}^{-\frac{1}{2}}\ ,
\label{eq:LS.5}\end{equation}
which satisfies the Lippmann-Schwinger equation \cite{MRS89}
\begin{eqnarray}
%\lefteqn{
&& (3,4|T|1,2) = (3,4|V|1,2) + \frac{\small{1}}{\small{(2\pi)^{3}}}
      \sum_{n}\int d^{3}k_{n}\cdot \nonumber \\[0.3cm]
 && \times (3,4|V|n_{1},n_{2})
   \frac{\small{2M_{n_{1},n_{2}}}}{\small{{\bf p}_{n}^{2}-{\bf k}_{n}^{2}+i\varepsilon}}
      (n_{1},n_{2}|T|1,2)\ ,   \nonumber \\ 
\label{eq:LS.6}\end{eqnarray}
and where analogously to Eq.~(\ref{eq:LS.5}) the potential $V$ is defined
as
\begin{equation}
  (f|V|i)=\{4M_{34}(E_{3}+E_{4})\}^{-\frac{1}{2}} \langle f|W|i \rangle
          \{4M_{12}(E_{1}+E_{2})\}^{-\frac{1}{2}}\ .
\label{eq:LS.7}\end{equation}
Above, we denoted the initial- and final-state 
CM-momenta by ${\bf p}_i$ and ${\bf p}_f$.
Using rotational invariance and parity conservation we expand
the $T$-matrix, which is a $4\times 4$-matrix in Pauli-spinor space,
 into a complete set of Pauli-spinor invariants (\cite{MRS89,SNRV71}) 
\begin{equation}
  T=\sum_{i=1}^{8}T_{i}({\bf p}_{f}^{2},{\bf p}_{i}^{2},
  {\bf p}_{i}.{\bf p}_{f})\ P_{i}\ .
\label{eq:LS.8}\end{equation}
Introducing
\begin{equation}
  {\bf q}=\frac{\small{1}}{\small{2}}({\bf p}_{f}+{\bf p}_{i}),\hspace{2em}
  {\bf k}={\bf p}_{f}-{\bf p}_{i},\hspace{2em}
  {\bf n}={\bf p}_{i}\times {\bf p}_{f}={\bf q}\times {\bf k}
\label{eq:LS.9}\end{equation}
with , of course, ${\bf n}={\bf q}\times {\bf k}$,
we choose for the operators $P_{i}$ in spin-space
\begin{subequations}
\label{eq:LS.10}
\begin{eqnarray}
  P_{1} &=& 1\ , \\ 
  P_{2} &=& \mbox{\boldmath $\sigma$}_1\cdot\mbox{\boldmath $\sigma$}_2\ , \\       
  P_{3} &=& (\mbox{\boldmath $\sigma$}_1\cdot{\bf k})
  (\mbox{\boldmath $\sigma$}_2\cdot{\bf k})
  -\frac{1}{3}(\mbox{\boldmath $\sigma$}_1\cdot\mbox{\boldmath $\sigma$}_2)
  {\bf k}^{2}\ , \\
  P_{4} &=& \frac{i}{2}(\mbox{\boldmath $\sigma$}_1+\mbox{\boldmath $\sigma$}_2)
  \cdot{\bf n}\ , \\[0.1cm]
  P_{5} &=& (\mbox{\boldmath $\sigma$}_1\cdot{\bf n})(\mbox{\boldmath $\sigma$}_2
  \cdot{\bf n})\ , \\
  P_{6} &=& \frac{i}{2} (\mbox{\boldmath $\sigma$}_1-\mbox{\boldmath $\sigma$}_2)
  \cdot{\bf n}\ ,  \\[0.1cm]
  P_{7} &=& (\mbox{\boldmath $\sigma$}_1\cdot{\bf q})(\mbox{\boldmath $\sigma$}_2
  \cdot{\bf k})
   +(\mbox{\boldmath $\sigma$}_1\cdot{\bf k}) (\mbox{\boldmath $\sigma$}_2\cdot{\bf q})\
  \\[0.1cm]
  P_{8} &=& (\mbox{\boldmath $\sigma$}_1\cdot{\bf q})(\mbox{\boldmath $\sigma$}_2
  \cdot{\bf k})
   -(\mbox{\boldmath $\sigma$}_1\cdot{\bf k}) (\mbox{\boldmath $\sigma$}_2\cdot{\bf q})\ .
\end{eqnarray}
\end{subequations}
Here we follow \cite{MRS89,SNRV71}, except that we have chosen
here $P_{3}$ to be a purely `tensor-force' operator.
 
Similarly to (\ref{eq:LS.9}) the potentials are expanded as
\begin{equation}
  V=\sum_{i=1}^{6}V_{i}({\bf k}\,^{2},{\bf q}\,^{2}) P_{i}\ .
\label{eq:LS.11}\end{equation}
The potentials in configuration space are described in Pauli-spinor space
as follows
\begin{eqnarray}
 V(r) &=& V_C(r) + V_\sigma(r) \bm{\sigma}_1\cdot\bm{\sigma}_2 + V_T(r)\ S_{12} 
 + V_{SLS}(r)\cdot\nonumber\\ && 
   \times{\bf L}\cdot{\bf S}_+ + V_{ALS}(r)\ {\bf L}\cdot{\bf S}_-
   + V_{Q}(r)\ Q_{12}\ ,
\label{eq:LS.12}\end{eqnarray}
where ${\bf S}_\pm = (\mbox{\boldmath $\sigma$}_1 \pm \mbox{\boldmath $\sigma$}_2)/2$,
and see e.g. \cite{MRS89} for a definition of the operators $S_{12}$ and $Q_{12}$.
%$Q_{12}= 
%(\mbox{\boldmath $\sigma$}_1\cdot{\bf L}\mbox{\boldmath $\sigma$}_2\cdot{\bf L}
%+\mbox{\boldmath $\sigma$}_2\cdot{\bf L}\mbox{\boldmath $\sigma$}_1\cdot{\bf L})\2$.
%-----------------------------------------------------------------------------

\section{Channels, Potentials, and $SU(3)$ Symmetry}
\label{sec:3}
 \subsection{Channels and Potentials}     
 \label{sec:3a}
On the physical particle basis, there are three charge NN-channels:
\begin{eqnarray}
   q=+2,+1,0: && pp\rightarrow\ pp\ \ ,\ pn \rightarrow pn\ \ ,\ 
  nn \rightarrow nn\ .
\label{eq:3.2}\end{eqnarray}
Similarly, there are four charge YN-channels:
\begin{eqnarray}
   q=+2:\ \  && \Sigma^+p\rightarrow\Sigma^+p,         \nonumber\\
   q=+1:\ \  && (\Lambda p, \Sigma^+n, \Sigma^0p)\rightarrow
             (\Lambda p, \Sigma^+n, \Sigma^0p),        \nonumber\\
   q=\ \ 0:\ \ && (\Lambda n, \Sigma^0n, \Sigma^-p)\rightarrow
             (\Lambda n, \Sigma^0n, \Sigma^-p),        \nonumber\\
   q=-1:\ \  && \Sigma^-n\rightarrow\Sigma^-n.
\label{eq:3.3}\end{eqnarray}
Like in \cite{MRS89,RSY99}, the potentials are calculated on the isospin basis.
For $S=0$ nucleon-nucleon systems there are two isospin-channels, namely 
$I=1$ and $I=0$. 
For $S=-1$ hyperon-nucleon systems there are also two isospin channels:
(i) $ I={\textstyle\frac{1}{2}}:\ \ (\Lambda N,\Sigma N\rightarrow
                                  \Lambda N,\Sigma N)$, and              
(ii) $ I={\textstyle\frac{3}{2}}:\ \ \Sigma N\rightarrow\Sigma N$. 

For the OBE-part of the potentials the treatment of $SU(3)$ for the BBM interaction
Lagrangians and the coupling coefficients of the OBE-graphs has been given in detail
in previous work of the Nijmegen group, e.g. \cite{MRS89} and \cite{RSY99},
For the TME- and the MPE-parts the calculation of the coupling coefficients
has been exposed in our paper on the ESC04-model \cite{Rij04b}. There we 
described the method of an automatic computerized calculation of these coefficients by
exploiting the 'cartesian-octet'-representation.

Also in this work we do not solve the Lippmann-Schwinger equation, but the 
multi-channel Schr\"{o}dinger equation in configuration space, completely 
analogous to \cite{MRS89}.
The multichannel Schr\"odinger equation for the configuration-space
potential is derived from the Lippmann-Schwinger equation through
the standard Fourier transform, and the equation for the radial
wave function is found to be of the form~\cite{MRS89}
\begin{equation}
   u^{\prime\prime}_{l,j}+(p_i^2\delta_{i,j}-A_{i,j})u_{l,j}
       -B_{i,j}u'_{l,j}=0,           
\label{eq:3.5}\end{equation}
where $A_{i,j}$ contains the potential, nonlocal contributions, and
the centrifugal barrier, while $B_{i,j}$ is only present when non-local
contributions are included. 
The solution in the presence of open and closed channels is given,
for example, in Ref.~\cite{Nag73}.
The inclusion of the Coulomb interaction in the
configuration-space equation is well known and included in the evaluation
of the scattering matrix.

The momentum space and configuration space potentials for the ESC-models
have been described in paper I \cite{Rij04a} for baryon-baryon in general.
Here, we will only give the new contributions to these potentials, both
in momentum and configuration space.

 \subsection{SU(3)-Symmetry and -Breaking, Form Factors} 
 \label{sec:3b}
The treatment of the mass differences among the baryons is handled 
in the same way as for ESC04, which is exactly that of 
other Nijmegen models \cite{NRS77,MRS89,RSY99}. 
Also, exchange potentials related to
strange meson exchange $K, K^*$ etc. , can be found in these references. 

The breaking of SU(3)-symmetry occurs in several places. The physical
masses of the baryons and mesons are used. Noticable is the SU(2) $\subset$ SU(3) 
breaking due to $\Lambda-\Sigma^0$-mixing \cite{Dal64}.
This $\Lambda-\Sigma^0$-mixing leads also to a non-zero coupling of the $\Lambda$
to the other $I=1$ mesons: $\rho(760), a_0(980), a_1(1270)$, as well as to the 
$I=1$-pairs. For the details of these OBE-couplings see e.g. \cite{RSY99}, 
equations (2.15)-(2.17). Like in ESC04, the corresponding so-called CSB-potentials are 
included in the ESC08-model for OBE, TME, and MPE.

\noindent The medium-strong SU(3)-symmetry breaking of the BBM-coupling constants
is not tried in ESC08. In the ESC04-model this was considered optional, and
regulated by the $^3P_0$-model by a differentiation between the $s\bar{s}$-quark pair
and the creation of a non-strange quark-antiquark pair. Of course, we could 
contemplate about such an option here, but we did not investigate this option.

The baryon mass differences in the intermediate states for TME- and MPE-
potentials have been neglected for YN-scattering. This, although possible
in principle, becomes rather laborious and is not expected to change the 
characteristics of the baryon-baryon potentials much.

%---------------------------------------------------------------------------------
Also in this work, like ESC04- \cite{Rij04a,Rij04b,Rij04c} and 
in the NSC97-models \cite{RSY99}, the form 
factors depend on the SU$(3)$ assignment of the mesons,  
In principle, we introduce form factor masses, i.e. cut-off's,
$\Lambda_{8}$ and $\Lambda_{1}$ for the $\{8\}$ and $\{1\}$ members of each 
meson nonet, respectively. In the application to $YN$ and $YY$, we could allow
for SU$(3)$-breaking, by using different cut-offs for the strange mesons
$K$, $K^{*}$, and $\kappa$. However, in the ESC08-model we do not exploit this
possible breaking, but assign for the strange $I=1/2$-mesons the same cut-off 
as for the $I=1$-mesons.
Moreover, for the $I=0$-mesons we assign the 
cut-offs as if there were no meson-mixing. For example we assign $\Lambda_1$ 
for the dominant singlet mesons
$\eta', \omega, \epsilon$, and $\Lambda_8$ for $\eta, \phi, S^*$, etc.
This means a very slight form of SU(3)-symmetry breaking.

%-------------------------------------------------------------------------------
\section{OBE-Potentials in ESC08}                    
\label{sec:OBE}
The OBE-potentials in ESC08 are those contained already in ESC04 \cite{Rij04a,Rij04b},
and some new additional contributions.
The additions to the OBE-potentials w.r.t. the ESC04-models consist of the 
following elements: (i) extension of the baryon-baryon-meson vertex of the
axial-vector mesons ($J^{PC}=1^{++}$) by adding the  derivative coupling,
(ii) inclusion of the axial-vector mesons of the 2nd kind, 
having $J^{PC}=1^{+-}$. 
In paper I \cite{NRY12a} the potentials for non-strange meson exchange 
have been given. Here, we list the additions and the basic potentials for 
meson exchange with non-zero strangeness.

\subsection{Additions to the OBE-Potentials in ESC08}                    
\label{sec:OBEA}
The interaction Hamiltonian densities for the new couplings are\\

%\begin{enumerate}
\noindent a)\ Axial-vector-meson exchange ( $J^{PC}=1^{++}$, 1$^{st}$ kind):
\begin{equation}
 {\cal H}_A = g_A[\bar{\psi}\gamma_\mu\gamma_5\psi] \phi^\mu_A + \frac{if_A}{{\cal M}}
 [\bar{\psi}\gamma_5\psi]\ \partial_\mu\phi_A^\mu\ .
\label{eq:OBE.1}\end{equation}
In ESC04 the $g_A$-coupling was included, but not the derivative $f_A$-coupling.\\[0.2cm]
\noindent b)\ Axial-vector-meson exchange ( $J^{PC}=1^{+-}$, 2$^{nd}$ kind):
\begin{equation}
 {\cal H}_B = \frac{if_B}{m_B}
 [\bar{\psi}\sigma_{\mu\nu}\gamma_5\psi]\ \partial_\nu\phi_B^\mu\ .
\label{eq:OBE.2}\end{equation}
In ESC04 this coupling was not included. Like for the axial-vector mesons of the
1$^{st}$-kind we include a SU(3)-nonet with members $b_1(1235), h_1(1170), h_1(1380)$.
In the quark-model they are $Q\Bar{Q}(^1P_1)$-states.\\[0.2cm]

\noindent The inclusion of the gaussian form factors is discussed in previous papers
\cite{MRS89} etc. For the approximations made in deriving the potentials
from the relativistic Born-Approximation we refer also to \cite{MRS89}. Due to these
approximations the dependence on ${\bf q}^2$ is linearized and we write
\begin{equation}
 V_i({\bf k}^2,{\bf q}^2) = V_{ia}({\bf k}^2) + V_{ib}({\bf k}^2)\ {\bf q}^2\ ,
\label{eq:OBE.5}\end{equation}
where $i=1-8$. It turns out that to order ${\bf q}^2$ only $V_{1b} \neq 0$.
The additional OBE-potentials are obtained in the standard way, see \cite{NRS78,MRS89}.
We write the potential functions $V_i$ of (\ref{eq:LS.11}) in the form
\begin{equation}
 V_i({\bf k}^2,{\bf q}^2) = \sum_X\Omega^{(X)}_i({\bf k}^2)\cdot
 \Delta^{(X)}({\bf k}^2,m^2,\Lambda^2)\ ,
\label{eq:OBE.6}\end{equation}
where $m$ denotes the mass of the meson, $\Lambda$ the cut-off in the gaussian
form factor, and $X=S,A,B$, and $O$ (S= scalar, A= axial-vector, B= axial-vector, 
and O = diffractive/odderon). For the additions when X=S,B the propagator is
\begin{equation}
 \Delta^{(X)}({\bf k}^2,m^2,\Lambda^2) = (1- {\bf k}^2/U^2) e^{-{\bf k}^2/\Lambda^2}/
 ({\bf k}^2+m^2)\ , 
\label{eq:OBE.7}\end{equation}
and for the additions in the cases X=A,O the propagator is
\begin{equation}
 \Delta^{(O)}({\bf k}^2,m^2,\Lambda^2) = \frac{1}{{\cal M}^2} e^{-{\bf k}^2/4m_O^2}\ .
\label{eq:OBE.8a}\end{equation}
Here, ${\cal M}$ is a universal scaling mass, taken to be the proton mass, 
which we also use in the derivative couplings above, 
as well as in the $f_V$-coupling of the vector-mesons.
We note that the pole for the derivative coupling of the axial-vector exchange is canceled
because of a factor $k^\mu\left(g_{\mu\nu}-k_\mu k_\nu/m_A^2\right)$.

\onecolumngrid
\subsection{Meson-exchange with Non-zero Strangeness $(\Delta Y \neq 0)$}
For the non-strange mesons the mass differences at the vertices are neglected,
we take at the $YYM$- and the $NNM$-vertex the average hyperon and the average
nucleon mass respectively. This implies that we do not include contributions
to the Pauli-invariants $P_7$ and $P_8$.
 For vector-, and diffractive OBE-exchange we  
refer the reader to Ref.~\cite{MRS89}, where the contributions to the different
$\Omega^{(X)}_{i}$'s for baryon-baryon scattering are given in detail.
%-------------------------------------------------------------------------
These exchanges lead to the so-called 'exchange-potentials'.
For the invariants $O_1, \ldots , O_6$, the expressions analogous to those for 
the non-strange mesons given above apply. This with the amendments that (i) 
in momentum and configuration space there is a
complete symmetric appearance of $M_Y$ and $M_N$, (ii) in confuguration space
there appears the baryon-echange operator 
${\cal P}=-{\cal P}_x\ {\cal P}_\sigma$ operator, and (iii) for the
antisymmetric spin-orbit potential ${\cal P} \rightarrow {\cal P}_x$.
The details are given in Appendix~\ref{app:H}.
Therefore, the $\Omega^{(X)}_i$ for these potentials can be obtained
from those given in paper I Eqs. (\ref{eq1})-(\ref{eq:bxi1}), by replacing both 
$M_Y$ and $M_N$ by $(M_YM_N)^{1/2}$, and adding a (-)-sign.
Furthermore, in the case of using the Proca-formalism \cite{IZ80},
we get non-negligible contributions from the second part of the
vector-meson propagator $(k_\mu k_\nu/m^2)$ of the $K^*$ meson giving
\begin{equation}
 -V^{K^*}_i = V^{(V)}_i -\frac{(M_3-M_1)(M_4-M_2)}{m^2}\ V^{(S)}_i,
\label{eq:strange.1} \end{equation}
where in $V^{(S)}_i$ the vector-meson-couplings have to be used, and 
$M_Y$ and $M_N$ must be replaced by $(M_YM_N)^{1/2}$.                        
In (\ref{eq:strange.1}) $M_1=M_4=M_Y$ and $M_2=M_3=M_N$.
Then the additional terms are
\begin{equation}
 -V^{K_A}_i = V^{(A)}_i -\frac{(M_3-M_1)(M_4-M_2)}{m^2}\ V^{(P)}_i,
\label{eq:strange.2} \end{equation}
For the axial-vector meson $K_B$ there is no contribution from the
second-term in the propagator.

For the mesons with non-zero strangeness, $K, K^*, \kappa, K_A$ and $K_B$, 
 the mass differences at the vertices are not neglected,
we take into account at the $YNM$-vertices the differences between 
the average hyperon and the average nucleon mass. 
This implies that we do include contributions
to the Pauli-invariants $P_8$. There do not occur contributions to $P_7$.\\

%-------------------------------------------------------------------------
%-------------------------------------------------------------------------
\begin{enumerate}
 \item[(a)]   Pseudoscalar K-meson exchange:
      \begin{equation} \begin{array}{lcr}
       \Omega^{(P)}_{2} & = & -g^P_{13}g^P_{24}\left( \frac{{\bf k}^{2}}
           {12M_{Y}M_{N}} \right) 
  \ \ ,\ \ 
%\\[0.5cm]
       \Omega^{(P)}_{3}  =  -g^P_{13}g^P_{24}\left( \frac{1}
           {4M_{Y}M_{N}}  \right).
         \end{array}       \label{eq1}   \end{equation}
%--------------------------------------------------------------------------------------
 \item[(b)]   Vector-meson $K^*$-exchange:
     \begin{eqnarray}  
       \Omega^{(V)}_{1a}&=&
  \left\{g^V_{13}g^V_{24}\left( 1-\frac{{\bf k}^{2}}{8M_{Y}M_{N}}\right)
           -\left(g^V_{13}f^V_{24}+f^V_{13}g^V_{24}\right)
      \frac{{\bf k}^{2}}{4{\cal M}\sqrt{M_{Y}M_N}}
 \vphantom{\frac{A}{A}}\right. \nonumber\\ && \left. \vphantom{\frac{A}{A}}
           +f^V_{13}f^V_{24}\frac{{\bf k}^{4}}
           {16{\cal M}^{2}M_{Y}M_{N}}\right\} \nonumber\\         
% && \\
    \Omega^{(V)}_{1b}&=& g^V_{13}g^V_{24}\left(
    \frac{3}{2M_{Y}M_{N}}\right)\ \ ,\ \
%\\&& \\
  \Omega^{(V)}_{2} = -\frac{2}{3} {\bf k}^{2}\,\Omega^{(V)}_{3} \nonumber\\
% && \\
    \Omega^{(V)}_{3}&=& \left\{
       (g^V_{13}g^V_{24}+\left(g^V_{13}f^V_{24}+g^V_{24}f^V_{13}\right)
       \frac{\sqrt{M_{Y}M_N}}{{\cal M}})
       +f^V_{13}f^V_{24}\frac{M_YM_N}{{\cal M}^2}
       \left(1-\frac{{\bf k}^2}{M_YM_N} \right)\right\}
            /(4M_{Y}M_{N})              \nonumber\\
%--------------------------------------------------------------------------------------
% && \\
    \Omega^{(V)}_{4}&=&-\left\{12g^V_{13}g^V_{24}+8(g^V_{13}f^V_{24}+f^V_{13}g^V_{24})
           \frac{\sqrt{M_{Y}M_{N}}}{{\cal M}} 
%    \right.  \\&&\\ &&\hspace*{1cm}\left. 
     - f^V_{13}f^V_{24}\frac{3{\bf k}^{2}}{{\cal M}^{2}}\right\}
            /(8M_{Y}M_{N})              \nonumber\\
% && \\
       \Omega^{(V)}_{5}&=& -\left\{
           g^V_{13}g^V_{24}+4(g^V_{13}f^V_{24}+f^V_{13}g^V_{24})
           \frac{\sqrt{M_{Y}M_{N}}}{{\cal M}}  
%     \right. \\&& \\ &&\hspace*{1cm}\left. 
           +8f^V_{13}f^V_{24}\frac{M_{Y}M_{N}}{{\cal M}^{2}}\right\}
          /(16M_{Y}^{2}M_{N}^{2})        \nonumber\\
% &&  \\
       \Omega^{(V)}_{6}&=& -\left\{ (g^V_{13}f^V_{24}-f^V_{13}g^V_{24})
      \frac{1}{\sqrt{{\cal M}^{2}M_{Y}M_{N}}}\right\}.
 \label{eq2}\end{eqnarray}
%\end{subequations}
%\newpage
 \item[(c)]   Scalar-meson $\kappa$-exchange:  \hspace{2em}
      \begin{eqnarray} 
% \begin{array}{lcr@{\ }l}
      \Omega^{(S)}_{1} & = & -g^S_{13} g^S_{24}
        \left( 1+\frac{{\bf k}^{2}}{8M_{Y}M_{N}} -\frac{{\bf q}^2}{2M_YM_N}\right)
 \ ,\ 
%--------------------------------------------------------------------------------------
      \Omega^{(S)}_{4}=
      -g^S_{13} g^S_{24} \frac{1}{2M_{Y}M_{N}}
       \nonumber\\ &&\nonumber\\
      \Omega^{(S)}_{5} &=&
      g^S_{13} g^S_{24} \frac{1}{16M_{Y}^{2}M_{N}^{2} } 
 \ ,\ 
      \Omega^{(S)}_{6}= 0.
% \end{array}  
       \label{Eq:scal} \end{eqnarray}
%-------------------------------------------------------------------------
\item[(d)] Axial-vector $K_1$-exchange $J^{PC}=1^{++}$:
%-----------------------------------------------------------------------
%     From Feynman-graph, without special treatment/approximation: 
      \begin{eqnarray} 
      \Omega^{(A)}_{2} & = & -g^A_{13}g^A_{24}\left[
         1-\frac{2{\bf k}^2}{3M_YM_N}\right]
         +\left[\left(\vphantom{\frac{A}{A}} g_{13}^A f_{24}^A
         +f_{13}^A g_{24}^A \right) \frac{\sqrt{M_YM_N}}{\cal M}
         -f_{13}^A f_{24}^A \frac{{\bf k}^2}{2{\cal M}^2}\right]\
         \frac{{\bf k}^2}{6M_YM_N}
       \nonumber\\ && \nonumber\\
      \Omega^{(A)}_{2b} &=& 
        -g^A_{13}g^A_{24} \left(\frac{3}{2M_{Y}M_{N}}\right) 
        \nonumber\\ && \nonumber\\
      \Omega^{(A)}_{3}&=&
        -g^A_{13}g^A_{24} \left[\frac{1}{4M_{Y}M_{N}}\right]
         +\left[\left(\vphantom{\frac{A}{A}} g_{13}^A f_{24}^A
         +f_{13}^A g_{24}^A \right) \frac{\sqrt{M_YM_N}}{\cal M}
         -f_{13}^A f_{24}^A \frac{{\bf k}^2}{2{\cal M}^2}\right]\
         \frac{1}{2M_YM_N}
       \nonumber\\ && \nonumber\\
	\Omega^{(A)}_{4}  &=&
     -g^A_{13}g^A_{24}   \left[\frac{1}{2M_{Y}M_{N}}\right] 
      \ \ ,\ \
      \Omega^{(A)'}_{5}  = 
     -g^A_{13}g^A_{24}   \left[\frac{2}{M_{Y}M_{N}}\right] 
      \ \ ,\ \
      \Omega^{(A)}_{6} = 0.
         \label{eq:axi1} \end{eqnarray}
%-----------------------------------------------------------------------
%    B-field formalism and NLT-treatment:
Here, we used the B-field description with $\alpha_r=1$, see 
paper I, Appendix A.             
 The detailed treatment of the potential proportional to $P_5'$, i.e. 
 with $\Omega_5^{(A)'}$, is given in paper I, Appendix B.
%-----------------------------------------------------------------------
%-----------------------------------------------------------------------
\item[(e)] Axial-vector mesons with $J^{PC}=1^{+-}$: 
      \begin{eqnarray} 
       \Omega^{(B)}_{2} & = & +f^B_{13}f^B_{24}\frac{(M_N+M_Y)^2}{m_B^2}
       \left[\left(1-\frac{{\bf k}^2}{4M_YM_N}\right)+
       3\frac{({\bf q}^2+{\bf k}^2/4}{2M_YM_N}\right]
       \left( \frac{{\bf k}^{2}}{12M_{Y}M_{N}} \right) 
     \nonumber\\ 
%  \ \ ,\ \ 
       \Omega^{(B)}_{3} & = & +f^B_{13}f^B_{24}\frac{(M_N+M_Y)^2}{m_B^2}
           \left[\frac{1}{4M_{Y}M_{N}}\right].            
     \label{eq:bxi1} \end{eqnarray}
%--------------------------------------------------------------------------------------
 \item[(f)]   Diffractive-exchange (pomeron, $K_{2}(J=0)$): \\
         The pomeron carries no strangeness. Therefore, the contribution 
         to the potentials comes from the J=0-part of $K_2$-exchange \cite{note.diffr}.
         The $\Omega^{D}_{i}$ are the same as for scalar-meson-exchange
         Eq.(\ref{Eq:scal}), but with
         $\pm g_{13}^{S}g_{24}^{S}$ replaced by
         $\mp g_{13}^{D}g_{24}^{D}$, and except for the zero in the form factor.
         Since in ESC08-models $g_{NN a_2}=0$ there is no contribution to the
         exchange with non-zero strangeness.
%-------------------------------------------------------------------------
\item[(g)] Odderon-exchange:              
         Since the gluons carry no strangeness, there is no contribution
         to the potentials.
%-------------------------------------------------------------------------

\end{enumerate}

As in Ref.~\cite{MRS89} in the derivation of the expressions for $\Omega_i^{(X)}$, 
given above, $M_{Y}$ and $M_{N}$ denote the mean hyperon and nucleon
mass, respectively \begin{math} M_{Y}=(M_{1}+M_{3})/2 \end{math}
and \begin{math} M_{N}=(M_{2}+M_{4})/2 \end{math},
 and $m$ denotes the mass of the exchanged meson.
Moreover, the approximation                            
        \begin{math}
              1/ M^{2}_{N}+1/ M^{2}_{Y}\approx
              2/ M_{N}M_{Y},
        \end{math}
is used, which is rather good since the mass differences
between the baryons are not large.\\

%--------------------------------------------------------------------------------------
\subsection{One-Boson-Exchange Interactions in Configuration Space I}
\label{sect.IIIb}
In configuration space the BB-interactions are described by potentials
of the general form
\begin{eqnarray}
 V &=& \left\{\vphantom{\frac{A}{A}} V_C(r) + V_\sigma(r)
\mbox{\boldmath $\sigma$}_1\cdot\mbox{\boldmath $\sigma$}_2
 + V_T(r) S_{12} + V_{SO}(r) {\bf L}\cdot{\bf S} + V_Q(r) Q_{12}
 \right.\nonumber\\ && \left.
 + V_{ASO}(r)\ \frac{1}{2}(\mbox{\boldmath $\sigma$}_1-
  \mbox{\boldmath $\sigma$}_2)\cdot{\bf L}
 -\frac{1}{2}\left(\vphantom{\frac{A}{A}} 
\mbox{\boldmath $\nabla$}^2 \phi(r) + \phi(r) \mbox{\boldmath $\nabla$}^2\right)
\right\}\cdot P,
 \label{eq:3b.a}\end{eqnarray}
where
\begin{subequations}
 \label{eq:3b.b}
\begin{eqnarray}
 S_{12} &=& 3 (\mbox{\boldmath $\sigma$}_1\cdot\hat{r})
 (\mbox{\boldmath $\sigma$}_2\cdot\hat{r}) -
 (\mbox{\boldmath $\sigma$}_1\cdot\mbox{\boldmath $\sigma$}_2), \\
 Q_{12} &=& \frac{1}{2}\left[\vphantom{\frac{A}{A}} 
 (\mbox{\boldmath $\sigma$}_1\cdot{\bf L})(\mbox{\boldmath $\sigma$}_2\cdot{\bf L})
 +(\mbox{\boldmath $\sigma$}_2\cdot{\bf L})(\mbox{\boldmath $\sigma$}_1\cdot{\bf L})
 \right], \\
 \phi(r) &=& \phi_C(r) + \phi_\sigma(r) 
 \mbox{\boldmath $\sigma$}_1\cdot\mbox{\boldmath $\sigma$}_2, 
\end{eqnarray}
\end{subequations}
For the basic functions for the Fourier transforms with gaussian form factors,
we refer to Refs.~\cite{NRS78,MRS89}.                           
%---------------------------------------------------------------------------------
For the details of the Fourier transform for the potentials with $P_5'$, which 
occur in the case of the axial-vector mesons with $J^{PC}=1^{++}$, we refer 
to paper I, Appendix B.            

%---------------------------------------------------------------------------------
\noindent (a)\ Pseudoscalar-meson K-exchange:
\begin{eqnarray}
  V_{PS}(r) &=& \frac{m}{4\pi}\left[ g^P_{13}g^P_{24}\frac{m^2}{4M_YM_N}
 \left(\frac{1}{3}(\mbox{\boldmath $\sigma$}_1\cdot\mbox{\boldmath $\sigma$}_2)\
 \phi_C^1 + S_{12} \phi_T^0\right)\right]\ {\cal P}.
 \label{eq:3b.1}\end{eqnarray}
%---------------------------------------------------------------------------------
\noindent (b)\ Vector-meson $K^*$-exchange:          
\begin{eqnarray}
&& V_{V}(r) = \frac{m}{4\pi}\left[\left\{ 
 g^V_{13}g^V_{24}\left[ \phi_C^0 +
 \frac{m^2}{2M_YM_N} \phi_C^1 -\frac{3}{4M_YM_N}\left(\mbox{\boldmath $\nabla$}^2 \phi_C^0 
 + \phi_C^0 \mbox{\boldmath $\nabla$}^2\right) \right]
\right.\right.\nonumber\\ && \left.\left.  \hspace{0cm} 
 +\left(g^V_{13}f^V_{24}+f^V_{13}g^V_{24}\right)
 \frac{m^2}{ 4{\cal M}\sqrt{M_YM_N} }\ \phi_C^1 
 + f^V_{13}f^V_{24} \frac{m^4}{16{\cal M}^2 M_YM_N} \phi_C^2\right\}
  \right.\nonumber\\ && \left.  \hspace{0cm} 
 +\frac{m^2}{6M_YM_N}\left\{\left[ g^V_{13}g^V_{24}
 +\left(g^V_{13}f^V_{13}+g^V_{24}f^V_{13}\right)
 \frac{\sqrt{M_YM_N}}{{\cal M}}
 +f^V_{13}f^V_{24}\frac{M_YM_N}{{\cal M}^2}\right] \phi_C^1 
 +f^V_{13} f^V_{24}\frac{m^2}{8{\cal M}^2} \phi_C^2\right\}
 (\mbox{\boldmath $\sigma$}_1\cdot\mbox{\boldmath $\sigma$}_2)\
  \right.\nonumber\\ && \left.  \hspace{0cm} 
 -\frac{m^2}{4M_YM_N}\left\{\left[ g^V_{13}g^V_{24}+
 \left(g^V_{13}f^V_{24}+g^V_{24}f^V_{13}\right)\frac{\sqrt{M_YM_N}}{{\cal M}}\right) \phi_T^0 
 +f^V_{13} f^V_{24}\frac{m^2}{8{\cal M}^2} \phi_T^1\right\} S_{12}
  \right.\nonumber\\ && \left.  \hspace{0cm} 
 -\frac{m^2}{M_YM_N}\left\{\left[ \frac{3}{2}g^V_{13}g^V_{24}
 +\left(g^V_{13}f^V_{24}+f^V_{13}g^V_{24}\right)
 \frac{\sqrt{M_YM_N}}{{\cal M}}\right] \phi_{SO}^0 
 +\frac{3}{8}f^V_{13} f^V_{24}\frac{m^2}{{\cal M}^2} \phi_{SO}^1\right\} {\bf L}\cdot{\bf S}
  \right.\nonumber\\ && \left.  \hspace{0cm} 
 +\frac{m^4}{16M_Y^2M_N^2}\left\{\left[ g^V_{13}g^V_{24}
 +4\left(g^V_{13}f^V_{24}+f^V_{13}g^V_{24}\right)
 \frac{\sqrt{M_YM_N}}{{\cal M}} 
 +8f^V_{13}f^V_{24}\frac{M_YM_N}{{\cal M}^2}\right]\right\} 
  \cdot\right.\nonumber\\ && \left.  \hspace{0cm} \times
\frac{3}{(mr)^2} \phi_T^0 Q_{12}
 +\frac{m^2}{M_YM_N}\left\{ 
  \left(g^V_{13}f^V_{24}-f^V_{13}g^V_{24}\right)\frac{\sqrt{M_YM_N}}{{\cal M}}\ \phi_{SO}^0
 \right\}\cdot\frac{1}{2}\left(
 \mbox{\boldmath $\sigma$}_1-\mbox{\boldmath $\sigma$}_2\right)\cdot{\bf L}\
 {\cal P}_\sigma\right]\ {\cal P}.
 \label{eq:3b.2}\end{eqnarray}

%---------------------------------------------------------------------------------
\noindent (c)\ Scalar-meson $\kappa$-exchange:          
\begin{eqnarray}
 V_{S}(r) &=& -\frac{m}{4\pi}\left[ g^S_{13}g^S_{24}\left\{\left[ \phi_C^0 
 -\frac{m^2}{4M_YM_N} \phi_C^1\right] + \frac{m^2}{2M_YM_N} \phi_{SO}^0\ {\bf L}\cdot{\bf S}
 +\frac{m^4}{16M_Y^2M_N^2}
 \cdot\right.\right.\nonumber\\ && \left.\left. \times
\frac{3}{(mr)^2} \phi_T^0 Q_{12} 
  +\frac{1}{4M_YM_N}\left(\mbox{\boldmath $\nabla$}^2 \phi_C^0 
 + \phi_C^0 \mbox{\boldmath $\nabla$}^2\right) \right\}\right]\ {\cal P}.
 \label{eq:3b.3}\end{eqnarray}
%---------------------------------------------------------------------------------
\noindent (d)\ Axial-vector $K_1$-meson exchange $J^{PC}=1^{++}$:
\begin{eqnarray}
&& V_{A}(r) = -\frac{m}{4\pi}\left[ 
 \left\{ g^A_{13}g^A_{24}\left(\phi_C^0 +\frac{2m^2}{3M_YM_N} \phi_C^1\right)
 +\frac{m^2}{6M_YM_N}\left(g^A_{13}f^A_{24}
 +f^A_{13}g^A_{24}\right)\frac{\sqrt{M_YM_N}}{{\cal M}}\phi_C^1
\right.\right.\nonumber\\ && \left.\left.
 +f^A_{13}f^A_{24}\frac{m^4}{12M_YM_N{\cal M}^2}\phi_C^2\right\}
 (\mbox{\boldmath $\sigma$}_1\cdot\mbox{\boldmath $\sigma$}_2)
  -\frac{3}{4M_YM_N}\left(\mbox{\boldmath $\nabla$}^2 \phi_C^0 
 + \phi_C^0 \mbox{\boldmath $\nabla$}^2\right) 
 (\mbox{\boldmath $\sigma$}_1\cdot\mbox{\boldmath $\sigma$}_2)
 \right.\nonumber\\ && \left. 
 - \frac{m^2}{4M_YM_N}\left\{\left[g^A_{13}g^A_{24}-2\left(g^A_{13}f^A_{24}
 +f^A_{13}g^A_{24}\right)\frac{\sqrt{M_YM_N}}{{\cal M}}\right] \phi_T^0
 -f^A_{13}f^A_{24}\frac{m^2}{2{\cal M}^2} \phi_T^1\right\} S_{12}
 \right.\nonumber\\ && \left. 
 +\frac{m^2}{2M_YM_N}g^A_{13}g^A_{24} \phi_{SO}^0\ {\bf L}\cdot{\bf S}
 \right]\ {\cal P}.
 \label{eq:3b.4}\end{eqnarray}
%---------------------------------------------------------------------------------
\noindent (e)\ Axial-vector-meson exchange $J^{PC}=1^{+-}$:
\begin{eqnarray}
 V_{B}(r) &=& -\frac{m}{4\pi}\frac{(M_N+M_Y)^2}{m^2}\left[ 
 f^B_{13}f^B_{24}\left\{\frac{m^2}{12M_YM_N}\left(\phi_C^1+\frac{m^2}{4M_YM_N} 
 \phi_C^2\right)
 \right.\right.\nonumber\\ && \left.\left. 
  -\frac{m^2}{8M_YM_N}\left(\mbox{\boldmath $\nabla$}^2 \phi_C^1 
 + \phi_C^1 \mbox{\boldmath $\nabla$}^2\right) 
 +\left[\frac{m^2}{4M_YM_N}\right]\phi_T^0\ S_{12}\right\}\right]\ {\cal P}.
 \label{eq:3b.5}\end{eqnarray}
%---------------------------------------------------------------------------------
\noindent (f)\ Diffractive-exchange:           
Since in the ESC08-model the diffractive pomeron and odderon exchanges are
SU(3) singlets there are no contribution to $S \neq 0$-exchange potentials.

\noindent Above, in Eq.'s~(\ref{eq:3b.1}-\ref{eq:3b.5}) the exchange operator is
defined as 
\begin{equation}
 {\cal P} = -{\cal P}_x {\cal P}_\sigma, 
 \label{eq:3b.7}\end{equation}
where ${\cal P}_x$ and ${\cal P}_\sigma$ are the space- and 
spin-exchange operators respectively.
The extra $(-{\cal P}_\sigma)$-operator in (\ref{eq:3b.2}) for the antisymmetric 
spin-orbit potential is explained in 
Appendix~\ref{app:H}. We note that $-{\cal P}_\sigma{\cal P} = {\cal P}_x$, 
which is well defined for the coupled singlet-triplet systems.

%--------------------------------------------------------------------------------------
%\onecolumngrid
\subsection{One-Boson-Exchange Interactions in Configuration Space II}
\label{sect.IIIc}
Here we give the extra potentials due to the 
zero's in the scalar and axial-vector form factors.
\begin{enumerate}
%----------------------------------------------------------------------

\item[a)] Again, for $X=V,D$ we refer to the configuration space potentials 
in Ref.~\cite{MRS89}. For $X=S$ we give here the additional terms w.r.t. those 
in \cite{MRS89}, which are due to the zero in the scalar form factor. They are

\begin{eqnarray}
 &&  \Delta V_{S}(r) = - \frac{m}{4\pi}\ \frac{m^2}{U^2}\
 \left[ g^S_{13} g^S_{24}\left\{
 \left[\phi^1_C - \frac{m^2}{4M_YM_N} \phi^2_C\right]
%% \right.\right. \nonumber\\ && \left.\left.
+\frac{m^2}{2M_YM_N}\phi^1_{SO}\ {\bf L}\cdot{\bf S}
 \right.\right. \nonumber\\ && \left.\left.
+\frac{m^4}{16M_Y^2M_N^2}\phi^1_T\ Q_{12}  
  \right\}\right]\ {\cal P}.
 \label{eq:3.15}\end{eqnarray}

\item[b)] For the axial-vector mesons, the configuration space potential 
 corresponding to (\ref{eq:axi1}) is             
\begin{eqnarray}
 &&  V_{A}^{(1)}(r) = - \frac{g_{A}^{2}}{4\pi}\ m  \left[
 \phi^{0}_{C}\ (\mbox{\boldmath $\sigma$}_1\cdot\mbox{\boldmath $\sigma$}_2) 
  -\frac{1}{12M_YM_N}
  \left( \nabla^{2} \phi^{0}_{C}+\phi^{0}_{C}\nabla^{2}\right)
 (\mbox{\boldmath $\sigma$}_1\cdot\mbox{\boldmath $\sigma$}_2) 
 \right. \nonumber \\ && \nonumber \\ & & \left. \hspace*{1.4cm}
   + \frac{3m^{2}}{4M_YM_N}\ \phi^{0}_{T}\ S_{12}
 +\frac{m^{2}}{2M_YM_N}\ \phi^{0}_{SO}\ {\bf L}\cdot{\bf S}
  \right]\ {\cal P}.
 \label{eq:3.16}\end{eqnarray}
%----------------------------------------------------------------------
The extra contribution to the potentials coming from the zero in the axial-vector
meson form factor are obtained from the expression (\ref{eq:3.16}) by making 
substitutions as follows
\begin{eqnarray}
   \Delta V_{A}^{(1)}(r) &=&  V_{A}^{(1)}\left(\phi_C^0 \rightarrow \phi_C^1,
 \phi_T^0 \rightarrow \phi_T^1, \phi_{SO}^0 \rightarrow \phi_{SO}^1\right)
 \cdot\frac{m^2}{U^2}\ .
\label{eq:3.17b}\end{eqnarray}
Note that we do not include the similar $\Delta V_A^{(2)}(r)$ since they involve
${\bf k}^4$-terms in momentum-space. 

\end{enumerate}

%\end{widetext}
\twocolumngrid

\subsection{PS-PS-exchange Interactions in Configuration Space}                       
\label{sect.IIId}
In Fig.'s 2 and 3 of paper I, the included two-meson exchange 
graphs are shown schematically. Explicit expressions for  
$K^{irr}(BW)$ and $K^{irr}(TMO)$ were derived \cite{Rij91}, where also the 
terminology BW and TMO is explained.
The TPS-potentials for nucleon-nucleon have been given in detail in \cite{RS96ab}.
The generalization to baryon-baryon is similar to that for the OBE-potentials.
So, we substitute $M \rightarrow \sqrt{M_YM_N}$, and include all PS-PS 
possibilities with coupling constants as in the OBE-potentials. 
As compared to nucleon-nucleon in \cite{RS96ab} we have here in addition 
the potentials with double K-exchange.  The masses
are the physical pseudo-scalar meson masses. For the intermediate two-baryon
states we take into account the effects of the different thresholds.
We have not included uncorrelated PS-vector, PS-scalar, or PS-diffractive 
exchange. This because the range of these potentials is similar to 
that of the vector-,scalar-,and axial-vector-potentials. Moreover, for 
potentially large potentials, in particularly those with scalar mesons involved,
there will be very strong cancellations between the planar- and crossed-box
contributions.

\subsection{MPE-exchange Interactions}
\label{sect.IIIe}
In Fig.~4 of paper I the pair graphs are shown.
In this work we include only the one-pair graphs. The argument for neglecting 
the two-pair graph is to avoid some 'double-counting'. Viewing the pair-vertex 
as containing heavy-meson exchange means that the contributions from $\rho(750)$
and $\epsilon=f_0(760)$ to the two-pair graphs is already accounted for by 
our treatment of the broad $\rho$ and $\epsilon$ OBE-potential.
The MPE-potentials for nucleon-nucleon have been given in Ref.~\cite{RS96ab}.
The generalization to baryon-baryon is similar to that for the TPS-potentials.
For the intermediate two-baryon
states we neglect the effects of the different two-baryon thresholds. The inclusion of these,
although in principle possible, would complicate the computation of the 
potentials considerably and the influence is not expected to be significant.
The generalization of the pair-couplings to baryon-baryon is described in 
Ref.~\cite{Rij04b}, section III.
Also here in $N\!N$, we have in addition to  \cite{RS96ab}  
included the pair-potentials with KK-, KK*-, and K$\kappa$-exchange.
The convention for the MPE coupling constants is the same as in Ref.~\cite{RS96ab}.

%-----------------------------------------------------------------------------
\subsection{Meson-Pair Potentials, Axial-Pairs (2$^{nd}$-kind, $J^{PC}=1^{+-}$) } 
\label{sec:MPE}
Recently we have completed the $1/M, 1/M^2$-corrections to the adiabatic approximation
for the pair-potentials. The main reason is the need for a careful evaluation of
the anti-symmetric spin-orbit terms for $\Lambda N$, in particular for 
pair-interactions involving strangeness-exchange like $\pi-K, \pi-K^*$ etc.
From this evaluation new contributions emerged, in particular for the 
axial pair-interactions $J^{PC}=1^{++},1^{+-}$, leading to a substantial improvement
w.r.t. experimental spin-orbit splittings \cite{HT06}. 
In our fitting procedure
for the YN-data the spin-orbit plays no role, therefore we will report on the
details of the new spin-orbit terms in a separate paper \cite{HMYR08}.
However, also new $1/M$-corrections for the spin-spin and tensor potentials
were obtained for the axial-pair interaction of the 2nd kind, 
i.e. $J^{PC}=1^{+-}$. These are relevant for the fits presented in this paper,
and will be given in this section. Below we give the full one-pair exchange
potential as used at present, because it has not been published before. In the
ESC04-models only the leading, i.e. the $(1/M)^0$-terms, were used.
For the derivation of the soft-core pair-interactions we refer the reader to 
\cite{RS96ab}. Below we report on this derivation for the axial-pair terms
of the 2nd kind. The used pair-interaction Hamiltonian for 
e.g. the $(\pi\omega)$-pair is
\begin{equation}
 {\cal H}_B = g_{(\pi\omega)}\bar{\psi}\gamma_5\sigma_{\mu\nu}
 \mbox{\boldmath $\tau$}\psi\cdot
 \partial^\nu\left(\mbox{\boldmath $\pi$}\ \phi_\omega^\mu\right)/(m_\pi{\cal M})\ ,
 \label{eq:MPE.a}\end{equation}
which gives the $BBm_1m_2$-vertex
\begin{equation}
 \bar{u}({\bf p}') \Gamma_B^{(2)}\ u({\bf p}) =  i\frac{g_{(\pi\omega)_1}}{m_\pi{\cal M}}       
 \left[\vphantom{\frac{A}{A}}(\pm\omega_1\pm\omega_2)\
 \mbox{\boldmath $\sigma$}\cdot\mbox{\boldmath $\omega$}+
 \mbox{\boldmath $\sigma$}\cdot{\bf k}\ \omega^0\right]\ .  
 \label{eq:MPE.2}\end{equation}
The full SU(3)-structure is given in \cite{Rij04b}, section IIIA. It is assumed 
that this pair-coupling is dominated by the SU(3)-octet symmetric coupling, and
is given by the  $SU(3)$-octet symmetric couplings Hamiltonian in terms of SU(2)-isospin
invariants and SU(3) isoscalar-factors:

\begin{eqnarray}
&& {\cal H}_{B_8VP} =
 \frac{g_{B_8VP}}{\sqrt{6}}\left\{\vphantom{\frac{A}{A}}\right. 
 \frac{1}{2}\left[\left({\bf B}_1^\mu\cdot\mbox{\boldmath $\rho$}_\mu\right) \eta_8 +
 \left({\bf B}_1^\mu\cdot\mbox{\boldmath $\pi$}_\mu\right) \phi_8 \right] 
  \nonumber \\ && \hspace{5mm}
 +\frac{\sqrt{3}}{4}\left[{\bf B}_1\cdot(K^{*\dagger}\mbox{\boldmath $\tau$}K)
 + h.c. \right] 
  \nonumber \\ && \hspace{5mm}
 +\frac{\sqrt{3}}{4}\left[(K_1^\dagger\mbox{\boldmath $\tau$} K^*)\cdot
 \mbox{\boldmath $\pi$}
 +(K_1^\dagger\mbox{\boldmath $\tau$} K)\cdot\mbox{\boldmath $\rho$} + h.c. \right] 
  \nonumber \\ && \hspace{5mm}
 -\frac{1}{4}\left[(K_1^\dagger\cdot K^*) \eta_8 + (K_1^\dagger\cdot K) \phi_8 
 + h.c. \right] 
  \nonumber \\ && \hspace{5mm}
 +\frac{1}{2}H^0\left[\mbox{\boldmath $\rho$}\cdot\mbox{\boldmath $\pi$} 
 -\frac{1}{2}\left(K^{*\dagger}\cdot K+ K^\dagger\cdot K^* \right)
 -\phi_8\eta_8 \right]
 \left.\vphantom{\frac{A}{A}}\right\} \nonumber\\                
 \label{eq:MPE.1b}\end{eqnarray}
Here, ${\bf B}_1 \sim \left[\bar{\psi}\gamma_5\bm{\tau}\sigma_{\mu\nu}\psi\right]$ etc.
See for a definition of the octet-fields $\eta_8,\phi_8$ in terms of the physical mesons 
\cite{Rij04b}. 
From the pair-interaction Hamiltonian (\ref{eq:MPE.1b}) 
one can easily read off the different meson-pairs
that occur from the $J^{PC}=1^{+-}$-vertex. 
In Appendix~\ref{app:MPE.ax.2nd} we give the 
explicit potentials generated by the pair-interaction (\ref{eq:MPE.1b}).
%---------------------------------------------------------------------------------

%----------------------------------------------------------------------------
\section{ Short-range Phenomenology}                               
\label{sec:5} 
It is well known that the most extensive study of the baryon-baryon interactions
using meson-exchange has difficulties to achieve sufficiently repulsive short-range
interactions in two channels. Namely, (i) the $\Sigma^+ p(I=3/2,^3S_1)$- and (ii) the
$\Sigma N(I=1/2,^1S_0)$-channel. The short-range repulsion in baryon-baryon comes
in principle from two sources: (a) meson- and multi-gluon-exchange, and (b) the 
occurrence of forbidden states by the Pauli-principle, henceforth referred to as
Pauli-repulsion or Quark-core. As for (a) in the ESC-model \cite{Rij04a,Rij04b} the short-range
repulsion comes from vector-meson exchange and  Pomeron/Odderon-exchange (i.e. multi-gluon).
The possibility of mechanism (b) has been explored in the Quark-Cluster model. See the
reviews \cite{Oka00,Fuj07}.

\noindent Analyzing the Pauli-repulsion in terms of the $SU_f(3)$-irreps
we find that the "forbidden" $L=0$ BB-states, which are classified in $SU_{fs}(6)$
by the $[51]$-irrep, indeed occur dominantly in the $SU_f(3)$-irreps $\{10\}$ and
$\{8_s\}$. These $SU(3)$-irreps  dominate the $\Sigma^+ p(I=3/2,^3S_1)$- and the
$\Sigma N(I=1/2,^1S_0)$-states respectively. These facts open the possibility to
incorporate the exceptionally strong Pauli-repulsion for these states by enhancing
the Pomeron coupling in the ESC-approach to baryon-baryon. For the other BB-states
the $[51]$-irrep is present also, but roughly with an equal weight as the $[33]$-irrep.
Only in a few S=-2 channels,e.g. $\Xi N(I=1,S=0)$
there is a stronger presence of the irrep $[51]$. Therefore a slightly 
moderated $SU_f(3)$-singlet
Pomeron-exchange can effectively take care of this Quark-core phenomenologically, 
together with multi-gluon-exchange effects.

% -----------------------------------------------------------------------
\subsection{Relation $SU_f(3)$-irreps and $SU_{fs}(6)$-irreps Classification YN-states}
\label{sec:0}  
In Table~\ref{tab.1} the $SU_f(3)$-contents of the NN and YN 
states are shown. 
%---------------------------------------------------------------------
\begin{table}
\renewcommand{\arraystretch}{2.0}
\begin{center} 
\caption{ $SU(3)_f$-contents of the various potentials\\
 \hspace*{3cm} on the isospin basis.} 
\begin{tabular}{ccl} \hline \hline
\multicolumn{3}{c}{Space-spin antisymmetric states $^{1}S_{0},
                    \ ^{3}P,\ ^{1}D_{2},...$} \\
\hline
$ NN \rightarrow NN $  & $ I=1 $
                    & $V_{NN}(I=1) = V_{27}$    \\ \hline
$ \Lambda N \rightarrow \Lambda N $  &
                  & $ V_{\Lambda\Lambda} \left( I = \frac{1}{2} \right) =
                      \left( 9V_{27}+V_{8_{s}} \right) /10 $        \\
$ \Lambda N \rightarrow \Sigma N  $ & $ I= \frac{1}{2} $
                  & $ V_{\Lambda\Sigma} \left( I= \frac{1}{2} \right) =
                      \left( -3V_{27}+3V_{8_{s}} \right) /10 $       \\
$ \Sigma N \rightarrow \Sigma N $  &
                  & $ V_{\Sigma\Sigma} \left( I = \frac{1}{2} \right) =
                      \left( V_{27}+9V_{8_{s}} \right) /10 $         \\ \hline
$ \Sigma N \rightarrow \Sigma N $ & $ I= \frac{3}{2} $
                  & $ V_{\Sigma\Sigma} \left( I = \frac{3}{2} \right) =
                      V_{27} $                                    \\
\hline \hline
\multicolumn{3}{c}{Space-spin symmetric states $^{3}S_{1},\ ^{1}P_{1},
                   \ ^{3}D,...$} \\
\hline
$ NN \rightarrow NN $  & $ I=0 $
                & $ V_{NN}(I=0) = V_{10^{\star}}  $ \\ \hline
$ \Lambda N \rightarrow \Lambda N $  &
                  & $ V_{\Lambda\Lambda\;} \left( I= \frac{1}{2} \right) =
                      \left( V_{10^{\star}}+V_{8_{a}} \right) /2   $ \\
$ \Lambda N \rightarrow \Sigma N $  & $ I = \frac{1}{2} $
                  & $ V_{\Lambda\Sigma} \left( I = \frac{1}{2} \right) =
                      \left( V_{10^{\star}}-V_{8_{a}} \right) /2 $   \\
$ \Sigma N \rightarrow \Sigma N $  &
                  & $ V_{\Sigma\Sigma} \left( I = \frac{1}{2} \right) =
                      \left( V_{10^{\star}}+V_{8_{a}} \right) /2  $  \\ \hline
$ \Sigma N \rightarrow \Sigma N $  & $ I = \frac{3}{2}  $
                  & $ V_{\Sigma\Sigma} \left( I = \frac{3}{2} \right) =
                      V_{10} $                                   \\
\hline \hline
\end{tabular}
\end{center}
\renewcommand{\arraystretch}{1.0}
\label{tab.1} 
\end{table}
%---------------------------------------------------------------------
In Table~\ref{tab.3} we show the the weights of the $SU(6)_{sf}$-irreps.
These are taken from \cite{Oka00} Table I, 
where the $SU(6)_{fs}$-classifications are given.
%---------------------------------------------------------------------
 \begin{table}
\renewcommand{\arraystretch}{2.0}
\begin{center}
\caption {$SU(6)_{fs}$-contents of the various potentials\\
 on the spin,isospin basis. }
\begin{tabular}{ccl} \hline \hline
   & $(S,I)$ & $V = a V_{[51]} + b V_{[33]}$ \\ \hline
$ NN \rightarrow NN $  & $ (0,1) $
                & $ V_{NN} = \frac{4}{9}V_{[51]}+\frac{5}{9}V_{[33]}  $ \\
$ NN \rightarrow NN $  & $ (1,0) $
                & $ V_{NN} = \frac{4}{9}V_{[51]}+\frac{5}{9}V_{[33]}  $ \\
\hline 
$ \Lambda N \rightarrow \Lambda N $  & $(0,1/2)$ &
                   $ V_{\Lambda\Lambda\;} = \frac{1}{2}V_{[51]}+\frac{1}{2}V_{[33]}$ \\
$ \Lambda N \rightarrow \Lambda N $  & $(1,1/2)$ &
                   $ V_{\Lambda\Lambda\;} = \frac{1}{2}V_{[51]}+\frac{1}{2}V_{[33]}$ \\
\hline 
$ \Sigma N \rightarrow \Sigma N $  & $(0,1/2)$ &
                   $ V_{\Sigma\Sigma}  = \frac{17}{18}V_{[51]}+\frac{1}{18}V_{[33]}$\\
$ \Sigma N \rightarrow \Sigma N $  & $(1,1/2)$ &
                   $ V_{\Sigma\Sigma}  = \frac{1}{2}V_{[51]}+\frac{1}{2}V_{[33]}$\\
$ \Sigma N \rightarrow \Sigma N $  & $(0,3/2)$ &
                   $ V_{\Sigma\Sigma}  = \frac{4}{9}V_{[51]}+\frac{5}{9}V_{[33]}$\\
$ \Sigma N \rightarrow \Sigma N $  & $(1,3/2)$ &
                   $ V_{\Sigma\Sigma}  = \frac{8}{9}V_{[51]}+\frac{1}{9}V_{[33]}$\\
\hline \hline
\end{tabular}
\end{center}
\renewcommand{\arraystretch}{1.0}
\label{tab.3} \end{table}
%---------------------------------------------------------------------
Analyzing now the $(\Lambda N, \Sigma N)$-system for $(S=0,I=1/2)$ we find from these 
tables
\begin{eqnarray}
\left(\begin{array}{c} 
V_{\Lambda N, \Lambda N} \\ V_{\Sigma N, \Sigma N} \end{array}\right) &=&
\left( \begin{array}{cc}
\frac{1}{2} & \frac{1}{2} \\ \frac{17}{18} & \frac{1}{18} \end{array}\right)
\left(\begin{array}{c} V_{[51]} \\ V_{[33]} \end{array}\right) \nonumber\\ 
 &=& \left( \begin{array}{cc}
\frac{9}{10} & \frac{1}{10} \\ \frac{1}{10} & \frac{9}{10} \end{array}\right)
\left(\begin{array}{c} V_{\{27\}} \\ V_{\{8_s\}} \end{array}\right)\ .          
\label{eq:2.1}\end{eqnarray}
\noindent 1.\ From (\ref{eq:2.1}) we obtain by simple matrix operations the relation between 
the $SU(6)_{fs}$-irreps and the $SU(3)_f$-irrreps, which read 
\begin{eqnarray}
\left(\begin{array}{c} V_{\{27\}} \\ V_{\{8_s\}} \end{array}\right) &=&         
\left( \begin{array}{cc}
\frac{4}{9} & \frac{5}{9} \\ 1 & 0 \end{array}\right)
\left(\begin{array}{c} V_{[51]} \\ V_{[33]} \end{array}\right)\ .          
\label{eq:2.2}\end{eqnarray}
\noindent 2.\ Also, we can read off from the tables the following relations
\begin{subequations}
\label{eq:2.3}
\begin{eqnarray}
 \hspace{-5mm}
 V_{NN}(I=1,S=0) &=& \frac{4}{9}V_{[51]}+\frac{5}{9}V_{[33]} = V_{\{27\}}, \\
 \hspace{-5mm}
 V_{NN}(I=0,S=1) &=& \frac{4}{9}V_{[51]}+\frac{5}{9}V_{[33]} = V_{\{10^*\}}, \\
 \hspace{-5mm}
 V_{\Lambda N}(I=\frac{1}{2},S=1) &=& \frac{1}{2}V_{[51]}+\frac{1}{2}V_{[33]} \nonumber\\
                         &=&  \frac{1}{2}V_{\{10^*\}}+\frac{1}{2}V_{\{8_a\}}\ .
\end{eqnarray}
\end{subequations}
From these equations we can solve the $SU(3)_f$-irreps $\{27\}, \{10^*\}$, and 
$\{8_a\}$ in terms of the $SU(6)_{fs}$-irreps.\\
Listing the results we now have
\begin{subequations}
\label{eq:2.4}
\begin{eqnarray}
 V_{\{27\}} &=& \frac{4}{9}V_{[51]}+\frac{5}{9}V_{[33]}\ , \\
 V_{\{10^*\}} &=& \frac{4}{9}V_{[51]}+\frac{5}{9}V_{[33]}\ , \\
 V_{\{10\}} &=& \frac{8}{9}V_{[51]}+\frac{1}{9}V_{[33]}\ , \\
 V_{\{8_a\}} &=& \frac{5}{9}V_{[51]}+\frac{4}{9}V_{[33]}\ , \\
 V_{\{8_s\}} &=& V_{[51]}\ . 
\end{eqnarray}
\end{subequations}
We see from these results that the $[51]$-irrep has a large weight in the
$\{10\}$- and $\{8_s\}$-irrep, which gives an argument for the presence of
a strong Pauli-repulsion in these $SU(3)_f$-irreps.\\

\noindent According to the study of the wide range of meson-exchange models in the
last decade using the ESC-approach, as a working hypothesis, we assume that
the apparent lack of an exceptionally strong repulsion in the ESC-model 
for the states in the $SU(3)_f$-irreps $\{10\}$ and $\{8_s\}$ 
cannot be cured by meson-exchange. The inclusion of this possible 
"forbidden state" effect can be done phenomenologically in the ESC-approach 
by  making an effective Pomeron potential as the sum of 'pure' Pomeron exchange
and of a Pomeron-like representation of the Pauli-repulsion. As a consequence
the effective Pomeron potential gets quite stronger in the $SU(3)_f$-irreps 
 $\{10\}$ and $\{8_s\}$. This way we incorporate the Pauli repulsion effect 
in the ESC-approach in this paper.
%---------------------------------------------------------------------
%----------------------------------------------------------------------------
\subsection{ Parametrization Quark-core effects}                      
\label{sec:5b} 
\noindent 1.\ Linear method: we split the repulsive 
short-range Pomeron-like NN potential as follows:
\begin{eqnarray}
 V_{PNN} &=& (1-a_{PB}) V_{PNN} + a_{PB} V_{PNN} \nonumber\\
 &\equiv& V_{NN}(POM)+V_{NN}(PB)
\label{eq:51.1}\end{eqnarray}
where $V_{NN}(POM)$ represents the genuine Pomeron and $V_{NN}(PB)$ the structural 
effects of the Quark-core forbidden [51]-configuration, 
i.e. a Pauli-blocking (PB) effect. So $a_{PB}$ denotes the Quark-core fraction 
of the total Pomeron-like potential. A similar relation holds for all BB channels.
\begin{eqnarray}
 V_{PBB} &=& (1-a_{PB}) V_{PBB} + a_{PB} V_{PBB} \nonumber\\
 &\equiv& V_{BB}(POM)+V_{BB}(PB)
\label{eq:51.2}\end{eqnarray}
Since the Pomeron is a unitary-singlet its contribution is the same
for all BB-channels (apart from some small baryon mass breaking effects). 
\noindent The PB-effect for the BB-channels is assumed
to be proportional to the relative weight of the forbidden
 [51]-configuration compared to its weight in NN 
\begin{equation}
 V_{BB}(PB) = \left(w_{BB}[51]/w_{NN}[51]\right)\cdot V_{NN}(PB)
\label{eq:51.3}\end{equation}
Then we have
\begin{equation}
 V_{PBB} = (1-a_{PB}) V_{PNN} + a_{PB} 
 \left(\frac{w_{BB}[51]}{w_{NN}[51]}\right)\cdot V_{PNN}
\label{eq:51.4}\end{equation}
For example, in the SU(3)-irrep $\{10\}$, e.g. the 
 $\Sigma^+p (^3S_1,T=3/2)$-channel, one has $w_{BB}[51]=8/9=2 w_{NN}[51]$
and therefore $V_{10}(PB)= 2 V_{NN}(PB)$. Consequently, the total short-range
repulsive potential, i.e. the 'effective pomeron', becomes
 $V_{10}=(1-a_{PB})V_{PNN}+2a_{PB} V_{PNN}$ $= (1+a_{PB}) V_{PNN}$.
In Table \ref{tab.4} we give the factors for the various S=0 and S=-1 BB channels
as well as the results in the ESC08c model.

 \begin{table}
\renewcommand{\arraystretch}{2.0}
\begin{center}
\caption {Effective Pomeron+PB contribution\\
  \hspace*{2.5cm} on the spin,isospin basis. }
\begin{tabular}{cccc} \hline \hline
   & $(S,I)$ & $V_{PBB}/V_{PNN}$ & $ESC08c$ \\ \hline
$ NN \rightarrow NN $  & $ (0,1) $
                & $ 1 $ & $ 1.000 $\\
$ NN \rightarrow NN $  & $ (1,0) $
                & $ 1 $ & $ 1.000 $ \\
\hline 
$ \Lambda N \rightarrow \Lambda N $  & $(0,1/2)$ &
                   $ 1+\frac{1}{8} a_{PB}$ & $ 1.022 $\\
$ \Lambda N \rightarrow \Lambda N $  & $(1,1/2)$ &
                   $ 1+\frac{1}{8} a_{PB}$ & $ 1.022 $ \\
\hline 
$ \Sigma N \rightarrow \Sigma N $  & $(0,1/2)$ &
                   $ 1+\frac{9}{8} a_{PB}$ & $ 1.200 $\\
$ \Sigma N \rightarrow \Sigma N $  & $(1,1/2)$ &
                   $ 1+\frac{1}{8} a_{PB}$ & $ 1.022 $\\
$ \Sigma N \rightarrow \Sigma N $  & $(0,3/2)$ &
                   $ 1 $ & $ 1.000 $\\
$ \Sigma N \rightarrow \Sigma N $  & $(1,3/2)$ &
                   $ 1+ a_{PB}$ & $ 1.178 $\\
\hline \hline
\end{tabular}
\end{center}
\renewcommand{\arraystretch}{1.0}
\label{tab.4} \end{table}
%---------------------------------------------------------------------
\noindent In principle one might choose a different mass for the Quark-core
repulsive potential. However, this extra parameter does not lead to better 
fits to NN and/or YN. Therefore we keep for the Pauli-blocking the same mass 
as the Pomeron mass.
\noindent The value of the PB factor $a_{PB}$ is searched in the fit 
to the YN-data. The S=-2 PB effects are then also entirely determined.
\noindent In the case of the models ESC08a' and ESC08b' only the 
channels where $w_{BB}[51]$ 
is conspicuously large are treated approximately this way, but with equal weights. These channels are:
 $\Sigma^+p(^3S_1,T=3/2), \Sigma N(^1S_0,T=1/2)$, and $\Xi N(^1S_0,T=1)$.
\noindent A subtle treatment of $all$ BB channels according to this linear scheme is 
characteristic for the ESC08c-model. The parameter $a_{PB}$ turns out 
to be about 27.5\%. This means that the Quark-core repulsion is
roughly 34\% of the genuine Pomeron repulsion. Around r=0 the Quark-core repulsion
comes out at 115 MeV, whereas the pure Pomeron repulsion is 304 MeV.

\noindent 2.\ Non-linear method: we introduce next to the fraction parameter 
 $a_{PB}$ a nonlinear function  $f(w_{BB}[51])$ in (\ref{eq:51.2}) by writing 
\begin{equation}
 V_{BB} = (1-a_{PB}) V_{PNN} + a_{PB} f(w_{BB}[51]) V_{PNN}
\label{eq:51.5}\end{equation}
\noindent with the requirement that $f(w_{NN}[51]) = 1$ .
Such a scheme is more flexible than the linear method. 
For example, by taking a steeper function than the linear function for arguments 
larger than $w_{NN}[51]$ one could reduce the PB-effects for small $w_{BB}[51]$
values like in $\Lambda N$ or $\Xi N)$ and at the same time one could preserve
a strong PB-effect in the $\{10\}$-irrep having $f(w_{10}[51])$ around 2. 
Therefore starting from the ESC08c solutions alternative YN/YY solutions can be
obtained easily.
\noindent As an example one could take the function  $f(w_{BB} [51]) = \tan(\varphi_{BB})$
with $\varphi_{NN}= 45^o$. Taking for 
 $\arctan(\varphi_{10})=2$ one recovers the same $V_{10}(PB)$ as in the linear scheme above. 
Exploiting the weights $w_{BB}[51]$ by parametrizing $\varphi_{BB}$
in the following way
\begin{equation}
 \varphi_{BB} = \left( \frac{w_{BB}[51] - w_{NN}[51]}              
 {w_{10}[51] - w_{NN}[51]}\right)\cdot\left(\varphi_{10}-\varphi_{NN}\right)
 + \varphi_{NN}.
\label{eq:51.6}\end{equation}
one gets the same PB-repulsion for $\Sigma^+p(^3S_1,T=3/2)$ as in the linear method,
but smaller PB-repulsions in the $\Lambda N$ and  $\Xi N$ channels.

%-endtheory------------------------------------------------------------------

\section{ ESC08: Fitting $NN\oplus YN\oplus YY$-data}                                
\label{sec:6} 
In this section we describe mainly the recent changes in the fitting process. 
For details on the standard NN$\oplus$YN-fitting we refer to \cite{RSY99}.

\noindent (i) As usual we fit to the 1993 Nijmegen representation of the $\chi^2$-hypersurface of the 
$NN$ scattering data below $T_{lab}=350$ MeV \cite{Sto93,Klo93}. The NN 
low-energy parameters are fitted along with the scattering data. In order to accomodate 
the differences between the $^1S_0$-waves for pp, np, and nn in the model, we introduce
some charge independence breaking by taking different electric $\rho$-couplings
 $g_{pp\rho} \neq g_{np\rho} \neq g_{nn\rho}$, 
where $g_{nn\rho}$ is considered to be the $SU(3)$-octet coupling.
With this phenomenological device we fit the difference between the $^1S_0(pp)$ 
and $^1S_0(np)$ phases, and the different NN scattering lengths and effective ranges very well.
 We have found $g_{pp\rho} = 0.6389,\ g_{np\rho} = 0.5889$, which are not far from
$g_{nn\rho} = 0.6446$ (cfr.Table~\ref{tab.su3par}).

\noindent (ii) Simultaneously we fit to 52 YN scattering data. These data consist
of the usual set of 35 low-energy $Y\!N$-data, as used in \cite{MRS89,RSY99} 
and \cite{Rij04b} plus 3 total $\Sigma^+p$ X-sections 
from the recent KEK-experiment E289 \cite{Kanda05} and some $\Lambda p$ elastic and
inelastic data Ref.~\cite{Kadyk71} and $\Sigma^-p$ elastic data 
at higher energies Ref.~\cite{Kondo00}.
For the total $\Sigma^+p$ and $\Sigma^-p$ elastic X-sections we have performed the
same redefinition as eq. (6.3) of \cite{RSY99}.
\noindent (iii) A novel feature in the simultaneous fitting procedure is the inclusion of 
constraints from information derived from hypernuclei and hypernuclear matter. 
For the $\Lambda N$-interaction this means not only
\noindent a)  the usual absence of bound states but also 
b) the requirement
of a sizable spin-splitting leading to $U_{\sigma \sigma} > 1$ (cfr. section~\ref{sec:5bb}).
c) Because of the experimental absence of $\Sigma$-hypernuclei we require the total
single particle $\Sigma$-potential in nuclear matter $U_\Sigma$ to be overall repulsive.
\noindent In the S=-2 channels there are two clear experimental indications:
d) from the analysis of the Nagara event \cite{Tak01} of the double-$\Lambda$ hypernucleus
$_{\Lambda \Lambda}^{6}$He it appears that the forces in the $\Lambda\Lambda(^1S_0)$-channel
are weakly attractive, indicated by a scattering 
length $a_{\Lambda\Lambda}(^{1}S_{0}) > -1$ fm \cite{Hiy02}. This evidence has been
incorporated in the fit in the form of 'pseudo-data' for the $(^{1}S_{0}) \Lambda \Lambda$ scattering
length $a_{\Lambda\Lambda} = -0.8 \pm (0.2-0.4)$. the error depending on the desired impact 
in the fitting process. 
\noindent e) Experimentally the $\Xi$-nucleus interaction
seems to be attractive from analyses of events with twin-$\Lambda$ hypernuclei in emulsion data,
where the initial $\Xi^-$ energies were determined after $\Xi^-p - \Lambda\Lambda$
conversion in nuclei. The $\Xi$-nucleus interaction can be described well with a Wood-Saxon 
potential with a depth of $\approx 14 MeV$ \cite{Khaus00}. As a consequence we require the total
single particle $\Xi$-potential in nuclear matter $U_\Xi$ to be overall attractive.
 We calculate $U_\Xi$ with the G-matrix formalism (cfr. section X). It appears that in the ESC08
model the $\Xi N(^3S_1,T=1)$ channel is quite sensitive to variation of the coupling constants, whereas
the other s- and p-wave contributions have rather shallow dependencies. The total sum of the 
contributions of the other $T=0$ and $T=1$ s- and p-waves to $U_\Xi$ is $\approx + 4$ MeV. 
In order to ensure  a clearly total attractive $U_\Xi$ 
we add 'pseudo-data' for the $\Xi N(^3S_1,T=1)$ contribution $U_\Xi (^3S_1,T=1) = -12 \pm (2-4) $,
the error depending on the desired impact in the fitting process. In practice this particular partial
wave channel plays a very important role in the simultaneous fitting of the total BB-description. 
The fit has resulted in an excellent simultaneous $NN\oplus YN \oplus YY$-fit.
We obtained for the NN-data $\chi^2/Ndata =1.081$ with also very good results for the NN low 
energy parameters: the deuteron binding energy and the pp, np and nn scattering lengths and 
effective ranges. For the YN-data $\chi^{2}/YN_{data} = 1.08$.
The ESC08 fits were achieved with only physical meson-coupling parameters, which are partial-wave 
independent. The quality of the NN-fit is at par with models of Reid-like potentials like the 
Nijm93, Nijm I, and II, which are effective NN-potentials with some meson parameters adjusted 
per partial wave \cite{Stoks93a, Stoks94}. 
\noindent Since the ESC08-model is an extension of the ESC04-model, it is not surprising that in the 
simultaneous $NN$-, $YN$ and $YY$ fit
the OBE-couplings could be kept in line with the 
'naive' predictions of the QPC-model \cite{Rij04a,LeY73}. 
Just as in ESC04 most of the 
$F/(F+D)$-ratios are fixed by QPC, both for the OBE and MPE couplings.
Once more we stress the fact that the in the simultaneous fit of the $NN$-, $YN$- and $YY$-data, 
a {\it single set of parameters} was used. Of course, the accurate and 
very numerous $NN$-data put strong constraints on the parameters. However, the  
YN-data, plus the constraints for the YN- and YY- channels, are also
quite relevant for the set of parameters finally obtained. In particular,
certain fitted $F/(F+D)$-ratios, are obviously influenced by the $YN$-data. The requirement
of an overall attractive $U_\Xi$ results in the model always in the occurrence of a bound state in
$\Xi N (^3S_1,T=1)$-channel with a binding energy of $\approx 2$ MeV.

\subsection{ Coupling Parameters and $NN\oplus YN\oplus YY$ Fit}                                  
\label{sec:5aa} 
The treatment of the broad mesons $\rho$ and $\epsilon$ is as
usual in the Nijmegen models. For the $\rho$-meson the same parameters are 
used in the OBE-models \cite{NRS78,MRS89}.  For the $\epsilon=f_0(760)$ assuming 
$m_\epsilon=760$ MeV and $\Gamma_\epsilon = 640$ MeV we use the
Bryan-Gersten parametrization \cite{Bry72}. For the choosen mass and width they are: 
$ m_1=496.39796$ MeV, $m_2=1365.59411$ MeV, and $\beta_1=0.21781, \beta_2=0.78219$.
For the 'diffractive' $0^{++}$-exchanges we restrict ourselves to the SU(3)-singlet
part, henceforth referred to as 'pomeron'. The possible J=0 part of the tensor-meson
exchange \cite{NRS78,MRS89} is not considered.
The 'mass' parameter of the pomeron is fitted to be $m_P = 227.05$ MeV.
The 'diffractive' $0^{--}$-exchange 'odderon' is also an SU(3)-singlet with a fitted mass
 $m_O = 273.35$ MeV.

Summarizing the fitted parameters in ESC08c we have:
\begin{enumerate}
\item Meson-couplings: $ f_{NN\pi}, f_{NN\eta'}, g_{NN\rho}, g_{NN\omega}$, \\
 $f_{NN\rho},f_{NN\omega}, g_{NNa_0}, g_{NN\epsilon}$,
 $g_{NNa_1}, f_{NNa_1}, g_{NNf'_1}, f_{NNf'_1}, f_{NNb_1}, f_{NNh'_1}$ 
\item Pair couplings: $g_{NN(\pi\pi)_1}, f_{NN(\pi\pi)_1}, g_{NN(\pi\rho)_1}$,\\
 $g_{NN\pi\omega}, g_{NN\pi\eta}, g_{NN\pi\epsilon}$
\item Diffractive-couplings/masses: $ g_{NNP}, g_{NNO}, f_{NNO}, a_{PB}, m_P, m_O $ 
\item $F/(F+D)$-ratio's: $\alpha_{V}^m, \alpha_{A}$
\item Cut-off masses:  $\Lambda_8^P=\Lambda_1^P,  \Lambda_8^V , \Lambda_1^V ,
  \Lambda_8^S , \Lambda_1^S , \\
 \Lambda_8^A=\Lambda_1^A $ 
\end {enumerate} 
These are in total 34 physical parameters, of which are (i) 14 meson-couplings, 
(ii) 2 $F/(F+D)$-ratio's, (iii) 4 'diffractive' couplings and 2 mass parameters, 
(iv) 6 meson-pair couplings, and (v) 6 cut-off mass parameters.

\noindent As compared to the ESC04-model, we have added in ESC08 the 
following fitting parameters:
(i) the derivative axial-couplings $f_{NNa_1}, f_{NNf'_1}$, 
(ii) the $1^{+-}$ axial-couplings $f_{NNb_1}, f_{NNh'_1}$, (iii) the odderon-couplings
$g_{NNO}, f_{NNO}$, and mass $m_O$  (iv) the pomeron Pauli-blocking parameter $a_{PB}$,
i.e. 8 new physical parameters. All new parameters have been explained above. 
They introduce new dynamical refinements/effects 
into the model, which have resulted in a quality of the combined NN+YN+YY
fit for the NN-phases equal to those of a purely NN-fit.
Some other parameters have been set, like e.g. many $F/(F+D)$-ratio's, see below,
and a few cut-off parameters.

The pair coupling $g_{NN(\pi\pi)_0}$ is set to be zero, which is motivated
in the Nijmegen soft-core models in view of the fact that in $\pi N$ it is
constrained by chiral-symmetry. 
In the fitting process we look for solutions which have meson-couplings which are
reasonably close to the 'naive' predictions of the QPC-model. This is also
the case for the $F/(F+D)$-ratio's, both for meson- and for pair-couplings.
During the fitting we experienced a rather shallow dependence on the $F/(F+D)$-ratio
$\alpha^{P}$ for the pseudoscalar octet. In fact we could obtain a very good YN fit
in a values range 0.33-0.40. Therefore we have fixed it on the value $\alpha^{P}=0.365$
obtained from the Cabibbo theory of semileptonic decay of baryons \cite{Clo93}.
Furthermore, the meson-pair couplings turn out to come out rather close to
predictions based on the 'heavy-meson-saturation'-model. So, the fit-parameters
are (i) physical parameters, i.e. they can be checked in other reactions, and
(ii) many are 'constraint' by the QPC-model.\\
In this work like in the ESC04-models \cite{Rij04a,Rij04b}, the form 
factors depend on the SU$(3)$ assignment of the mesons, In principle, 
we introduce form factor masses 
$\Lambda_{8}$ and $\Lambda_{1}$ for the $\{8\}$ and $\{1\}$ members of each 
meson nonet, respectively. Moreover, for the $I=0$-mesons we assign the 
$\{1\}$ cut-off to the dominant singlet meson and the $\{8\}$ cut-off to the 
dominant octet meson, as if there were no meson-mixing. For example we assign $\Lambda_1$ 
to $\eta', \omega, \epsilon$, and $\Lambda_8$ to $\eta, \phi, S^*$, etc.
Notice that the strange octet-mesons $K$ etc. are given the same $\{8\}$ form factors
as their non-strange companions.
For the cut-off masses $\Lambda$ we used as free search parameters 
$\Lambda_8^P=\Lambda_1^P$ for the pseudoscalar mesons, $ \Lambda_8^V$ and $\Lambda_1^V$
for the vector mesons and $ \Lambda_8^S$ and $ \Lambda_1^S $ for the scalar mesons.
Furthermore, we finally used $\Lambda_8^A=\Lambda_1^A$ for the axial-mesons 
with $J^{PC}=1^{++}$. For the axial-mesons with $J^{PC}=1^{+-}$ (B-mesons) the cut-off
masses have been set equal to those of the pseudoscalar mesons 
$\Lambda_8^B=\Lambda_8^P$ and $\Lambda_1^B=\Lambda_1^P$. 
Some of the previous $\{8\}$ and $\{1\}$ form factors have been chosen 
to be equal as a consequence
of the impossibility to distinguish them in the fitting process. \\
Similar to ESC04 we introduce a zero in the form factors of mesons, which are
P-wave bound states in a $q\bar{q}$-picture. These are the scalar mesons ($^3P_0$) 
and the axial-vector ($^3P_1, ^1P_1$) mesons. Like in ESC04, we use 
a fixed zero by taking $U = 750$ MeV in (\ref{eq:3.15}) and (\ref{eq:3.17b}).
%------------------------------------------------------------------------
\onecolumngrid
%----------------------------------------------------------------------
\section{ Coupling Constants, $F/(F+D)$ Ratios, and Mixing Angles}             
\label{sec:7} 
%----------------------------------------------------------------------
% 3P0 material:
Like in ESC04, we constrained the OBE-couplings
by the 'naive' predictions of the QPC-model \cite{Mic69}.  
We kept during the 
searches all OBE-couplings in the neighborhood of these predictions, but 
less tight than in ESC04.
The same holds for the searched $\alpha=F/(F+D)$-ratios, i.e. for
the $BBM$-couplings and the $BB$-Pair-couplings. 
In fact only two meson-coupling $F/(F+D)$-ratio's were allowed to vary 
during the searches:
$\alpha_V^m$ for the vector mesons, and  $\alpha_A$ for the axial-vector mesons. 
As mentioned above $\alpha_P $ was kept at the fixed value $\alpha_P = 0.365$.
Furthermore we kept $\alpha_V^E=1$ and  $\alpha_S=1$ at their QPC values.
The input and fitted values are displayed in Table \ref{tab.su3par}.

\noindent The mixing angles for the various meson nonets are discussed in paper I.
The used values can be found in Table \ref{tab.su3par}.
For completeness we reproduce in Table \ref{table5} the fitted ESC08c NN
meson couplings and cut-off masses from paper I.
%--------------------------------------------------------------------
\begin{table}
\caption{Meson couplings and parameters employed in the ESC08c-potentials.
  Coupling constants are at ${\bf k}^{2}=0$.
  An asterisk denotes that the coupling constant is constrained via SU(3).
  The masses and $\Lambda$'s are given in MeV. 
  Note that the B-meson couplings are scaled with $m_{B_1}$.  
  The mesons with strangeness are the (i) pseudoscalar $K(495.8)$, 
  (ii) vector $K^*(892.6)$, (iii) scalar $\kappa(841.0)$, 
  (iv) axial $K_A(1273.0)$ and $K_B(1400.0)$,
  with masses as indicated in the parentheses.
}
\label{table4}
\begin{center}
\begin{ruledtabular}
\begin{tabular}{crccr} 
meson & mass & $g/\sqrt{4\pi}$ & $f/\sqrt{4\pi}$ & \multicolumn{1}{c}{$\Lambda$}\\
\hline
 $\pi$         &  138.04 &           &   0.2687   &   1056.13\    \\
 $\eta$        &  547.45 &           & \hspace{2mm}0.1265$^\ast$   & ,, \hspace{5mm} \\
 $\eta'$       &  957.75 &           &   0.2309   &  ,, \hspace{5mm} \\ \hline
 $\rho$        &  768.10 &  0.6446   &   3.7743   &    695.67\    \\
 $\phi$        & 1019.41 &--1.3390$^\ast$ & \hspace{2mm}3.1678$^\ast$ & 
 ,, \hspace{5mm}  \\
 $\omega$      &  781.95 &  3.4570   & --0.8575   &    758.58\\ \hline
 $a_1 $        & 1270.00 &--0.7895   & --0.8192   &   1051.80\    \\
 $f_1 $        & 1420.00 &  \hspace{3mm}0.7311$^\ast$ &\hspace{2mm}  0.3495$^\ast$  &      ,,  \hspace{5mm} \\
 $f_1'$        & 1285.00 &--0.7613   & \hspace{2mm}--0.4467 &      ,, \hspace{5mm}  \\ \hline
 $b_1 $        & 1235.00 &           & --1.8088   &   1056.13    \\
 $h_1 $        & 1380.00 &           & \hspace{2mm}--0.5553$^\ast$   &     
 ,,  \hspace{5mm} \\
 $h_1'$        & 1170.00 &           & --0.3000   &      ,, \hspace{5mm}  \\ \hline
 $a_{0}$       &  962.00 &  0.5853   &            &    994.89\    \\
 $f_{0}$       &  993.00 &\hspace{0mm}--1.6898$^\ast$   &            &  
 ,, \hspace{5mm}  \\
 $\varepsilon$ &  760.00 &  4.1461   &            &   1113.57 \\ \hline
 Pomeron       &  220.50 &  3.5815   &            &              \\
 Odderon       &  273.35 &  4.6362   & --4.7602   &              \\
%\hline
\end{tabular}
\end{ruledtabular}
\end{center}
\label{table5}
\end{table}
% \end{wraptable}
%--------------------------------------------------------------------------
Here we discuss only aspects specific for the YN-channels.    
%----------------------------------------------------------------------
In Table~\ref{tab.su3par} the ESC08 $SU(3)$ singlet and octet couplings
 $g/\protect\sqrt{4\pi}$ are 
listed, the $F/(F+D)$-ratios and the used mixing angles.
 \begin{table}[hbt]
\begin{center}
\caption{ESC08c SU(3) coupling constants, $F/(F+D)$-ratio's, mixing angles etc.
 The values with $\star )$ have
% been determined in the fit to the $YN$-data. The other parameters
 are theoretical input or determined by the fitting and
 the constraint from the $YN$-analysis. }
\label{tab.su3par} 
\begin{ruledtabular}
\begin{tabular}{ccccll} & & & & &  \\
%\hline\hline 
  mesons    &   & $\{1\}$  & $\{8\}$  &  $F/(F+D)$   &    angles   \\
            &   &          &          &           &             \\
\hline
            &   &          &          &           &             \\
  ps-scalar  & f & 0.2534   &  0.2687 & $\alpha_{P}=0.365^{\ast)}$
             & $\theta_{P} = -13.00^{0~\ast)}$ \\
            &   &          &          &           &             \\
  vector    & g & 3.5351   &  0.6446  & $\alpha^{e}_{V}=1.0^{\ast)}$ &
 $\theta_{V} =\ \ 38.70^{0~\ast)}$   \\
            &   &          &          &           &             \\
            & f &--2.6499  &  3.7743  & $\alpha^{m}_{V}=0.4721^{\ast)}$ &  \\
            &   &          &          &           &             \\
  axial(A)  & g &--1.0494  &--0.7895  & $\alpha_{A}=0.3121$ &
 $\theta_{A} =+50.00^{0~\ast)}$ \\
            &   &          &          &           &             \\
            & f &--0.5548  &--0.8192  & $\alpha^{p}_{A}=0.3121^{\ast)}$ \\
            &   &          &          &           &             \\
  axial(B)  & f &  0.0760  &--1.8088  & $\alpha_{B}=0.40^{\ast)}$ &
 $\theta_{B} = 35.26^{0~\ast)}$ \\
            &   &          &          &           &             \\
  scalar    & g &  4.3610  &  0.5853  & $\alpha_{S}=1.00^{\ast)}$ &
 $\theta_{S} =\ \ 35.26^{0~\ast)}$ \\
            &   &          &          &           &             \\
  diffractive   & $g_P$ &  3.5815  &     &   &
 $a_{PB}=\ \ 0.275^{\ast)}$\\
%$\psi_{D}=\ \ 0.0^{0~\ast)}$\\
                & $g_O$ &  4.6362  &    &   & \\
                & $f_O$ &--4.7602  &    &   & \\
            &   &          &          &           &             \\
%\hline
\end{tabular}
\end{ruledtabular}
\end{center}
 \end{table}
%--------------------------------------------------------------------------

\noindent In Table \ref{tab.cop08a} we list the couplings of the physical mesons to the 
nucleons $(Y=1)$, and to the hyperons with $Y=0$ or $Y=-1$. These were 
calculated using unbroken SU(3)-symmetry. Next to the values in the
table we have incorporated, like in the ESC04 model \cite{Rij04b},
Charge Symmetry Breaking (CSB) between $\Lambda p$ and $\Lambda n$ 
with nonzero $\Lambda$-couplings of the I=1 mesons and I=1 pairs 
due to $\Lambda - \Sigma^0$ mixing.
%----------------------------------------------------------------------
% parameters: parbbsc.ESC08c.best14june
\begin{table}[hbt]
\caption{Coupling constants for model ESC08c, divided by
         $\protect\sqrt{4\pi}$. $M$ refers to the meson.
         The coupling constants are listed in the order pseudoscalar,
         vector ($g$ and $f$), axial vector A ($g$ and $f$), scalar, axial vector B, 
         and diffractive.}
\label{tab.cop08a}
\begin{center}
\begin{ruledtabular}
\begin{tabular}{ccrrrrcrrrr} 
      &\multicolumn{1}{c}{$M$} & $N\!N\!M$ & 
 $\Sigma\Sigma M$ & $\Sigma\Lambda M$ & $\Xi\Xi M$
 & $M$ & $\Lambda N\!M$ & $\Lambda\Xi M$ & $\Sigma N\!M$ & $\Sigma\Xi M$ \\
%\colrule
\hline 
 $f$ & $\pi$    &   0.2687  &   0.1961  &   0.1970  & --0.0725 
     & $K$      & --0.2683  &   0.0714   &   0.0725  & --0.2687  \\
 $g$ & $\rho$   &   0.6446  &   1.2892  &   0.0000  &   0.6446  
     & $K^*$    & --1.1165  &   1.1165   & --0.6446  & --0.6446   \\
 $f$ &          &   3.7743  &   3.5639  &   2.3006  & --0.2104  
     &          & --4.2367  &   1.9362   &   0.2104  & --3.7743   \\
 $g$ & $a_1$    & --0.7895  & --0.4929  & --0.6271  &   0.2967  
     & $K_{1A}$ &   0.7404  & --0.1133   & --0.2967  &   0.7895   \\
 $f$ &          & --0.8192  & --0.5114  & --0.6507  &   0.3078  
     &          &   0.7683  & --0.1175   & --0.3078  &   0.8192   \\
 $g$ & $a_0$    &   0.5853  &   1.1705  &   0.0000  &   0.5853  
     & $\kappa$ & --1.0137  &   1.0137   & --0.5852  & --0.5852   \\
 $f$ & $b_1$    & --1.8088  & --1.4470  & --1.2532  &   0.3618  
     & $K_{1B}$ &   1.8798  & --0.6266   & --0.3618  &   1.8088   \\
% $f$ & $b_1$    & --0.2022  & --0.1617  & --0.1401  &   0.0404  
%     & $K_{1B}$ &   0.2101  & --0.0700   & --0.0404  &   0.2022   \\
 $g$ & $a_2$    &   0.00000 &   0.00000 &   0.00000 &   0.00000
     & $K^{**}$ &   0.00000 &   0.00000 &   0.00000 &   0.00000 \\[2mm]
      & $M$ & $N\!N\!M$ & $\Lambda\Lambda M$ & $\Sigma\Sigma M$ & $\Xi\Xi M$
 & $M$ & $N\!N\!M$ & $\Lambda\Lambda M$ & $\Sigma\Sigma M$ & $\Xi\Xi M$ \\
%\tableline
%\colrule
 \hline 
 $f$ & $\eta$        &   0.1265  & --0.1349  &   0.2490  & --0.2045  
     & $\eta'$       &   0.2309  &   0.2912  &   0.2026  &   0.3073   \\
 $g$ & $\omega$      &   3.4570  &   2.7589  &   2.7589  &   2.0608  
     & $\phi$        & --1.3390  & --2.2103  & --2.2103  & --3.0816   \\
 $f$ &               & --0.8574  & --3.5064  & --0.6296  & --4.7170  
     &               &   3.1678  & --0.1386  &   3.4522  & --1.6497   \\
 $g$ & $f'_1$        & --0.7613  & --0.1942  & --1.1549  & --0.1074  
     & $f_1$         &   0.7311  &   1.2070  &   0.4008  &   1.2798   \\
 $f$ &               & --0.4467  &   0.1418  & --0.8551  &   0.2319  
     &               &   0.3495  &   0.8433  &   0.4008  &   1.2798   \\
 $g$ & $\varepsilon$ &   4.1461  &   3.5609  &   3.5609  &   2.9758  
     & $f_0$         & --1.6898  & --2.5176  & --2.5176  & --3.3453   \\
 $f$ & $h'_1$        & --0.3000  &   0.7852  & --0.6617  &   1.1469  
     & $h_1$         & --0.5553  &   0.9796  & --1.0669  &   1.4913   \\
 $g$ & $P$           &   3.5815  &   3.5815  &   3.5815  &   3.5815 
     & $f_2$         &   0.0000  &   0.0000  &   0.0000  &   0.0000  \\
 $g$ & $O$           &   3.6362  &   3.6362  &   3.6362  &   3.6362    
     &               &           &           &           &           \\
 $f$ &               & --4.7602  & --4.7602  & --4.7602  & --4.7602  
     &               &           &           &           &           \\
\end{tabular}
\end{ruledtabular}
\end{center}
\end{table}
%--------------------------------------------------------------------------

%--------------------------------------------------------------------------
\vspace*{7mm}
%--------------------------------------------------------------------------

\noindent In Table~\ref{tab.gpair} we present the fitted Pair-couplings for the MPE-potentials.
We recall that only One-pair graphs are included, in order to avoid double
counting, see paper I. The $F/(F+D)$-ratios are all fixed, assuming heavy-boson 
domination of the pair-vertices. The ratios are taken from the QPC-model for 
$Q\bar{Q}$-systems with the same quantum numbers as the dominating boson.
Only the ratio in the system with the pseudoscalar quantum numbers deviates slightly
from QPC, since it has been set equal to the value of $\alpha_P = 0.365$.
The $BB$-Pair couplings are calculated, assuming unbroken $SU(3)$-symmetry, from the 
$NN$-Pair coupling and the $F/(F+D)$-ratio using $SU(3)$.
 
%--------------------------------------------------------------------------
% parameters: parbbsc.ESC08c.best14june 
\begin{table}[hbt]
\caption{Pair-meson coupling constants employed in the ESC08c MPE-potentials.     
         Coupling constants are at ${\bf k}^{2}=0$.
         The F/(F+D)-ratio are QPC-predictions, except that 
 $\alpha_{(\pi\omega)}=\alpha_{pv}$, which is very close to QPC.}
\label{tab.gpair}
\begin{center}
\begin{ruledtabular}
\begin{tabular}{cclrc} 
 $J^{PC}$ & $SU(3)$-irrep & $(\alpha\beta)$  &\multicolumn{1}{c}{$g/4\pi$} & $F/(F+D)$ \\
%\colrule
 \hline \\
 $0^{++}$ & $\{1\}$  & $g(\pi\pi)_{0}$   &  ---    &  ---    \\
 $0^{++}$ & ,,       & $g(\sigma\sigma)$ &  ---    &  ---    \\
 $0^{++}$ &$\{8\}_s$ & $g(\pi\eta)$      & -1.2371 &  1.000  \\ \hline
%$0^{++}$ &          & $g(\pi\eta')$     &  ---    &  ---    \\
 $1^{--}$ &$\{8\}_a$ & $g(\pi\pi)_{1}$   &  0.2703 &  1.000  \\
          &          & $f(\pi\pi)_{1}$   &--1.6592 &  0.400  \\ \hline
 $1^{++}$ & ,,       & $g(\pi\rho)_{1}$  &  5.1287 &  0.400  \\
 $1^{++}$ & ,,       & $g(\pi\sigma)$    &--0.2989 &  0.400  \\
 $1^{++}$ & ,,       & $g(\pi P)$        &  ---    &  ---    \\ \hline
 $1^{+-}$ &$\{8\}_s$ & $g(\pi\omega)$    &--0.2059 &  0.365  \\
%\hline
\end{tabular}
\end{ruledtabular}
\end{center}
\end{table}
% \end{wraptable}     
%--------------------------------------------------------------------------

\noindent Unlike in \cite{RS96ab}, we did not fix pair couplings using
a theoretical model, based on heavy-meson saturation and chiral-symmetry.
So, in addition to the 14 coupling parameters used in \cite{RS96ab} we now have
6 pair-coupling fit parameters. 
In Table~\ref{tab.gpair} the fitted pair-couplings are given, and in
Appendix~\ref{app:MPE.SU3} the SU(3)-structure of the pair-couplings.
Note that the $(\pi\pi)_0$-pair coupling gets contributions from the $\{1\}$ and
the $\{8_s\}$ pairs as well, giving in total $g_{(\pi\pi)}=-0.4876/2=-0.2438$, which has an 
 opposite sign as in \cite{RS96ab}. Also the $f_{(\pi\pi)_1}$-pair coupling has an opposite
sign as compared to \cite{RS96ab}. In a model with a more complex and realistic
meson-dynamics \cite{SR97} this coupling is predicted as found in the present 
ESC-fit. The $(\pi\rho)_1$-coupling agrees nicely with $A_1$-saturation, see 
\cite{RS96ab}. 
The pair-couplings are used in a phenomelogical way in the ESC-approach. 
They are in general not yet quantitatively 
understood, and certainly deserve more study in the future.

The ESC-model described here, is fully consistent with $SU(3)$-symmetry  
using a straightforward extension of the NN-model to YN and YY.
For example $g_{(\pi\rho)_1} = g_{A_8VP}$, and
besides $(\pi\rho)$-pairs one sees also that $K K^*(I=1)$- and 
$K K^*(I=0)$-pairs contribute to the $NN$ potentials.
All $F/(F+D)$-ratio's are taken fixed with heavy-meson saturation in mind.
The approximation we have made in this paper is to neglect the baryon mass
differences, i.e. we put $m_\Lambda = m_\Sigma = m_N$. This because we
have not yet worked out the formulas for the inclusion of these mass 
differences, which is straightforward in principle.

%-----------------------------------------------------------------------------------
\subsection{ Parameters and Hyperon-nucleon Fit}                                  
\label{sec:5bb} 
All 'best' low-energy YN-data are included in the fitting, This is a selected set of   
35 low-energy $Y\!N$-data, the same set has been used in \cite{MRS89} and \cite{RSY99}.
We added (i) 3 total $\Sigma^+p$ X-sections 
from the  KEK-experiment E289 \cite{Kanda05}, (ii) 7 elastic and 4 inelastic 
$\Lambda p$ X-sections from Berkeley \cite{Kadyk71}, and (iii) 3 elastic $\Sigma^- p$ 
X-sections \cite{Kondo00}.
In section~\ref{sec:8} these are given together with the results.
Next to these we added 'pseudo-data'
for the $\Lambda p$ and $\Sigma^+ p$ scattering length's in order to ensure that
 the $\Lambda p(^1S_0)$ forces are sufficiently stronger than the $\Lambda p(^3S_1)$.
 In the construction of the ESC04-models, the experience 
 with the NSC97 models was used in hypernuclear calculations. Technically  
 'favored' values of the s-wave scattering lengths for $\Lambda N$ were imposed
 as pseudo-data during the fitting procedures, in order to get the 
 right spin-splitting for the $\Lambda N$-interaction in hypernuclei.
 In nuclear matter this implies $U_{\sigma \sigma}>1$. 
 In this succeeding model ESC08, however, the $\Lambda N$ behavior is slightly different
 leading to smaller values of $U_{\sigma \sigma}$.
 This time, we impose instead a larger difference between the s-wave scattering lengths
 with, of course,  $|a_s| > |a_t|$
 in order to fulfill the constraint $U_{\sigma \sigma} > 1$. Furthermore we added 
 'pseudo-data' for the $\Sigma^+ p(^3S_1)$ scattering length with the goal
 to get enough repulsion in this wave in order to reach a total
repulsive $U_\Sigma$. For the pseudo-data in the S=-2 channels we refer to
section~\ref{sec:6}. In the final stages af the fitting process all pseudo-data
were turned off.
\noindent in fm: 
\begin{eqnarray}
 \hat{a}_{\Lambda p}(^1S_0) &=& -2.60 \pm (0.10-0.20)\ \ ,\ \   
\nonumber\\
 \hat{a}_{\Lambda p}(^3S_1) &=& -1.60 \pm (0.10-0.20)\ \ ,\ \   
\nonumber\\
 \hat{a}_{\Sigma^+ p}(^3S_1) &=& +0.65 \pm (0.10-0.20)\ , 
\label{eq:5.1}\end{eqnarray}
Also, during the fitting process checks were done to 
prevent the occurrence of bound $\Lambda p$ states. 
Parameters, typically strongly influenced by the YN-data, are 
\begin{enumerate}
\item $F/(F+D)$-parameters: $\alpha_V^m$ and to a less extent $\alpha_A$, For the
sensitivity of $\alpha^P$ see section~\ref{sec:7}.
\item Pauli-blocking fraction parameter $a_{PB}$.
\end{enumerate}
The dependence of $a_{PB}$ in the fit to YN
and YY is rather shallow in a range $0.20 - 0.30$. The final value has been determined
by a minimal value of $\chi^2$ of the set of the 52 YN data, while simultaneously 
providing a repulsive $U_\Sigma$.
This implies that the S=-2 (and -3, -4) 
results are completely determined.
\noindent Finally we want to mention that in the fitting process we have, if necessary,
accounted for the vast difference in quality of the data. The abundance of the 4313 precise
NN data is to be contrasted to the 52 less precise YN data. In the simultaneous fit we
require for both the NN and for the YN that the quality of the partial fit is comparable, i.e.
$\chi^2/NN_{data} \approx \chi^{2}/YN_{data}$. If necessary we add weight factors to
the partial sums in the total $\chi^2$. It turned out that in the last stages of the
fitting process the weight factors are equal.
%----------------------------------------------------------------------
\section{ ESC08-model , $Y\!N$-Results}                                  
\label{sec:8} 
%\subsection{Fit to $\protect\bbox{Y\!N}$ total cross sections}
\subsection{Hyperon-nucleon (S=-1) X-sections, phases, etc.}
\label{sec:8a}
The used $Y\!N$ scattering data from Refs.~\cite{Ale68}-\cite{Ste70}
in the combined $N\!N$ and $Y\!N$ fit are shown in Table~\ref{tab.reslts1}.
%----------------------------------------------------------------------
The $N\!N$ interaction puts very strong constraints on most
of the parameters, and so we are left with only a limited set of
parameters which have some freedom to steer the YN channels as compared
to the NN-channels. 
The fitted parameters are given in paper I, Table's III-VI and X.

The aim of the present study was to construct a realistic potential model 
for baryon-baryon systems with parameters that are optimal theoretically, but at
the same time describes the baryon-baryon scattering data very satisfactory.

This model can then be used with a great deal of confidence  
in calculations of hypernuclei and in their predictions for the $S=-2$,
$-3$, and $-4$ sectors. Especially for the latter application, these
models will be the first models for the $S = -2, -3, -4$ sectors to have
their theoretical foundation in the $N\!N$ and $Y\!N$ sectors.
 
% esc08c.yn10:
% parameters: parbbsc.ESC08c.best14june
\begin{table}[ht]
\begin{center}
\begin{ruledtabular}
\begin{tabular}{cccccc} 
%\hline\hline \\
\multicolumn{2}{c}{$\Lambda p\rightarrow \Lambda p$} & $\chi^{2}=3.6$    &
\multicolumn{2}{c}{$\Lambda p\rightarrow \Lambda p$} & $\chi^{2}=3.8$   \\
$p_{\Lambda}$ & $\sigma^{RH}_{exp}$ & $\sigma_{th}$ &
$p_{\Lambda}$ & $\sigma^{M }_{exp}$ & $\sigma_{th}$ \\
\hline & & & & &\\
145 &   180$\pm$22 &197.0    & 135 &   187.7$\pm$58 &215.6\\
185 &   130$\pm$17 &136.3    & 165 &   130.9$\pm$38 &164.1\\
210 &   118$\pm$16 &107.8    & 195 &   104.1$\pm$27 &124.1\\
230 &   101$\pm$12 & 89.3    & 225 &    86.6$\pm$18 & 93.6\\
250 &    83$\pm$ 9 & 73.9    & 255 &    72.0$\pm$13 & 70.5\\
290 &    57$\pm$ 9 & 50.6    & 300 &    49.9$\pm$11 & 46.0\\
       & & & & &\\
\multicolumn{2}{c}{$\Lambda p\rightarrow \Lambda p$} & $\chi^{2}=12.1$    & & & \\
\hline & & & & &\\
350 &    17.2$\pm$8.6 & 28.7  & 
750 &    13.6$\pm$4.5 & 10.2  \\
450 &    26.9$\pm$7.8 & 11.9  & 
850 &    11.3$\pm$3.6 & 11.4  \\
550 &     7.0$\pm$4.0 &  8.6  & 
950 &    11.3$\pm$3.8 & 12.9  \\
650 &     9.0$\pm$4.0 & 18.5  &     &        &      \\
       & & & & &\\
\multicolumn{2}{c}{$\Lambda p\rightarrow \Sigma^0 p$} & $\chi^{2}= 6.9$    & & & \\
\hline & & & & &\\
667 &    2.8 $\pm$2.0 & 3.3  & 
850 &    10.6$\pm$3.0 & 4.1  \\ 
750 &    7.5$\pm$2.5 & 4.0  & 
950 &    5.6$\pm$5.0 & 3.9  \\
       & & & & &\\
\multicolumn{2}{c}{$\Sigma^+ p\rightarrow \Sigma^+ p$} & $\chi^{2}=12.4$ &
\multicolumn{2}{c}{$\Sigma^- p\rightarrow \Sigma^- p$} & $\chi^{2}=5.2$ \\
$p_{\Sigma^+}$ & $\sigma_{exp}$ & $\sigma_{th}$ &
$p_{\Sigma^-}$ & $\sigma_{exp}$ & $\sigma_{th}$ \\
\hline & & & & &\\
 145 & 123.0$\pm$62 & 136.1 & 142.5 & 152$\pm$38 & 152.8 \\
 155 & 104.0$\pm$30 & 125.1 & 147.5 & 146$\pm$30 & 146.9 \\
 165 &  92.0$\pm$18 & 115.2 & 152.5 & 142$\pm$25 & 141.4 \\
 175 &  81.0$\pm$12 & 106.4 & 157.5 & 164$\pm$32 & 136.1 \\
     &            &       & 162.5 & 138$\pm$19 & 131.1 \\
     &            &       & 167.5 & 113$\pm$16 & 126.3 \\
 400 &  93.5$\pm$28.1 & 35.1  & 450.0 & 31.7$\pm$8.3& 28.5  \\
 500 &  32.5$\pm$30.4 & 30.9  & 550.0 & 48.3$\pm$16.7 & 19.8  \\
 650 &  64.6$\pm$33.0 & 28.2  & 650.0 & 25.0$\pm$13.3 & 15.1 \\
     & & & & &\\
\multicolumn{2}{c}{$\Sigma^- p\rightarrow \Sigma^0 n$} & $\chi^{2}=5.7$ &
\multicolumn{2}{c}{$\Sigma^- p\rightarrow \Lambda n$} & $\chi^{2}=4.8$ \\
$p_{\Sigma^-}$ & $\sigma_{exp}$ & $\sigma_{th}$ &
$p_{\Sigma^-}$ & $\sigma_{exp}$ & $\sigma_{th}$ \\
\hline & & & & &\\
110 & 396$\pm$91 & 200.6  &110& 174$\pm$47& 241.3 \\
120 & 159$\pm$43 & 175.8  &120& 178$\pm$39& 207.2 \\
130 & 157$\pm$34 & 155.9  &130& 140$\pm$28& 180.1 \\
140 & 125$\pm$25 & 139.7  &140& 164$\pm$25& 158.1 \\
150 & 111$\pm$19 & 126.2  &150& 147$\pm$19& 140.0 \\
160 & 115$\pm$16 & 114.9  &160& 124$\pm$14& 125.0 \\
       & & & & &\\
\multicolumn{3}{l}{$r_{R}^{exp}=0.468\pm 0.010$}  &
\multicolumn{2}{l}{$r_{R}^{th }=0.455$} & $\chi^{2}=1.7$ \\
%\hline\hline \\
\end{tabular}
\end{ruledtabular}
\end{center}
\caption{Comparison of the calculated ESC08 and experimental values for the
         52 $YN$-data that were included in the fit. The superscipts
         $RH$ and $M$ denote, respectively,
         the Rehovoth-Heidelberg Ref.~\protect\cite{Ale68}
         and Maryland data Ref.~\protect\cite{Sec68}.
         Also included are (i) 3 $\Sigma^+p$ X-sections at 
         $p_{lab}=400, 500, 650$ MeV from Ref.~\protect\cite{Kanda05},
         (ii) $\Lambda p$ X-sections from Ref.~\protect\cite{Kadyk71}: 
         7 elastic between $ 350 \leq p_{lab} \leq 950$, and 4 inelastic 
         with $p_{lab}= 667, 750, 850, 950$ MeV, and (iii) 3 elastic 
         $\Sigma^-p$ X-sections at $p_{lab}= 450, 550, 650$ MeV from
         Ref.~\cite{Kondo00}. The laboratory momenta
         are in MeV/c, and the total cross sections in mb.}
\label{tab.reslts1}
\end{table}
%-----------------------------------------------------------------
  \begin{figure}[hbt]
  \resizebox{7.5cm}{!}       
% {\includegraphics[200, 50][400,400]{ynplot/lpfits3.ps}}
  {\includegraphics[200, 50][400,400]{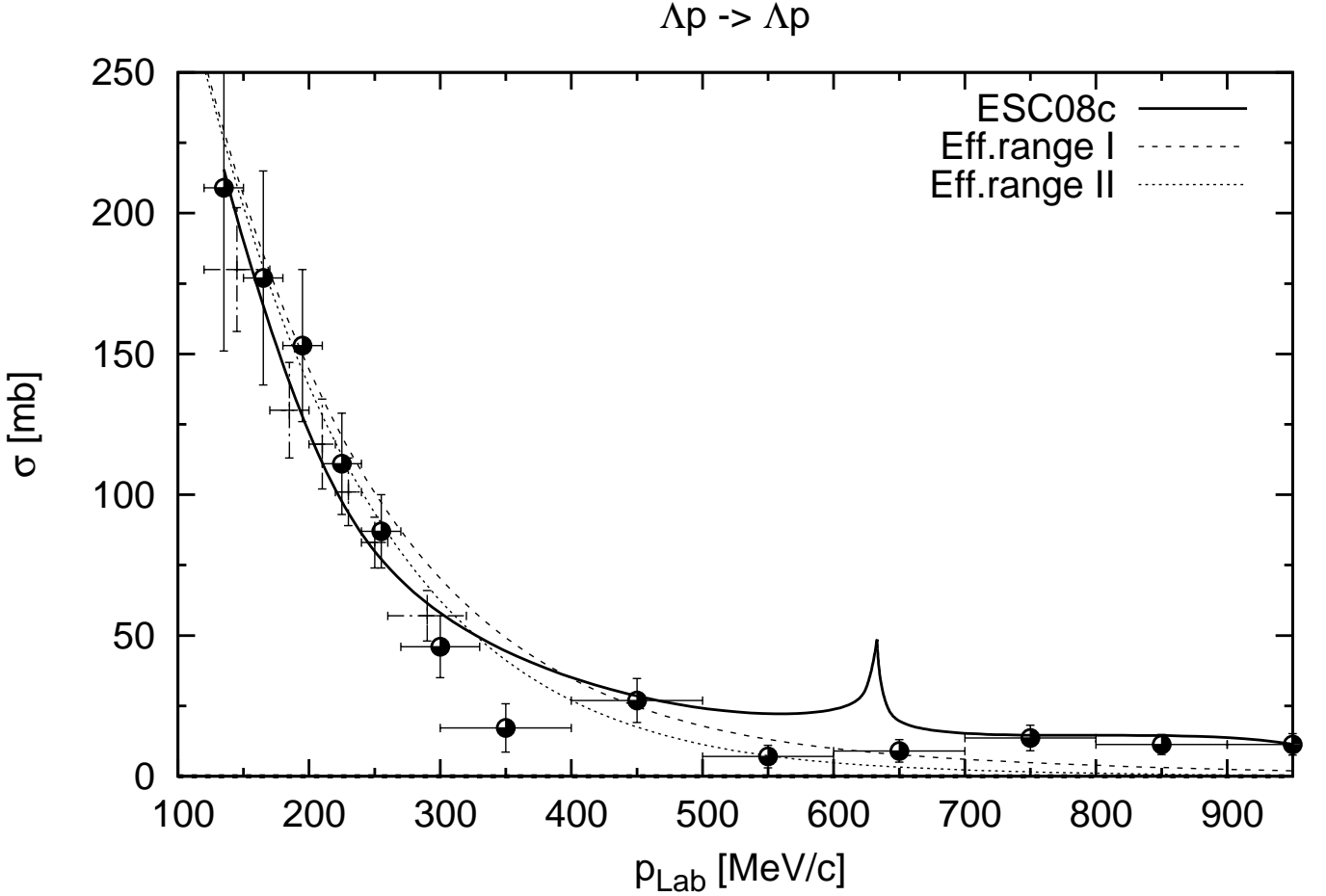}}
 \caption{Model fits total X-sections $\Lambda p$. Rehovoth-Heidelberg-,
 Maryland-, and Berkeley-data }                                  
 \label{fig.lpsig1}
  \end{figure}
%%-----------------------------------------------------------------
  \begin{figure}[hbt]
  \resizebox{3.5cm}{!}       
% {\includegraphics[200,000][400,850]{ynplot/snel2.fits.ps}}
  {\includegraphics[200,000][400,850]{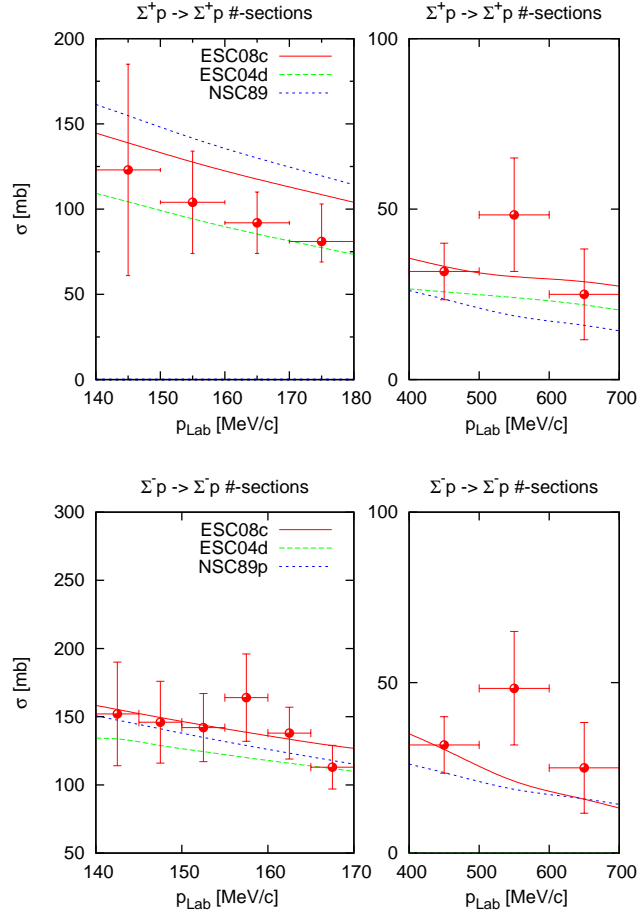}}
 \caption{Model fits total elastic X-sections $\Sigma^{\pm} p$. 
  Rehovoth-Heidelberg-, KEK-data }                                      
 \label{fig.spsig1}
  \end{figure}
%-----------------------------------------------------------------
  \begin{figure}[hbt]
  \resizebox{3.5cm}{!}       
% {\includegraphics[200,000][400,850]{ynplot/sminel.fits.ps}}
  {\includegraphics[200,000][400,850]{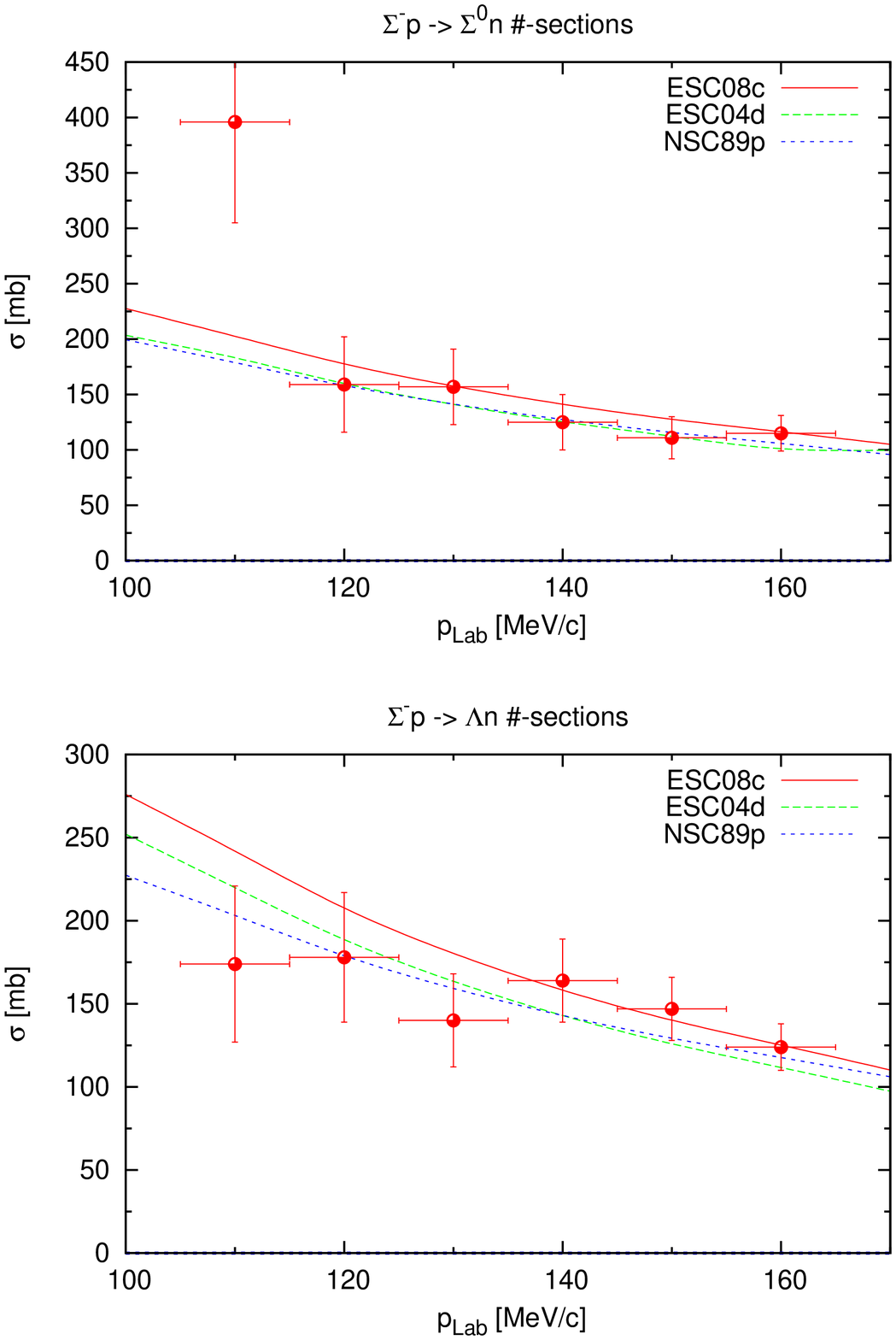}}
 \caption{Model fits total inelastic X-sections 
 $\Sigma^{-} p \rightarrow \Sigma^0 n, \Lambda n$. 
  Rehovoth-Heidelberg-data }                                      
 \label{fig.spsig2}
  \end{figure}
%-----------------------------------------------------------------

% potential figures plot04:
 
The $\chi^2$ on the 52 $Y\!N$ scattering data for the ESC08     
model is given in Table~\ref{tab.reslts1}.                            
The $\Lambda N$ total cross sections have been calculated with $L \leq 2$,
and the $\Sigma N$ total cross sections with $L \leq 1$.
The capture ratio at rest, given in the last
row of the table, for its definition see e.g. \cite{RSY99}.
This capture ratio turns out to be rather constant in the momentum
range from 100 to 170 MeV/$c$. Obviously, for very low momenta the
cross sections are almost completely dominated by $s$ waves, and so
the capture ratio in flight converges to the capture ratio at rest.
For more details on the evaluation of these observables, we refer to earlier
Nijmegen work on this subject.

The $\Sigma^+ p$ nuclear-bar phase shifts as a function of energy are given in
Table~\ref{spphas}. Notice that the $^3S_1$-phase shows repulsion. 
%except for very low energies. This means that the the potential has 
%a weak long range attractive tail.

The $\Lambda N$ nuclear-bar phase shifts as a function of energy are given in
Table~\ref{lpphas}. 
In Fig.~\ref{fig.lpsig1} the $\Lambda p$ total X-sections are shown for ESC08c
together with the data. 
At the $\Sigma N$-threshold the cross section shows a large cusp with a large
D-wave nuclear-bar phaseshift $\delta(^3D_1) = 73.2^o$.                       
This signals the fact that in the $\Sigma N (^3S_1,I=1/2)$-state there is a strong attraction,
with a deuteron-like virtual bound-state on the unphysical sheet.
Also, in Fig.~\ref{fig.lpsig1} we show the cross sections in the effective range 
approximation, dashed lines I and II. Line II is including the shape parameter 
in the effective range expansion. the two-term effective range expansion
with the a and r parameters describes the s-wave phases well up to
$p_\Lambda \approx 400$ MeV/c.

%------------------------------------------------------------------
% parameters: parbbsc.ESC08c.best14june
\begin{table}
\caption{ESC08c nuclear-bar $\Sigma^{+} p$ phases in degrees.}
\begin{ruledtabular}
\begin{tabular}{crrrrrrrrrr} & & & & & & & & & & \\ 
%\hline\hline
 $p_{\Sigma^{+}}$ & 100 & 200 & 300 & 400 & 500 &  600 & 700 & 
 800 & 900 & 1000 \\ \hline
     &    &     &     & & & &  \\
 $T_{\rm lab}$ & 4.2 & 16.7 & 37.3& 65.5 &100.8& 142.8 &190.7& 
 244.0&302.1 & 364.5 \\ \hline
     &    &     &     & &  & &  &\\
 $^{1}S_{0}$ & 32.21 & 38.49 & 33.16 & 25.30 & 16.89 &
 8.57 & 0.55 &--7.10 & -14.38 &--21.29\\
 $^{3}S_{1}$ & -5.44 &--11.95 & -18.40 &--24.84 & -30.98 &
 --36.54 & -41.38 & --45.51& -49.04 &--52.11\\
 $\epsilon_{1}$ &-0.36 &--1.74 & -3.22 &--4.32 &--4.96 &--5.23 &
 --5.23  &--5.09 &--4.86 & --4.60  \\
 $^{3}P_{0}$ & 0.93 & 4.86 & 8.49 &  9.37 & 7.61 & 4.12 &
 --0.25 &--4.96 & --9.70 &--14.27 \\
 $^{1}P_{1}$ & 0.42 & 2.27 & 4.68 & 6.64 & 7.53  & 7.20 & 
 5.79 & 3.59 & 0.81 &--2.33 \\
 $^{3}P_{1}$ & --0.56 &--2.88 & --5.91 &--9.20 & --12.76 &--16.57&
 --20.54 & --24.55& --28.50 &--32.33 \\
 $^{3}P_{2}$ & 0.11 & 0.92 & 2.65 &  4.76  & 6.71 &  8.25 &
 9.37 & 10.09 & 10.41 &10.31 \\
 $\epsilon_{2}$ & --0.03&--0.35&--1.02&--1.77 &--2.38 &--2.78&
 --2.95 & --2.91& --2.73 &--2.46 \\
 $^{3}D_{1}$ & 0.02 & 0.29 & 0.83 & 1.33  & 1.40 &  0.76 &
 --0.68 & --2.85 & --5.60 &--8.76  \\
 $^{1}D_{2}$ & 0.02 & 0.30& 0.97 &  2.01 & 3.38 & 4.93  & 
 6.40 & 7.57 & 8.24 &  8.33\\
 $^{3}D_{2}$ & --0.03 &--0.42& --1.25 &--2.27 & --3.41 &--4.74 &
 --6.34 & --8.21 & --10.33 &--12.65 \\
 $^{3}D_{3}$ & 0.00 & 0.05 & 0.25 & 0.63  & 1.05 &  1.32 & 
 1.34 & 1.09  & 0.64 &  0.08\\
     &    &     &     & & & & & &  & \\
\end{tabular}
\end{ruledtabular}
\label{spphas} 
\end{table}
 
% parameters: parbbsc.ESC08c.best14june
\begin{table}
\caption{ESC08c nuclear-bar $\Lambda p$ phases in degrees.}
\begin{ruledtabular}
\begin{tabular}{crrrrrrr} & & & & & & & \\ 
%\hline\hline
 $p_{\Lambda}$ & 100 & 200 & 300 & 400 & 500 & 600 &633.4 \\
 \hline    &    &     &     & & & & \\
 $T_{\rm lab}$ & 4.5 & 17.8 & 39.6 & 69.5 & 106.9& 151.1& 167.3 \\ 
 \hline    &    &     &     & & & & \\
 $^{1}S_{0}$ & 25.14 & 30.86 & 27.49 & 21.11 & 13.88 &  6.72 &  4.72 \\
 $^{3}S_{1}$ & 18.89 & 25.04 & 23.33 & 18.55 & 12.80 &  7.27 & 6.14  \\
 $\epsilon_{1}$ & 0.04 & 0.12 & 0.12 &--0.01 &--0.09 & 1.73 &  9.29  \\
 $^{3}P_{0}$ & 0.04 & 0.22  &  0.21 &--0.58 & --2.37 &--4.89 &--5.74 \\
 $^{1}P_{1}$ &--0.08 &--0.63 & --2.02 & --4.40 &--7.61 &--11.27 & --12.50  \\
 $^{3}P_{1}$ &  0.02 &  0.00 &--0.34 &--1.22 & --2.59&--3.88 &--3.89  \\
 $^{3}P_{2}$ & 0.12 &  0.77 & 1.98 & 3.33 & 4.48 & 5.31 & 5.52  \\
 $\epsilon_{2}$ & 0.00 &--0.00 &--0.04 &--0.15 & --0.32 & --0.53 &--0.63 \\
 $^{3}D_{1}$ & 0.00 & 0.08  & 0.57  & 2.21 & 6.74 & 24.38  & 73.17  \\
 $^{1}D_{2}$ & 0.00 & 0.06 & 0.38 & 1.17 & 2.48 & 4.19 & 4.81  \\
 $^{3}D_{2}$ & 0.00 & 0.08 & 0.44 & 1.25 & 2.53 & 4.10 & 4.66   \\
 $^{3}D_{3}$ & 0.00 & 0.05 & 0.25 & 0.71 & 1.38 & 2.10 & 2.32   \\
     &    &     &     &  & & & \\
%\hline\hline
\end{tabular}
\end{ruledtabular}
\label{lpphas} 
\end{table}

\begin{widetext}
%%--- CSB----------------------------------------------------------------
% CSB, parameters: comparison ESC08c, NSC97e, NSC89, HC-D
\begin{table}[hbt]
\caption{ Comparison $\Lambda p$ and $\Lambda n$ scattering lengths and 
effective ranges in fm for different Nijmegen models.}
\label{tab.csb.lp-ln2}                
\begin{center}
%\begin{tabularx}{10.5cm}{l@{\hspace{5mm}}|
 \begin{tabular*}{10.5cm}{l@{\hspace{5mm}}|
 c@{\hspace{1cm}}c@{\hspace{1cm}}c@{\hspace{1cm}}|
 c@{\hspace{1cm}}c@{\hspace{1cm}}c@{\hspace{1cm}}|}
 \hline\hline
  \multicolumn{1}{l|}{}
 &\multicolumn{1}{c}{}
 &\multicolumn{2}{c|}{\hspace{-5mm}$\Lambda p$}  
 &\multicolumn{1}{c}{}
 &\multicolumn{2}{c|}{\hspace{-5mm}$\Lambda n$}\\
 Model && $a_s$ & $a_t$ && $a_s$ & $a_t$ \\  
 \hline
 ESC08c&&-2.46  & -1.73  &&-2.62  & -1.72 \\
 NSC97e&&-2.10  & -1.86  &&-2.24  & -1.83  \\
 NSC97f&&-2.51  & -1.75  &&-2.68  & -1.67  \\
 NSC89 &&-2.73  & -1.48  &&-2.86  & -1.24  \\
 HC-D  &&-1.77  & -2.06  &&-2.03  & -1.84  \\
% && & && & \\  
 \hline
 Model && $r_s$ & $r_t$ && $r_s$ & $r_t$ \\  
 ESC08c&& 3.14  &  3.55  && 3.17  &  3.50 \\
 NSC97e&& 3.19  &  3.19  && 3.24  &  3.14  \\
 NSC97f&& 3.03  &  3.32  && 3.07  &  3.34  \\
 NSC89 && 2.87  &  3.04  && 2.91  &  3.33  \\
 HC-D  && 3.78  &  3.18  && 3.66  &  3.32  \\
 \hline
 \hline
 \end{tabular*}
%\end{tabularx}
\end{center}
\end{table}
%--- CSB----------------------------------------------------------------
% CSB, parameters: parbbsc.ESC08c.best14june
%--- CSB----------------------------------------------------------------
%-------------------------------------------------------------------------
In Table~\ref{tab.csb.lp-ln2} the low-energy parameters for 
$\Lambda p$ and $\Lambda n$ are shown. 
%In the first row these parameters 
%are given when the $\Lambda\Sigma^0$-mixing \cite{Dal64} is turned off.
The singlet and triplet parameters are displayed
with the $\Lambda\Sigma^0$-mixing turned on for pseudoscalar-, vector-,
scalar-, meson-pairs-, and ps-ps- exchanges.               
%In the last line the final values are given with all CSB-effects turned on.
Notice that the effect for the scalar mesons of the $\Lambda\Sigma^0$-mixing is 
zero because $\alpha_s=1.00$. It is clear from these tables that the total effect
of the $\Lambda\Sigma^0$-mixing is about given by pseudoscalar and
vector exchanges.
The differences in the scattering lengths are 
\begin{subequations}
\label{eq:scatl.dif} 
\begin{eqnarray}
 \Delta a_s &=& a_s(\Lambda p)-a_s(\Lambda n) = +0.164 fm, \\
 \Delta a_t &=& a_t(\Lambda p)-a_t(\Lambda n) = -0.010 fm.
\end{eqnarray}
\end{subequations}
These differences are comparable to those for the soft-core OBE models
\cite{MRS89,RSY99}, and therefore predict a too small binding energy
difference in the A=4 hypernuclei, which is 
$\Delta B_\Lambda(exp)= B_\Lambda(^4_\Lambda He)-B_\Lambda(^4_\Lambda H) =
(0.29 \pm 0.06)$ MeV.
This in contrast to the HC-model D,
which has a much larger $\Delta a_t$ \cite{NRS77}.
It appeared that CSB via meson-mixing, like $\pi^0-\eta, \rho^0-\omega$ etc., 
is small and does not improve the CSB for ESC08, 
which is understandable in view of the large cancellations. 
However, as a consequence of the ESC-models there is a three-body force 
produced by the MPE-interactions, which are fixed by the BB-fit.
Therefore, the CSB in the $\Lambda NN$-potential may improve the
CSB predictions significantly. 

%--- CSB----------------------------------------------------------------

In Table~\ref{tab.ESC04a-d.4} we list the $\Sigma^+p$ and $\Lambda \Lambda$ 
scattering lengths and effective ranges. 
Here, $(a_s,r_s)$ are these quantities for $\Sigma^+p(^1S_0)$ and
$(a_t,r_t)$ for $\Sigma^+p(^3S_1)$. 
Notice that the difference between ESC08a$^{\prime\prime}$ and ESC08c is small.
This is because with SU(3) the $^1S_0$-wave is constrained by NN, because the 
$^1S_0$-states in NN and $\Sigma^+p$ are both in the $\{27\}$-irrep,
and so there is 
little room for variations in the $^3S_1$-wave because of the X-section fit.
Therefore, much extra repulsion in the triplet wave is impossible.
%----------------------------------------------------------------------------------
% parameters: parbbsc.ESC08c.best14p   
\begin{table}[hbt]
\caption{$\Sigma^+p$ scattering lengths and effective ranges in fm.}
\label{tab.ESC04a-d.4}
\begin{center}
\begin{ruledtabular}
 \begin{tabular}{|l|cc|cc}
%\hline
 Model & $a_s$ & $a_t$ & $r_s$ & $r_t$ \\  
 \hline
%ESC04a &-4.09 & -0.020 & 3.49 & -3356  \\
%ESC04b &-2.87 & +0.179 & 4.10 & -34.20  \\
%ESC04c &-3.87 & +0.077 & 3.72 & -253.5  \\
ESC04d &-3.43 & +0.217 & 3.98 & -28.94  \\
% \hline
ESC08a$^{\prime\prime}$ &-3.85 & +0.62 & 3.40 & -2.13  \\
ESC08c &-3.91  & +0.61  & 3.41  & -2.35   \\
%\hline
 \end{tabular}
\end{ruledtabular}
\end{center}
\end{table}
%----------------------------------------------------------------------------------

%-----------------------------------------------------------------
 In Fig.~\ref{totfig1} we plot the total potentials for the S-wave channels
$\Lambda N \rightarrow \Lambda N$, $\Lambda N \rightarrow \Sigma N$,
and $\Sigma N  \rightarrow \Sigma N$. Note the for the soft-core model typical
structure of the $\Sigma^{+}p(^3S_1)$-potential. Most contributions to the 
spin-spin potentials are proportional to ${\bf k}^2$, and hence have zero 
volume integral. This causes the attraction in the inner region.

\noindent Figures for the OBE-, TME-, and MPE-contributions are similar to those for
the ESC04-model and have been displayed in Ref.~\cite{Rij04b} and we refer the 
interested reader to this reference. Likewise for the contributions of the 
various types of mesons to the OBE-potentials, ans also for the contributions of the
different kind of pair-potentials to MPE.

  \begin{figure}[hbt]
  \resizebox{5.5cm}{!}       
% {\includegraphics[200,000][400,850]{ynfig/plot08/yntot08.ps}}
  {\includegraphics[200,000][400,850]{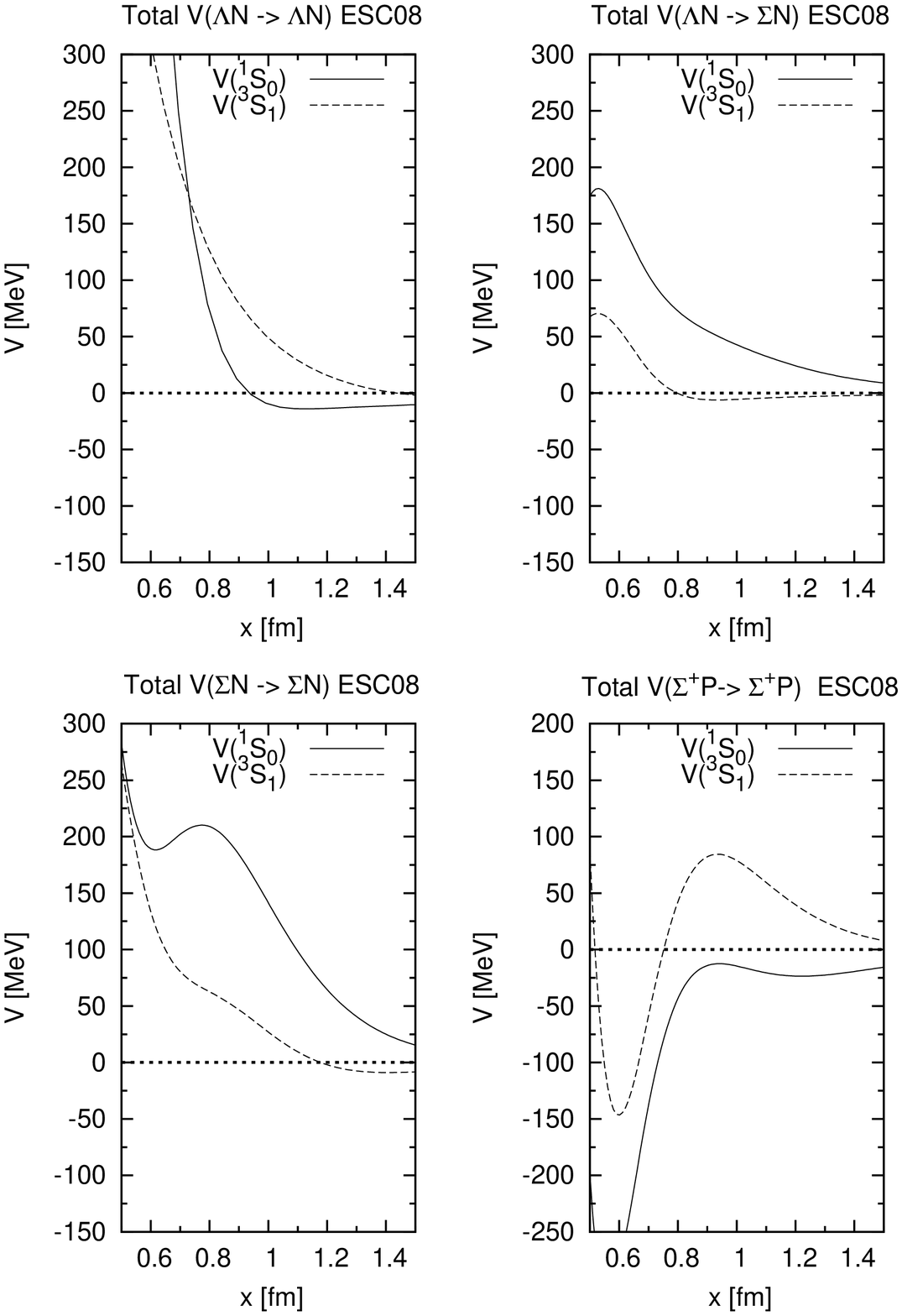}}
 \caption{Total potentials in the partial waves $^1S_0$ and $^3S_1$, for 
 $I=1/2$- and $I=3/2$-states.}
 \label{totfig1}
  \end{figure}

%%---------------------------------------------------------------------------------
%% plots su(3)-irreps:
%%---------------------------------------------------------------------------------
\subsection{Potentials in SU(3)-irreps}
\label{sec:8b}
In Fig.~\ref{irrep01.fig} the potentials $V_{\{\mu\}}$[GeV] in the SU(3) 
representations for
BB-channels are shown. The solid/red curves include SU(3)-breaking and 
the dashed/green ones
are the SU(3)-symmetric curves. In the latter average  masses are used for the
SU(3)-nonets. The curves resemble strongly (qualitatively) those
obtained in lattice QCD, except for the $\{1\}$-irrep \cite{Inou12}.
In the ESC-model the behavior in the typical for potentials with a strong 
spin-spin part, because the spin-spin potentials from pseudoscalar-
and vector-exchange have zero volume integral forcing them to change sign
for $r \sim 0.5$ fm.

{The similarity between the meson-exchange and QCD-lattice potentials 
shows that with the ESC realization of the program starting from the 
nuclear force, using SU$_f$(3)-symmetry and the QM, 
a realistic generalization to the BB-force is achieved.}\\
%%---------------------------------------------------------------------------------
\begin{center}
 \begin{figure}   
   \vspace*{15mm}
  \begin{center}
 \resizebox{10.cm}{ 7.43cm}        
%{\includegraphics[-50,-20][454,700]{ynplot/esc08c.irrep0.all.ps}}
%{\includegraphics[100,00][554,690]{ynplot/esc08c.irrep01.ps}}
 {\includegraphics[100,00][554,690]{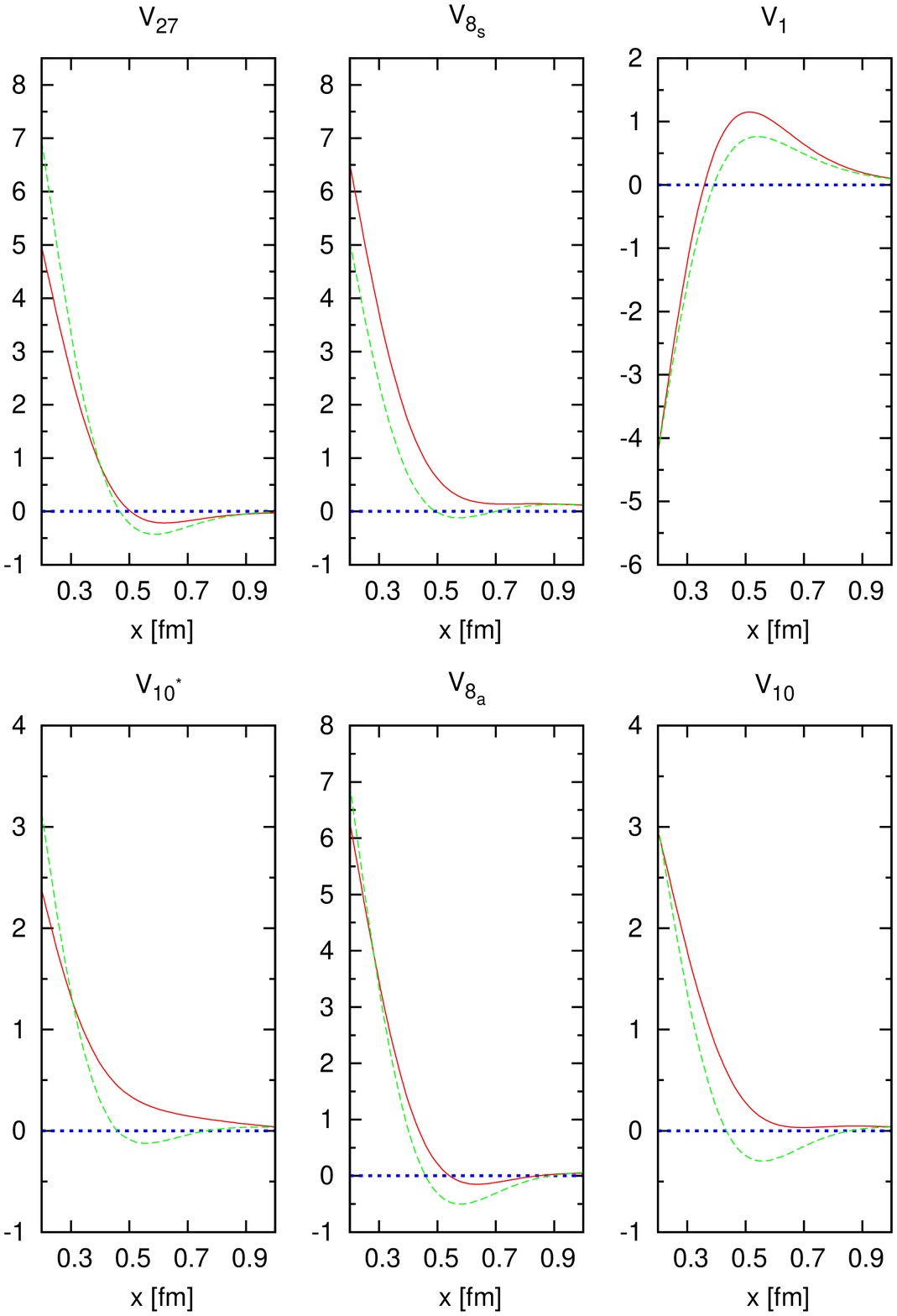}}
   \vspace*{-5mm}
\caption{Exact flavor SU(3)-symmetry: $m_\pi=m_K=m_\eta= 450$ MeV}
\label{irrep01.fig}
  \end{center}
 \end{figure}
\end{center}
%%---------------------------------------------------------------------------------

%-----------------------------------------------------------------
% NEW G-MATRIX BEGIN: for parbbsc.ESC08c.best14june: 
%-----------------------------------------------------------------
%\onecolumngrid

\section{Analyses with G-matrix interactions}
\label{sec:10} 
%--------------------------------------------------------------------------

\subsection{$\Lambda$ and $\Sigma$ in nuclear matter}

The G-matrix theory gives a good starting point 
for studies of hyperonic many-body systems on the basis of
free-space $Y\!N$ interaction models~\cite{Yam85,Yam94,Yam10}.
Here, the correlations induced by hyperonic coupling interactions 
such as $\Lambda\!N$-$\Sigma\!N$ ones are renormalized into 
single-channel G-matrices.
These G-matrix interactions are considered as effective interactions 
used in models of hypernuclei. 
Thus, the hypernuclear phenomena and the underlying $Y\!N$ interaction 
models are linked through the $Y\!N$ G-matrix interactions, and the 
hypernuclear information gives a feedback to the interaction models.
Here, the properties of $\Lambda\!N$ and $\Sigma\!N$ sectors of ESC08c
in nuclear medium are studied on the basis of the G-matrix theory.

In Refs.\cite{YFYR13,YFYR14}, the three-body interaction is added on 
ESC08c, being composed of the multi-pomeron exchange repulsive potential
(MPP) and the phenomenological three-baryon attraction (TBA).
The effective two-body potential derived from MPP is given as
\begin{eqnarray}
V_{MPP}(r;\rho)
= g_P^{(3)} (g_P)^3 \frac{\rho}{{\cal M}} \cdot
\frac{1}{4\pi} \frac{4}{\sqrt{\pi}}
\left(\frac{m_P}{\sqrt{2}}\right)^3
\exp\left(-\frac12 m_P^2 r^2 \right) \ ,
\label{eq:tbf.3}
\end{eqnarray}
where the pomeron mass $m_P$ and the pair pomeron coupling $g_P$ 
are fitted to the NN-data etc.
In a similar way, one can obtain an effective two-body potential 
with a quartic pomeron coupling $g_P^{(4)}$.
TBA also is given by a density-dependent two-body potential
\begin{eqnarray*}
V_{TBA}(r;\rho)= V^0_{TBA}\, \exp(-(r/2.0)^2)\, \rho\, 
\exp(-\eta \rho)\, (1+P_r)/2 \ ,
\end{eqnarray*}
$P_r$ being a space-exchange operator.
The values of $g_P^{(3)}$, $g_P^{(4)}$ and $V^0_{TBA}$ are adjusted 
to reproduce the angular distribution of $^{16}$O$+^{16}$O elastic 
scattering at $E/A=70$ MeV with use of the G-matrix folding potential 
and the value $E \sim -16$ MeV of the energy per nucleon in 
normal-density nuclear matter.
Among three sets given in Refs.\cite{YFYR13,YFYR14},
we adopt here the set MPa ($g_P^{(3)}=2.34$, $g_P^{(4)}=30.0$,
$V^0_{TBA}=-32.8$ MeV, $\eta=3.5$ fm$^3$), which gives rise to 
the stiff EoS of neutron matter to reproduce a maximum mass 
$2 M_{\odot}$ of a neutron star.

MPP works universally in all baryon-baryon channels according to 
its definition. The above values of $g_P^{(3)}$ and $g_P^{(4)}$
are adopted in this work.
Assuming here that TBA works also in $Y\!N$ channels, 
its strength is adjusted to reproduce well energy spectra of
$\Lambda$ hypernuclei. As explained later, we take 
$V^0_{TBA}=-21.0$ MeV differently from the above value.
Hereafter, the interaction ESC08c+MPP+TBA is denoted as ESC08c$^+$.

We start from the channel-coupled G-matrix equation 
for the baryon pair $B_1 B_2$ in nuclear matter~\cite{Yam85}, 
where $B_1 B_2 = \Lambda N$ and $\Sigma N$: 
\begin{eqnarray}
G_{cc_0}=v_{cc_0} + \sum_{c'} 
 v_{cc'} {Q_{y'} \over \omega -\epsilon_{B'_1}-\epsilon_{B'_2} +\Delta_{yy'}}
G_{c' c_0} \ ,
\label{eq:GM1}
\end{eqnarray}
where $c$ denotes a $Y\!N$ relative state $(y, T, L, S, J)$ with $y=(B_1,B_2)$. 
$S$ and $T$ are spin and isospin quantum numbers, respectively.
Orbital and total angular momenta are denoted by $L$ and $J$,
respectively, with ${\bf J}={\bf L}+{\bf S}$.
Then, a two-particle state is represented as $^{2S+1}L_J$.
In Eq.~(\ref{eq:GM1}), $\omega$ gives the starting energy in 
the starting channel $c_0$.
$\Delta_{yy'}= M_{B_1}+M_{B_2}-M_{B'_1}-M_{B'_2}$ denotes 
the mass difference between two baryon channels.
The Pauli operator $Q_y$ acts on intermediate nucleon states
in a channel $y=(B_1,B_2)=(\Lambda N$,$\Sigma N$).
We adopt here the continuous (CON) choice for intermediate 
single particle potentials in the G-matrix equation.
The G-matrix equation~(\ref{eq:GM1}) is represented in the
coordinate space, whose solutions give rise to G-matrix elements.
The hyperon single particle (s.p.) energy $\epsilon_Y$ 
in nuclear matter is given by
\begin{eqnarray}
\epsilon_Y(k_Y)={\hbar^2k_Y^2 \over 2M_Y} + U_Y(k_Y) \ ,
\label{eq:GM4}
\end{eqnarray}
where $k_Y$ is the hyperon momentum.
The potential energy $U_Y$ is obtained self-consistently
in terms of the G-matrix as
\begin{eqnarray}
U_Y(k_Y) &=& \sum_{|{\bf k}_N|} \langle {\bf k}_Y {\bf k}_N
\mid G_{YN}(\omega=\epsilon_Y(k_Y)+\epsilon_N(k_N)) \mid
{\bf k}_Y {\bf k}_N \rangle \ .
\label{eq:GM5}
\end{eqnarray}
Here, we need not only on-shell single particle potentials
but also off-shell ones because of adopting the CON choice.

First, let us calculate $\Lambda$ binding energies in nuclear matter. 
In Table~\ref{Gmat-ULam1} we show the potential energies $U_\Lambda(\rho_0)$
for a zero-momentum $\Lambda$ and their partial-wave contributions
in $^{2S+1}L_J$ states at normal density $\rho_0$ ($k_F$=1.35 fm$^{-1}$),
where a statistical factor $(2J+1)$ is 
included in each contributin in $^{2S+1}L_J$ state. 
The value specified by $D$ gives the sum of $^{2S+1}D_J$ contributions.
Results for ESC08c and ESC08c$^+$ are found to be not so different from
each other, because repulsive contributions of MPP are rather cancelled 
by attractive TBA contributions in normal and lower density regions.
It should be noted here that the important role of MPP is to stiffen the 
EOS remarkably by strongly repulsive contributions in high density regions.

The contributions to $U_\Lambda$ from $S$-state spin-spin components 
can be seen qualitatively in values of 
$U_{\sigma \sigma}=(U_\Lambda(^3S_1)-3U_\Lambda(^1S_0))/12$.
These values of $U_{\sigma \sigma}$ also are given in Table~\ref{Gmat-ULam1}.
In the same treatment, we obtain $U_{\sigma \sigma}$=1.54 and 0.92 MeV
for NSC97f and NSC97e, respectively.
Various analyses suggest that the reasonable value of 
$U_{\sigma \sigma}$ is between these values~\cite{Yam10}.
Then, one should note that the $U_{\sigma \sigma}$ values for 
ESC08c/c$^+$ are reasonable.

\begin{table}[ht]
\caption{Values of $U_\Lambda(\rho_0)$ and partial wave
contributions in $^{2S+1}L_J$ states from the G-matrix
calculations (in MeV).
The value specified by $D$ gives the sum of $^{2S+1}D_J$ contributions.
Contributions from $S$-state spin-spin interactions are given by
$U_{\sigma \sigma}=(U_\Lambda(^3S_1)-3U_\Lambda(^1S_0))/12$.}
\label{Gmat-ULam1}
\vskip 0.2cm
\begin{ruledtabular}
 \begin{tabular}{l|ccccccc|c|c}
% \hline\hline
& $^1S_0$ & $^3S_1$ & $^1P_1$ & $^3P_0$ & $^3P_1$ & $^3P_2$ & $D$ 
& $U_\Lambda$ & $U_{\sigma \sigma}$  \\
 \hline
ESC08c &$-$13.1& $-$26.5& 2.4 & 0.1  & 1.1 & $-$3.1 & $-$1.6 & $-$40.8 &1.07  \\
ESC08c$^+$ &$-$12.6& $-$25.4& 2.9 & 0.3  & 1.6 & $-$2.1 & $-$2.3 & $-$37.6 &1.03  \\
% \hline
 \end{tabular}
\end{ruledtabular}
\end{table}

Next, $\Sigma$ binding energies in nuclear matter are obtained by solving 
the $\Sigma\!N$ starting channel G-matrix equation for ESC08c/c$^+$. 
In Table~\ref{Gsig1} we show the potential energies $U_\Sigma(\rho_0)$
for a zero-momentum $\Sigma$ and their partial-wave contributions in
$(^{2S+1}L_J,T)$ states for ESC08c/c$^+$. 
It should be noted here that 
the strongly repulsive contributions in $^3S_1$ $T=3/2$ and
$^1S_0$ $T=1/2$ states are due to the Pauli-forbidden effects
in these states, being taken into account by strengthening the 
pomeron coupling in the ESC08 modeling.
Experimentally, the repulsive $\Sigma$-nucleus potentials are suggested
in the observed $(\pi^-,K^-)$ spectra.~\cite{Nou02,Harada05,Kohno06}
It is a future problem to calculate $(\pi^-,K^-)$ spectra
with use of G-matrix folding potentials, and to select out
a reasonable $\Sigma N$ interaction model.

\begin{table}[ht]
\caption{Values of $U_\Sigma(\rho_0)$ at normal density and partial 
wave contributions in $(^{2S+1}L_J,T)$ states for ESC08c/c$^+$ (in MeV).} 
\label{Gsig1}
\vskip 0.2cm
\begin{ruledtabular}
 \begin{tabular}{|l|c|rrrrrrr|r|}
% \hline\hline
model & $T$ & $^1S_0$ & $^3S_1$ & $^1P_1$ & $^3P_0$ & $^3P_1$ & $^3P_2$ 
& $D$ & $U_\Sigma$  \\
 \hline
ESC08c & $1/2$ &11.1 & $-$22.0 & 2.4 & 2.1 & $-$6.1 & $-$1.0 & $-$0.7 &  \\
&$3/2$& $-$12.8& 30.7 & $-$4.8 & $-$1.8 & 6.0 & $-$1.4 & $-$0.2 
&   +1.4   \\
 \hline
ESC08c$^+$ & $1/2$ &11.1 & $-$20.4 & 2.6 & 2.1 & $-$5.8 & $-$0.6 & $-$0.8 &  \\
&$3/2$& $-$11.9& 31.8 & $-$4.2 & $-$1.6 & 6.4 & $-$0.4 & $-$0.6
&   +7.9    \\
%  \hline
 \end{tabular}
\end{ruledtabular}
\end{table}

\begin{figure}[ht]
\begin{center}
\includegraphics*[width=13cm,height=8cm]{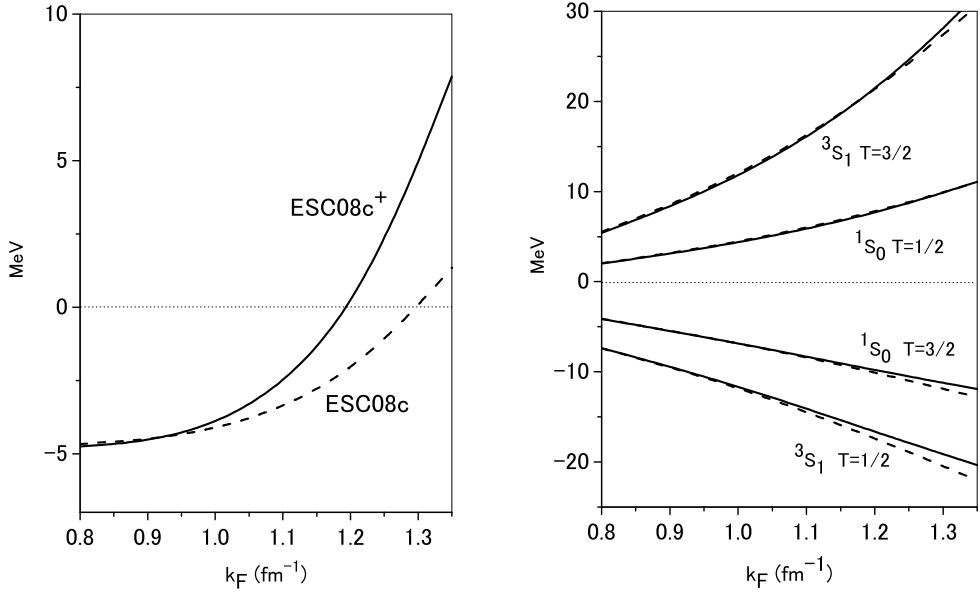}
\caption{In the left (right) panel, the values of $U_\Sigma$
(partial-wave contributions) are drawn as a function of $k_F$
by solid and dashed curves for ESC08c$^+$ and ESC08c, respectively.
}
\label{Ufig}
\end{center}
\end{figure}

In the left (right) panel of Fig.~\ref{Ufig},
$U_\Sigma$ values (their $S$-state contributions)
are drawn as a function of $k_F$ for ESC08c$^+$ and ESC08c
by solid and dashed curves, respectively.
It is demonstrated that the repulsive $U_\Sigma$ values are
due to $T=3/2$ $^3S_1$ and $T=1/2$ $^1S_0$ contributions,
and the repulsions are enhanced by the MPP contributions.

\subsection{$\Lambda N$ G-matrix interactions}
For applications to various hypernuclear problems, it is convenient
to construct $k_F$-dependent effective local potentials ${\cal G}(k_F;r)$
which simulate the G-matrices.
Here we parameterize them in a three-range Gaussian form 
\begin{eqnarray}
{\cal G}(k_F,r)= \sum^3_{i=1}\, (a_i+b_i k_F +c_i k_F^2) \,
\exp {(-r^2/\beta_i^2)} \ .
\label{eq:GM7}
\end{eqnarray}
The parameters $(a_i, b_i, c_i)$ are determined so as to simulate 
the calculated G-matrix for each $^{2S+1}L_J$ state. 
The procedures to fit the parameters are given in Ref.~\cite{Yam10}.
The obtained parameters for ESC08c are shown in Table~\ref{Gmat-L1}. 
For ESC08c$^+$, contributions from MPP+TBA are represented by
modifying the second-range parts of ${\cal G}(k_F,r)$ for ESC08c
by $\Delta {\cal G}(k_F,r)=(a+b k_F+c k_F^2) \exp -(r/0.9)^2$.
The parameters for $\Delta {\cal G}(k_F,r)$ are given 
in Table~\ref{Gmat-L2}.

Here, it is worthwhile to comment about a qualitative feature 
of $\Delta {\cal G}(k_F,r)$.
The MPP contributions increase rapidly with matter density:
In high (low) density region, they are very large (small), and
rather cancelled by TBA at normal-density region. Then, 
net contributions of MPP+TBA given by $\Delta {\cal G}(k_F,r)$  
are attractive for smaller values of $k_F$ than 1.35 fm$^{-1}$.

The solved G-matrices include not only $\Lambda\!N$-$\Lambda\!N$
diagonal parts but also $\Lambda\!N$-$\Sigma\!N$ coupling parts,
and it is possible to extract such coupling parts to treat
$\Lambda\!N$-$\Sigma\!N$ mixing problems.
The $\Lambda N$-$\Sigma N$ coupling interaction 
is determined so that its matrix elements in $k$ space simulate
the corresponding G-matrix elements and its radial form
tend to that of the bare interaction in the outermost region.
In Table~\ref{Gmat-CP1} (Table~\ref{Gmat-CP2}), the parameters 
of the central (tensor) parts of $\Lambda N$-$\Sigma N$ and 
$\Sigma N$-$\Sigma N$ interactions in $S$ states
are given in a three-range Gaussian ($r^2$-Gaussian) form.
Here, the $k_F$ dependences are represented in the same form
as the above diagonal parts.
These coupling interactions can be used for $\Lambda N$-$\Sigma N$
mixing problems together with the $\Lambda N$-$\Lambda N$ diagonal
interactions in the Table~\ref{Gmat-L1}.\\

\end{widetext}
\newpage
 \twocolumngrid
\noindent The SLS interactions ${\cal G}_{SLS}(r)$ are derived from 
G-matrices ${\cal G}^{JS}_{LL'}(r)$ with $S=1$ by 
the linear transformation.
The ALS G-matrix interaction ${\cal G}_{ALS}$ between 
$^3$P$_1$ and $^1$P$_1$ states is given so that its matrix elements 
in $k$ space simulate the corresponding G matrix elements 
$\langle ^3P_1 \mid G \mid ^1P_1 \rangle$.
Because $\langle ^3P_1 \mid G \mid ^1P_1 \rangle$ and
$\langle ^1P_1 \mid G \mid ^3P_1 \rangle$ are different
from each other, we derive ${\cal G}_{ALS}$ from their averaged values.
The SLS and ALS G-matrix interactions obtained as a function of $k_F$ 
are represented in three-range Gaussian forms, the parameters of which 
are given for ESC08c in Table~\ref{Gmat-SO}.

%--------------------------------------------------------------------
\begin{table}[ht]
\begin{center}
\caption{Parameters of YNG-ESC08c Continuous choice : 
${\cal G}(k_F;r)= \sum_{i=1}^3 (a_i+b_i k_F+c_i k_F^2) \exp -(r/\beta_i)^2$}
\label{Gmat-L1}
\vskip 0.2cm
%\begin{ruledtabular}
\begin{tabular}{|ccrrr|}\hline
&  $\beta_i$ & 0.50 &  0.90  &  2.00  \\
\hline
      & a  & $-$3434.  &    396.0  & $-$1.708  \\
$^1E$ & b  &    6937.  & $-$1057.  &   0.0    \\
      & c  & $-$2635.  &    415.9  &   0.0    \\
\hline
      & a  & $-$1933.  &    195.4  & $-$1.295  \\
$^3E$ & b  &    4698.  & $-$732.8  &   0.0    \\
      & c  & $-$1974.  &    330.1  &   0.0    \\
\hline
      & a  &    206.1  &    67.89  & $-$.8292  \\
$^1O$ & b  & $-$30.52  &    34.11  &   0.0    \\
      & c  &    16.23  &    2.471  &   0.0    \\
\hline
      & a  &    2327.  & $-$254.1  & $-$.9959  \\
$^3O$ & b  & $-$2361.  &    202.6  &   0.0    \\
      & c  &    854.3  & $-$43.71  &   0.0    \\
\hline
\end{tabular}
%\end{ruledtabular}
\end{center}
\end{table}

\begin{table}[ht]
\begin{center}
\caption{$\Delta {\cal G}(k_F;r)=  (a+b k_F+c k_F^2) \exp -(r/0.9)^2$}
\label{Gmat-L2}
\vskip 0.2cm
%\begin{ruledtabular}
\begin{tabular}{|crrrr|}
\hline
   &  $^1E$   & $^3E$    &  $^1O$   &  $^3O$  \\
\hline
 a &    20.71 &    19.16 &    26.31 &    24.95  \\
 b & $-$51.74 & $-$49.31 & $-$73.58 & $-$71.92  \\
 c &    28.84 &    27.30 &    64.01 &    66.73  \\
\hline
\end{tabular}
%\end{ruledtabular}
\end{center}
\end{table}

\begin{table}[ht]
\begin{center}
\caption{Central coupling parts of G-matrix interactions
for ESC08c, 
represented in a Gaussian form
$\sum_{i=1}^3 (a_i+b_i k_F+c_i k_F^2) \exp(-(r/\beta_i)^2$.
}
\label{Gmat-CP1}
\vskip 0.2cm
%\begin{ruledtabular}
\begin{tabular}{|cc|rrr|}\hline\hline
&  $\beta_i$ & 0.50 & 0.90 & 2.00 \\
\hline
                & a &     4596. & $-$702.3  &    8.525  \\
$\Lambda N$-$\Sigma N$ $^1S_0$  & b & $-$7150.  &   1160. &  0.0 \\
                & c &     2760. & $-$438.4  &     0.0   \\
\hline
                & a  &  $-$311.9 &    171.0  &    8.713 \\
$\Sigma N$-$\Sigma N$ $^1S_0$ & b &   861.6  &  $-$79.17 &  0.0 \\
                & c  &  $-$355.3 &    70.65  &     0.0  \\
\hline
                & a & $-$1525.  &    211.4  & $-$2.749  \\
$\Lambda N$-$\Sigma N$ $^3S_1$ & b &  2775. &  $-$442.0 &  0.0 \\
                & c  &  $-$1253. &   208.0  &   0.0   \\
\hline
                & a &    899.5  &  $-$158.0 & $-$4.252  \\
$\Sigma N$-$\Sigma N$ $^3S_1$  & b &  240.3 & $-$31.34  &  0.0 \\ 
                & c &  $-$199.9  &   41.30  &  0.0  \\
\hline
\end{tabular}
%\end{ruledtabular}
\end{center}
\end{table}

\begin{table}[ht]
\begin{center}
\caption{Tensor coupling parts of G-matrix interactions
for ESC08c, represented in a $r^2$-Gaussian form
$\sum_{i=1}^3 (a_i+b_i k_F+c_i k_F^2) r^2 \exp(-(r/\beta_i)^2$.
}
\label{Gmat-CP2}
\vskip 0.2cm
%\begin{ruledtabular}
\begin{tabular}{|cc|rrr|}\hline\hline
&  $\beta_i$ & 0.50 & 0.90 & 2.00  \\
\hline
          & a & $-$39470. &   470.3  & $-$.7443 \\
$\Lambda N$-$\Sigma N$ $^3S_1$  & b  &  61860. &  $-$901.1 & 0.0 \\
          & c  & $-$23750. &  343.3  &   0.0  \\
\hline
          & a & $-$209.3  & $-$1.836  & $-$.0218 \\
$\Lambda N$-$\Lambda N$ $^3S_1$  & b  &  367.6 & 8.251  & 0.0 \\
          & c &    334.4  & $-$14.09  &   0.0  \\ 
\hline
\end{tabular}
%\end{ruledtabular}
\end{center}
\end{table}
%\onecolumngrid

In order to compare clearly the $SLS$ and $ALS$ components,
it is convenient to derive the strengths of the 
$\Lambda$ $l$-$s$ potentials in hypernuclei. 
%In the same way as in Refs.~\cite{NSC97,ESC04},
In the same way as in Refs.~\cite{RSY99,Rij04b},
the expression can be derived with 
the Scheerbaum approximation~\cite{Sch76} as
$U^{ls}_\Lambda(r) = K_\Lambda\, \frac{1}{r} \, \frac{d\rho}{dr}\,
      {\bf l}\cdot{\bf s}$.
The values of $K_\Lambda$ can be calculated with use of
${\cal G}_{SLS}(r)$ and ${\cal G}_{ALS}(r)$:
The obtained value at $k_F=1.0$ fm$^{-1}$ is 3.6 MeV fm$^5$.
This value is far (slightly) smaller than those for 
NSC97e/f (ESC08a/b)~\cite{Yam10}.

\begin{table}[h]
%\begin{center}
\caption{Parameters of SLS and ALS G-matrix interactions  represented by
three-range Gaussian forms \ 
${\cal G}(r;k_F)= \sum_{i} (a_i+b_i k_F+c_i k_F^2) \exp -(r/\beta_i)^2$
in the cases of ESC08c.
 } 
\label{Gmat-SO}
\vskip 0.2cm
\begin{tabular}{|cc|rrr|}\hline\hline
&  $\beta_i$ & 0.40 & 0.80 & 1.20  \\
\hline
      & a  &  $-$12920. &    372.4   & $-$2.030  \\
SLS   & b  &     24580. &  $-$840.0  &   0.0     \\ 
      & c  &  $-$10180. &    337.1   &   0.0     \\
\hline
      & a  &     1985. &    12.73  &    2.109 \\
ALS   & b  &  $-$1828. &    41.30  &    0.0   \\
      & c  &     679.8 &  $-$17.58 &    0.0   \\
\hline
\end{tabular}
%\end{center}
\end{table}

%%------------------------------------------------------------------------------
\subsection{$\Lambda$ hypernuclei by G-matrix folding potentials}
The $\Lambda\!N$ G-matrix interaction given by Table~\ref{Gmat-L1}
is expressed as ${\cal G}^{S}_{(\pm)}(r)$, 
$S$ and $(\pm)$ denoting spin and party quantum numbers, respectively.
A $\Lambda$-nucleus potential in a finite system is derived from
this $\Lambda\!N$ interaction by the expression
\begin{eqnarray}
U_\Lambda({\bf r},{\bf r'}  ) &=& U_{dr} +U_{ex} \ ,
\nonumber \\
U_{dr} &=& \delta({\bf r}-{\bf r'})  
\int d{\bf r''} \rho({\bf r''}) \,
V_{dr}(|{\bf r}-{\bf r''}|;k_F  )
\nonumber \\
U_{ex} &=& \rho({\bf r},{\bf r'}) 
V_{ex}(|{\bf r}-{\bf r'}|;k_F  ) \ ,
             \label{eq:GM8}
%\nonumber
\end{eqnarray}

\begin{eqnarray}
\left(\begin{array}{c}
V_{dr} \\
V_{ex}
\end{array}\right)
&=&\frac14\,
%&=&\frac{1}{2(2t_Y+1)(2s_Y+1)}\,
\sum_{S=0,1} (2S+1) [{\cal G}^{S}_{(\pm)} \pm {\cal G}^{S}_{(\mp)}] \ .
             \label{eq:GM9}
\end{eqnarray}
Here, densities $\rho(r)$ and mixed densities $\rho(r,r')$ are 
obtained from spherical Skyrme-HF wave functions.

An important problem is how to treat $k_F$ values included in
G-matrix interactions. 
We use here the following Averaged-Density 
Approximation (ADA), where 
an averaged value $\langle \rho \rangle$ is calculated by
$\langle \phi_\Lambda(r)|\rho(r)|\phi_\Lambda(r) \rangle$ for each 
$\Lambda$ state $\phi_\Lambda(r)$, and $\langle k_F\rangle$ is obtained by
$(1.5 \pi^2\langle \rho \rangle)^{1/3}$. 

Let us calculate the energy spectra of $\Lambda$ hypernuclei systematically
($^{13}_{\ \Lambda}$C, $^{28}_{\ \Lambda}$Si, $^{51}_{\ \Lambda}$V, 
$^{89}_{\ \Lambda}$Y, $^{139}_{\ \Lambda}$La, $^{208}_{\ \Lambda}$Pb).
In calculations, since the G-matrix interaction for ESC08c gives rise to
larger values of $B_\Lambda$ systematically compared to experimental data,
a factor 1.033 is multiplied on core parts ($\beta=0.5$ fm).

\newpage
%------------------------------------------------------------------------------
\begin{widetext}
\onecolumngrid
\begin{figure*}[hbt]
\begin{center}
\includegraphics*[width=14cm,height=14cm]{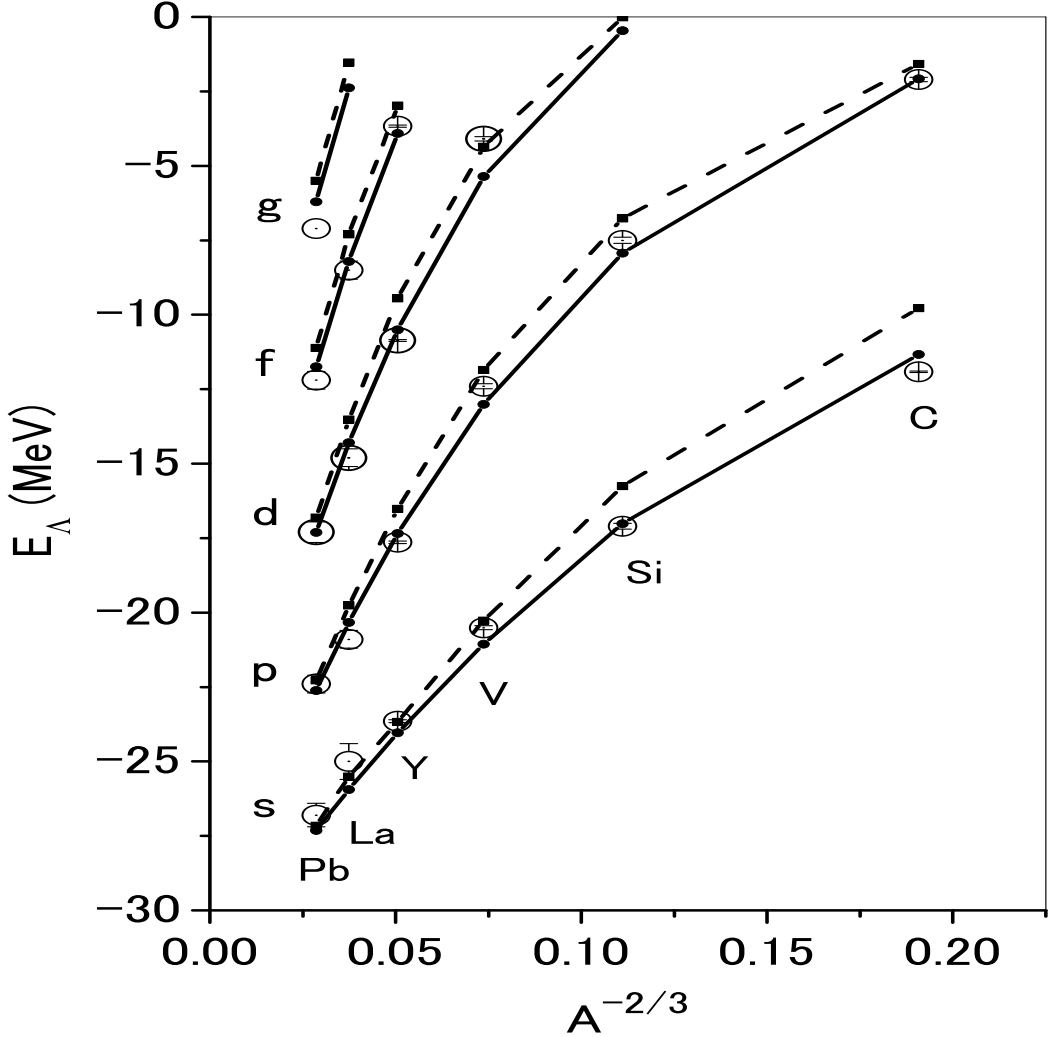}
\caption{Energy spectra of $^{13}_{\ \Lambda}$C, $^{28}_{\ \Lambda}$Si, 
$^{51}_{\ \Lambda}$V, $^{89}_{\ \Lambda}$Y, $^{139}_{\ \Lambda}$La and 
$^{208}_{\ \Lambda}$Pb are given as
a function of $A^{-2/3}$, $A$ being mass numbers of core nuclei.
Solid (dashed) lines show calculated values by the G-matrix folding model 
derived from ESC08c$^+$ (ESC08c). Open circles denote the experimental
values taken from Ref.~\cite{TamHashi}. }
\label{Lam1}
\end{center}
\end{figure*}
\end{widetext}
\twocolumngrid
%------------------------------------------------------------------------------
On the other hand, $V^0_{TBA}=-21.0$ MeV in ESC08c$^+$ is chosen so as 
to give best fitting to the experimental spectrum of $^{89}_{\ \Lambda}$Y.
In such an approach, the choice of $V^0_{TBA}$ depends on
the adopted two-body interaction. In~\cite{YFYR14}, the previous version
of ESC08c was used, giving $U_\Lambda(\rho_0)=-39.4$ MeV shallower than
the value $-40.8$ MeV in the present work. Then, $V^0_{TBA}$ was taken
as $-32.8$ MeV in the same as that in $NN$ channels.

In Fig.~\ref{Lam1}, 
the calculated values are compared with 
the experimental values marked by open circles, 
the horizontal axis being given as $A^{-2/3}$, where
solid and dashed curves are for ESC08c$^+$ and ESC08c, respectively.
Here, the experimental data are shifted by 0.5 MeV from the values 
given in~Ref.\cite{TamHashi}, which has been recently proposed
according to the improved calibration~\cite{Gogami}.
The difference between ESC08c$^+$ and ESC08c is due to the extra terms
$\Delta {\cal G}(k_F,r)$ originated from MPP+TBA. 
Especially, MPP plays an essential role to reproduce the nuclear saturation
property and the stiffness of the EoS of neutron-star matter~\cite{YFYR13,YFYR14}.
Then, it is very important that ESC08c$^+$ gives better fitting than ESC08c: 
The density-dependent attraction $\Delta {\cal G}(k_F,r)$ 
in low-density region works to reproduce better the energy spectra of 
heavy systems and $B_\Lambda$ values of light systems. 
In high-density region, this extra term leads to
the stiff EoS of the hyperon-mixed neutron-star matter~\cite{YFYR14}.
The present result suggests that such an effect of MPP+TBA can be 
tested in terrestrial data of $B_\Lambda$ values. 

Finally, it is commented 
that the $\Lambda$ s.p. energies in finite systems are not related simply 
to the $U_\Lambda(\rho_0)$ values given in Table~\ref{Gmat-L1}.
The $U_\Lambda(\rho_0)$ values of $-40.8$ MeV ($-37.6$ MeV) for 
ESC08c (ESC08c$^+$) are very attractive compared to the experimental 
value of $-30$ MeV, which is the depth $U_{WS}$ of the 
$\Lambda$ Woods-Saxon (WS) potential suitable to the data of 
$\Lambda$ hypernuclei. However, it is misleading to compare the 
$U_\Lambda(\rho_0)$ value directly to the $U_{WS}$ one.
The $\Lambda$-nucleus folding potential depends not only on the 
strengths of $\Lambda N$ G-matrices but also on their $k_F$ dependences.
Then, the depth $U_{WS}$ of the phenomenological Woods-Saxon potential 
of $\Lambda$ cannot be considered as the $\Lambda$ potential depth 
in nuclear matter.

\section{Discussion, Conclusions and Outlook}
We have again shown in this paper that the ESC-approach to the nuclear force 
problem is able to make a connection between on the one hand the at 
present available baryon-baryon data and on the other hand the underlying
quark structure of the baryons and mesons. Namely, a very successful description
of both the $N\!N$- and $Y\!N$-scattering data is obtained with meson-baryon
coupling parameters which are almost all explained by the QPC-model.
This at the same time in obediance of the strong constraint of no bound states 
in the $S=-1$-systems.
Therefore, 
the ESC08c model of this paper are an important further step in the determination of    
the baryon-baryon interactions for low energy scattering and the description
of hypernuclei in the context of broken SU$(3)$-symmetry.
The values for many parameters, which in previous work were considered to be free to a
large extend, are now limited strongly, and tried to be made consistent with the present 
theoretical view on low energy hadron physics. 
This is in particularly the case for the $F/(F+D)$-ratios of the MPE-interactions.
These ratio's for the vector- and scalar-mesons are rather close to the QPC-model
predictions. 

In analyzing the effect of the Pauli-blocking repulsion the conclusion is that from 
the standpoint of the BB scattering-data fitting such a repulsion is not strong.
This conclusion is in line with arguments from strong-coupling QCD (SCQCD). 
Namely, it has been argued in \cite{GAMil89} that quark-exchange effects are small.\\
The G-matrix results show that basic features of hypernuclear data are reproduced
nicely by ESC08c, improving the weak points of the soft-core OBE-models NSC89 \cite{MRS89},
NSC97 \cite{RSY99}, and ESC04-models \cite{Rij04a,Rij04b,Rij04c}.
The ESC08-models are superior for hypernuclear data and many aspects 
of the effective (two-body) interactions in hypernuclei can be obtained using the 
ESC08-model. For example, this is the case for the well depth $U_\Sigma$.

Experience has shown that a good fit to the scattering data not necessarily
means success in the G-matrix results. To explain this one can think of two reasons:     
(i) the G-matrix results are sensitive to the two-body interactions below 1 fm, 
whereas the present YN-scattering data are not, (ii) other than two-body forces play an
important role. 
However, since the $NN+YN$-fit is so much superior for ESC04- than for OBE-models, 
we are inclined to look for solutions to the remaining problems 
outside the two-body forces. 
A natural possibility is the presence of three-body forces (3BF) in hypernuclei
which can be viewed as generating effective two-body forces, 
which could solve the well-depth issues. 
In the case of the $\Delta B_{\Lambda\Lambda}$ also 3BF could be operating. 
This calls for an evaluation of the 3BF's $NNN$, $\Lambda NN$, $\Sigma NN$, 
$\Lambda\Lambda N$, etc. for the soft-core ESC-model, consistent 
with its two-body forces.

The $\Lambda N$ p-waves seem to be better, which is the result of the truly 
simultaneous $NN+YN$-fitting. This is also reflected in the better $K_\Lambda$-value,
making the well-known small spin-orbit splitting smaller.\\
%----------------------------------------------------------------------------------------
\noindent In the course of the development of the ESC-model for baryon-baryon,  
up to and including ESC06 \cite{HYP06} it was tried to solve all problems for NN and YN,
both for scattering and hypernuclear well-depth's, by 
keeping the potentials restricted to meson-exchange. 
For that purpose, in ESC06 a 'super-extended' the ESC-approach was studied
by including the second generation of the mesons, i.e. the heavy pseudo-scalar,
vector, and scalar meson nonets. In the Quark-Model they would correspond to
the first radially excited $Q\bar{Q}$-states, with masses in the range 
$ 1 GeV/c^2 < M < 1.7 GeV/c^2$.
With this extension it is possible to produce extra repulsion in the $\Sigma^+p(^3S_1)$, 
but correlated
with this was an exremely strong attraction in the $\Sigma^+p(^1P_1)$ partial-wave. 
Although the ESC06-approach is not ruled out by the data, we think that the solutions 
presented with ESC08 are much more superior.
In the future, such a 'super-extended' ESC08-model may be explored. For example,    
the axial-vector and heavy pseudoscalar ($\pi(1300)$) meson sectors
can be studied more closely. furthermore, for the scalar mesons the inclusion
of a finite width for the broad $\kappa$-meson can be investigated.
%--------------------------------------------------------------------------------

 \appendix

%---------------------------------------------------------------------------------
\section{ MPE interactions and $SU(3)$ }
\label{app:MPE.SU3}       
Below, $\sigma, {\bf a}_0, {\bf A}_1, \ldots $ are short-hands for 
respectively  the baryon SU(3)-singlet and -octet densities $\bar{\psi}\psi$,
$\bar{\psi}\mbox{\boldmath $\lambda$}\psi$,
$\bar{\psi}\gamma_5\gamma_\mu\mbox{\boldmath $\lambda$}\psi, \ldots $.
Here, $\lambda_i,\ i=0,1,...,8$ are the Gell-Mann SU(3)-matrices.

For the pseudoscalar-, vector-, scalar-, and axial-vector mesons 
The $SU(3)$ octet and singlet states appearing in the meson-pairs, denoted by the subscript $8$
respectively $1$, are in terms of the physical ones defined as follows:
\begin{enumerate}
\item[(i)] \underline{Pseudo-scalar-mesons}:
\begin{eqnarray*}
   \eta_1 &=& \cos\theta_{P} \eta' - \sin\theta_{P} \eta \\       
   \eta_8 &=& \sin\theta_{P} \eta' + \cos\theta_{P} \eta              
\end{eqnarray*}
Here, $\eta'$ and $\eta$ are the physical pseudo-scalar mesons 
 $\eta(957)$ respectively $\eta(548)$.
\item[(ii)] \underline{Vector-mesons}:         
\begin{eqnarray*}
   \phi_1 &=& \cos\theta_{V} \omega  - \sin\theta_{V} \phi \\       
   \phi_8 &=& \sin\theta_{V} \omega + \cos\theta_{V} \phi              
\end{eqnarray*}
Here, $\phi$ and $\omega$ are the physical vector mesons 
 $\phi(1019)$ respectively $\omega(783)$.
%------------------------------------------------------------------------------
\end{enumerate}
Then, one has the following $SU(3)$-invariant pair-interaction 
Hamiltonians:\\
 1.\ $J^{PC}=0^{+-}$: $SU(3)$-singlet couplings $S^\alpha_\beta = 
\delta^\alpha_\beta \sigma/\sqrt{3}$,
\begin{eqnarray*}
 {\cal H}_{S_1PP} &=& \frac{g_{S_1PP}}{\sqrt{3}}\left\{
\mbox{\boldmath $\pi$}\cdot\mbox{\boldmath $\pi$} + 
  2 K^\dagger K + \eta_8\eta_8\right\}\cdot \sigma
\end{eqnarray*}
 2.\ $J^{PC}=0^{++}$: $SU(3)$-octet symmetric couplings I, 
 $S^\alpha_\beta = (S_8)^\alpha_\beta \Rightarrow (1/4) Tr\{ S[P,P]_+\}$,
\begin{eqnarray*}
 {\cal H}_{S_8PP} &=& 
\frac{g_{S_8PP}}{\sqrt{6}}\left\{\vphantom{\frac{A}{A}}\right.
 ({\bf a}_0\cdot\mbox{\boldmath $\pi$})\eta_8 + 
 \frac{\sqrt{3}}{2}{\bf a}_0\cdot(K^\dagger \mbox{\boldmath $\tau$}K) 
 \nonumber \\
 && +\frac{\sqrt{3}}{2}
 \left\{(K_0^\dagger\mbox{\boldmath $\tau$}K)\cdot\mbox{\boldmath $\pi$}+ 
 h.c.\right\} \nonumber\\ &&
 -\frac{1}{2}\left\{(K_0^\dagger K)\eta_8 + h.c. \right\} 
 \nonumber \\ && 
 + \frac{1}{2}f_0\left(\mbox{\boldmath $\pi$}\cdot\mbox{\boldmath $\pi$} 
 - K^\dagger K -\eta_8\eta_8\right) 
 \left.\vphantom{\frac{A}{A}}\right\}
\end{eqnarray*}
 3.\ $J^{PC}=1^{+-}$: $SU(3)$-octet symmetric couplings II, 
 $S^\alpha_\beta = (B_8)^\alpha_\beta \Rightarrow (1/4) Tr\{ B^\mu [V_\mu, P]_+\}$,
\begin{eqnarray*}
 {\cal H}_{B_8VP} &=&
 \frac{g_{B_8VP}}{\sqrt{6}}\left\{\vphantom{\frac{A}{A}}\right. 
 \frac{1}{2}\left[\left({\bf B}_1^\mu\cdot\mbox{\boldmath $\rho$}_\mu\right) \eta_8 +
 \left({\bf B}_1^\mu\cdot\mbox{\boldmath $\pi$}_\mu\right) \phi_8 \right] 
  \nonumber \\ &&
 +\frac{\sqrt{3}}{4}\left[{\bf B}_1\cdot(K^{*\dagger}\mbox{\boldmath $\tau$}K)
 + h.c. \right] 
  \nonumber \\ &&
 +\frac{\sqrt{3}}{4}\left[(K_1^\dagger\mbox{\boldmath $\tau$} K^*)\cdot
 \mbox{\boldmath $\pi$}
 +(K_1^\dagger\mbox{\boldmath $\tau$} K)\cdot\mbox{\boldmath $\rho$} + h.c. \right] 
  \nonumber \\ &&
 -\frac{1}{4}\left[(K_1^\dagger\cdot K^*) \eta_8 + (K_1^\dagger\cdot K) \phi_8 
 + h.c. \right] 
  \nonumber \\ &&
 +\frac{1}{2}H^0\left[\mbox{\boldmath $\rho$}\cdot\mbox{\boldmath $\pi$} 
 -\frac{1}{2}\left(K^{*\dagger}\cdot K+ K^\dagger\cdot K^* \right)
 -\phi_8\eta_8 \right]
 \left.\vphantom{\frac{A}{A}}\right\}                            
\end{eqnarray*}
 4.\ $J^{PC}=1^{--}$: $SU(3)$-octet a-symmetric couplings I, 
 $A^\alpha_\beta = (V_8)^\alpha_\beta \Rightarrow 
 (-i/\sqrt{2}) Tr\{ V^\mu [P,\partial_\mu P]_-\}$,
\begin{eqnarray*}
 {\cal H}_{V_8PP} &=& g_{A_8PP}\left\{\vphantom{\frac{A}{A}}\right.
 \frac{1}{2}\mbox{\boldmath $\rho$}_\mu\cdot\mbox{\boldmath $\pi$}\times
 \stackrel{\leftrightarrow}{\partial^\mu}\!\!
 \mbox{\boldmath $\pi$}+\frac{i}{2}\mbox{\boldmath $\rho$}_\mu\cdot(K^\dagger
 \mbox{\boldmath $\tau$}\!\!\stackrel{\leftrightarrow}{\partial^\mu}\!\! K) 
 \nonumber \\
 && +\frac{i}{2}\left(\vphantom{\frac{A}{A}} K^{* \dagger}_\mu \mbox{\boldmath $\tau$}
(K\!\!\stackrel{\leftrightarrow}{\partial^\mu}\!\!\mbox{\boldmath $\pi$}) 
 - h.c. \right)
 +i\frac{\sqrt{3}}{2}\left(\vphantom{\frac{A}{A}} K^{* \dagger}_\mu\cdot
 \right.\nonumber\\  && \left. (K\cdot\stackrel{\leftrightarrow}
{\partial^\mu}\!\! \eta_8) - h.c. \vphantom{\frac{A}{A}}\right) 
 +\frac{i}{2}\sqrt{3} \phi_\mu (K^\dagger\stackrel{\leftrightarrow}{\partial^\mu}\!\! K)
 \left.\vphantom{\frac{A}{A}}\right\}
%\mbox{\Large /}2
\end{eqnarray*}
 5.\ $J^{PC}=1^{++}$ $SU(3)$-octet a-symmetric couplings II, 
 $A^\alpha_\beta = (A_8)^\alpha_\beta \Rightarrow 
 (-i/\sqrt{2}) Tr\{ A^\mu [P,V_\mu]_-\}$:
\begin{eqnarray*}
 {\cal H}_{A_8VP} &=& g_{A_8VP}\left\{\vphantom{\frac{A}{A}}\right.
 {\bf A}_1\cdot\mbox{\boldmath $\pi$}\times\mbox{\boldmath $\rho$}
 \nonumber\\ &&
 +\frac{i}{2}{\bf A}_1\cdot\left[(K^\dagger\mbox{\boldmath $\tau$} K^*)  
  -(K^{*\dagger}\mbox{\boldmath $\tau$} K)\right] \nonumber \\
 &&
 -\frac{i}{2}\left(\left[(K^\dagger\mbox{\boldmath $\tau$}K_A)\cdot
 \mbox{\boldmath $\rho$}
 + (K_{A}^\dagger\mbox{\boldmath $\tau$}K^*)\cdot\mbox{\boldmath $\pi$}
 \right] - h.c.\right) \nonumber \\
 &&
 -i\frac{\sqrt{3}}{2}\left(\left[(K^\dagger\cdot K_A)\phi_8
 +(K_A^\dagger\cdot K^*)\eta_8\right] - h.c. \right) \nonumber \\
 &&
 +\frac{i}{2}\sqrt{3} f_1 \left[K^\dagger\cdot K^*-K^{*\dagger}\cdot K\right]  
 \left.\vphantom{\frac{A}{A}}\right\}
\end{eqnarray*}
The relation with the pair-couplings used in this paper and paper I, see also \cite{RS96ab}, I is 
 $g_{S_1PP}/\sqrt{3}= g_{(\pi\pi)_0}/m_\pi$, 
$g_{A_8VP}= g_{(\pi\rho)_1}/m_\pi$ etc.
%---------------------------------------------------------------------------------

%----------------------------------------------------------------------------------------
\section{$J^{PC}=1^{+-}$ Axial-pair Potentials}
\label{app:MPE.ax.2nd}
In this appendix we document the $J^{PC}=1^{+-}$-axial $(\pi\omega)$ 1-pair potentials, 
which have not been reported elsewhere yet. The involved meson-pairs can be read off from 
the SU(2) structure of the interaction Hamiltonian (\ref{eq:MPE.1b}).\\

\noindent Below, we denote the type of potentials by writing $V^{(n)}_{\sigma+T}$, where 
$n=0,1$ refers to the $(1/M)$-order, and the subscript $\sigma+T$ indicates
that only the spin-spin and tensor contributions are given here and not the
spin-orbit potentials.
 
%---------------------------------------------------------------------------------
\subsection{NN-Potentials $(S=0,I=1)$-exchange, $(\pi\omega_1)$ etc. } 
To be specific, consider $(\pi\omega)_1$-exchange for NN and elastic $\Sigma$N 
potentials. One obtains:\\

\noindent 1. The leading, i.e. $(1/M)^0$-terms in momentum and configuration space are
 
\onecolumngrid
\begin{subequations}\label{app:MPE.1a}  
\begin{eqnarray}
\widetilde{V}^{(0)}_{\sigma+T}({\bf q},{\bf k}) &=& 
 +g_{(\pi\omega)_1;NN} f_{NN\pi} G_{NN\omega}       
 \left(\vphantom{\frac{A}{A}} \bm{\sigma}_1\cdot{\bf k}\bm{\sigma}_2\cdot{\bf k}_1
 +\bm{\sigma}_1\cdot{\bf k}_1\bm{\sigma}_2\cdot{\bf k}\right)    
 \times\frac{1}{\omega_1^2\omega_2^2}\cdot\frac{1}{m_\pi^2{\cal M}}\ , \\
  V_{\sigma+T}^{(0)}(r) &=& 
  -2g_{(\pi\omega;NN)} f_{NN\pi} G_{NN\omega}      
 \left[ \vphantom{\frac{A}{A}} F_{B,\sigma}^{(0)}(r) \bm{\sigma}_1\cdot\bm{\sigma}_2 +
  F_{B,T}^{(0)}(r)\ S_{12}\right]\cdot\frac{1}{m_\pi^2{\cal M}}\ , 
\end{eqnarray}
\end{subequations}
where 
\begin{subequations}\label{app:MPE.1b}  
\begin{eqnarray}
&& F_{B,\sigma}^{(0)}(r) = \frac{1}{3}
 \left(\frac{2}{r}F' G + F' G' + F^{\prime\prime} G\right)\ \ ,\
 F_{B,T}^{(0)}(r) = \frac{1}{3}
 \left(-\frac{1}{r}F' G + F' G' + F^{\prime\prime} G\right)\ .    
\end{eqnarray}
\end{subequations}
Above $\omega_1 = \sqrt{{\bf k}^1_1+m_\pi^2}$ and 
$\omega_2 = \sqrt{{\bf k}^2_1+m_\omega^2}$.    
For the Fourier transforms of the momentum pair-exchange potentials with
gaussians form factors, we refer to the basic papers \cite{RS96ab}.
The superscript for the functions $F_{B,\sigma,T}$ refers to the denominators
$1/\omega_1^2\omega_2^2$ in (\ref{app:MPE.1a}).
For these denominators, in the notation of \cite{RS96ab}, the functions F and G are
\begin{equation}
 F(r) = I_2\left(r,m_\pi,\Lambda_\pi\right)\ \ ,\ 
 G(r) = I_2\left(r,m_\omega,\Lambda_\omega\right)\ .  
\label{app:MPE.1c}\end{equation}

Similar formulas apply to e.g. $\Sigma$N-potentials, and also to $(K^*K)_1$-pair exchange.

\noindent 2. The non-leading, i.e. $(1/M)$-terms, are
\begin{subequations}\label{app:MPE.4a}  
\begin{eqnarray}
\widetilde{V}_{\sigma+T}^{(1)}({\bf q},{\bf k}) &=& 
 -g_{(\pi\omega)_1;NN} f_{NN\pi} G_{NN\omega}\frac{1}{2M_N}
 \left(\vphantom{\frac{A}{A}} \bm{\sigma}_1\cdot{\bf k}\bm{\sigma}_2\cdot{\bf k}_2
 +\bm{\sigma}_1\cdot{\bf k}_2\bm{\sigma}_2\cdot{\bf k}\right)    
 \times\frac{1}{\omega_1\omega_2(\omega_1+\omega_2)}\cdot\frac{1}{m_\pi^2{\cal M}}\ , \\
 V_{1/M,\sigma+T}^{(1)}(r) &=& 
  +2g_{(\pi\omega)_1;NN} f_{NN\pi} G_{NN\omega}\frac{m_\pi}{2M_N}
 \left[ \vphantom{\frac{A}{A}} F_{B,\sigma}^{(1)}(r) \bm{\sigma}_1\cdot\bm{\sigma}_2 +
  F_{B,T}^{(1)}(r)\ S_{12}\right]\cdot\frac{1}{m_\pi^3{\cal M}}\ , 
\end{eqnarray}
\end{subequations}
where now superscript for the functions $F_{B,\sigma,T}^{(1)}$ refers to the denominators
$1/\omega_1\omega_2(\omega_1+\omega_2)$ in (\ref{app:MPE.4a}).
For this denominator the basic Fourier transform is \cite{RS96ab}
\begin{equation}
  F_{B}^{(1)}(r) = \frac{2}{\pi}\int_0^\infty d\lambda\ F(\Lambda,r)\ G(\lambda,r)\ ,
\label{app:MPE.4b}\end{equation}
where the functions F and G are
\begin{equation}
 F(r) = I_2\left(r,m_\pi(\lambda),\Lambda_\pi\right)\ \ ,\ 
 G(r) = I_2\left(r,m_\omega(\lambda),\Lambda_\omega\right)\ ,  
\label{app:MPE.4c}\end{equation}
with the understanding that under the $\lambda$-integral in (\ref{app:MPE.4b})
there occur the combinations 
\begin{subequations}\label{app:MPE.4d}  
\begin{eqnarray}
&& F_{B,\sigma}^{(1)}(r) = \frac{1}{3}
 \left(\frac{2}{r}F G' + F' G' + F\ G^{\prime\prime}\right)\ \ ,\
 F_{B,T}^{(1)}(r) = \frac{1}{3}
 \left(-\frac{1}{r}F G' + F' G' + F G^{\prime\prime}\right)\ .    
\end{eqnarray}
\end{subequations}

\noindent 3. The symmetric spin-orbit $(1/M)^2$-terms, are
\begin{subequations}\label{app:MPE.5}  
\begin{eqnarray}
\widetilde{V}_{SLS}^{(2)}({\bf q},{\bf k}) &=& 
 -g_{(\pi\omega)_1;NN} f_{NN\pi} G_{NN\omega}\frac{1}{M_N^2}\
\frac{i}{2}(\bm{\sigma}_1+\bm{\sigma}_2)\cdot{\bf q}\times{\bf k}_2
\times\frac{1}{\omega_2^2}\ , \\
 V_{SLS}^{(2)}(r) &=& 
 -g_{(\pi\omega)_1;NN} f_{NN\pi} G_{NN\omega}\frac{1}{m_\pi^2M_N^2}\
 I_0(m_\pi,r)\ \left(-\frac{1}{r}\frac{d}{dr} I_2(m_\omega,\Lambda_V,r)\right)\
 {\bf L}\cdot{\bf S}\ ,              
\end{eqnarray}
\end{subequations}
where 
\begin{equation}
 I_0(\Lambda_P,r) = \frac{1}{4\pi} \frac{1}{2\sqrt{\pi}}
 \left(\frac{\Lambda_P}{m_\pi}\right)^3 
 \exp\left(-\frac{1}{4}\Lambda_P^2 r^2\right)\ .
\label{app:MPE.6}\end{equation}
We note that important contributions to the anti-symmetric spin-orbit potentials
are proportional to $(1/M_N-1/M_Y) \sim 1/M^2$. Also, spin-orbit potentials from
OBE are order $1/M^2$. Therefore, we included this SLS-potential in the ESC08-model.

\subsection{ YN-potentials, (S=0,I=0)-Exchange, $(\bm{\pi}\bm{\rho})_0$ etc.} 
%---------------------------------------------------------------------------------
The above given potentials also occur in YN- and YY-channels, of course.
In this subsection we give as an illustration only the $1/M$-contribution for
the spin-spin and tensor.
Again, to be specific, now we consider $(\pi\rho)_0$-exchange for $\Lambda$N 
potentials. We obtain:\\
\begin{subequations}\label{app:MPE.13a}  
\begin{eqnarray}
\widetilde{V}_{\sigma+T}^{(1)}({\bf q},{\bf k}) &=& 
 -2g_{\Lambda\Lambda;(\pi\rho)_0} f_{NN\pi} G_{NN\rho}
 \frac{1}{2M_N}\left[ \bm{\sigma}_1\cdot{\bf k}\bm{\sigma}_2\cdot{\bf k}_2
 \right]    
 \times\frac{1}{\omega_1\omega_2(\omega_1+\omega_2)}\ , \\
\widetilde{V}_{\sigma+T}^{(1)}({\bf q},{\bf k}) &=& 
 -2g_{NN;(\pi\rho)_0} f_{\Lambda\Sigma\pi} G_{\Lambda\Sigma\rho}          
 \frac{1}{M_\Lambda+M_\Sigma}\left[
 \bm{\sigma}_1\cdot{\bf k}_2\bm{\sigma}_2\cdot{\bf k}
 \right]    
 \times\frac{1}{\omega_1\omega_2(\omega_1+\omega_2)}\ .    
\end{eqnarray}
\end{subequations}
In configuration space we get
\begin{subequations}\label{app:MPE.13b}  
\begin{eqnarray}
 V_{\sigma+T}^{(1)}(r) &=& 
 +2g_{\Lambda\Lambda;(\pi\rho)_0} f_{NN\pi} G_{NN\rho}
 \frac{1}{2M_N}\left[ G_{B,\sigma}^{(1)}(r)\bm{\sigma}_1\cdot\bm{\sigma}_2
 + G_{B,T}^{(1)}(r)\ S_{12} \right]\ ,\\
 V_{\sigma+T}^{(1)}(r) &=& 
 +2g_{NN;(\pi\rho)_0} f_{\Lambda\Sigma\pi} G_{\Lambda\Sigma\rho}          
 \frac{1}{M_\Lambda+M_\Sigma}\left[
 G_{B,\sigma}^{(1)}(r) \bm{\sigma}_1\cdot\bm{\sigma}_2 + 
 G_{B,T}^{(1)}(r)\ S_{12}\right]\ ,                     
\end{eqnarray}
\end{subequations}
where 
\begin{subequations}\label{app:MPE.13c}  
\begin{eqnarray}
&& G_{B,\sigma}^{(1)}(r) = \frac{1}{3}
 \left(\frac{2}{r}F_\pi\otimes F'_\omega + F'_\pi\otimes F'_\omega 
 + F_\pi\otimes F^{\prime\prime}_\omega\right)\ , \\ 
&& G_{B,T}^{(1)}(r) = \frac{1}{3}
 \left(-\frac{1}{r}F_\pi\otimes F'_\omega + F'_\pi\otimes F'_\omega 
 + F\otimes_\pi F^{\prime\prime}_\omega\right)\ .    
\end{eqnarray}
\end{subequations}
Here, again the superscript on the G-functions refers to the denominator in 
momentum space. For the denominators in (\ref{app:MPE.13a}) the functions 
$F\otimes g$ are given by\cite{RS96ab}
\begin{equation}
 F_\alpha\otimes F_\beta(r) = \frac{2}{\pi}\int_0^\infty d\lambda\ 
 F_\alpha(\lambda,r) F_\beta(\lambda,r)\ , 
\label{app:MPE.13d}\end{equation}
where 
\begin{equation}
 F_\alpha(\lambda,r) = e^{-\lambda^2/\Lambda_\alpha^2} I_2(\sqrt{m_\alpha^2+\lambda^2},r)\ .
\label{app:MPE.13e}\end{equation}

%--------------------------------------------------------------------------
\subsection{ YN-potentials, $(S=\pm 1,I=1/2$)-Exchange, $(\pi K^*)_{1/2}$ etc.} 
%---------------------------------------------------------------------------------
Again, to be specific, consider $(\pi K^*)_{1/2}$-exchange for $\Lambda$N 
potentials. One obtains:\\
The leading, i.e. $(1/M)^0$-potentials
\begin{subequations}\label{app:MPE.7a}  
\begin{eqnarray}
\widetilde{V}^{(1)}_{\sigma+T}({\bf q},{\bf k}) &=& 
 +g_{(\pi K^*);\Lambda N} f_{NN\pi} G_{N\Lambda K^*}
 \left(\vphantom{\frac{A}{A}} \bm{\sigma}_1\cdot{\bf k}\bm{\sigma}_2\cdot{\bf k}_1
 +\bm{\sigma}_1\cdot{\bf k}_1\bm{\sigma}_2\cdot{\bf k}\right)    
 \times\frac{1}{\omega_1^2\omega_2^2}\cdot{\cal P}_f\ , \\
\widetilde{V}^{(1)}_{\sigma+T}({\bf q},{\bf k}) &=& 
 +g_{(\pi K^*);\Lambda N} f_{\Lambda\Sigma\pi} G_{N\Sigma K^*} 
 \left(\vphantom{\frac{A}{A}} \bm{\sigma}_1\cdot{\bf k}\bm{\sigma}_2\cdot{\bf k}_1
 +\bm{\sigma}_1\cdot{\bf k}_1\bm{\sigma}_2\cdot{\bf k}\right)    
 \times\frac{1}{\omega_1^2\omega_2^2}\cdot{\cal P}_f\ .    
\end{eqnarray}
\end{subequations}
The configuration space potentials are:
\begin{subequations}\label{app:MPE.7b}  
\begin{eqnarray}
 V^{(1)}_{\sigma+T}(r) &=& 
 -2g_{(\pi K^*);\Lambda N} f_{NN\pi} G_{N\Lambda K^*}
 \left(\vphantom{\frac{A}{A}} F_{B,\sigma}^{(0)}(r) 
\bm{\sigma}_1\cdot\bm{\sigma}_2 + F_{B,T}^{(0)}\ S_{12}\right]\cdot{\cal P}_f\ , \\
 V^{(1)}_{\sigma+T}(r) &=& 
 -2g_{(\pi K^*);\Lambda N} f_{\Lambda\Sigma\pi} G_{N\Sigma K^*} 
 \left(\vphantom{\frac{A}{A}} F_{B,\sigma}^{(0)}(r)                             
\bm{\sigma}_1\cdot\bm{\sigma}_2 + F_{B,T}^{(0)}(r)\ S_{12}\right]\cdot{\cal P}_f\ .    
\end{eqnarray}
\end{subequations}
%---------------------------------------------------------------------------------

The non-leading, i.e. $(1/M)^1$-potentials are
\begin{subequations}
\label{appeq:MPE.8a}  
\begin{eqnarray}
&& \widetilde{V}_{\sigma+T}^{(1)}({\bf q},{\bf k}) = 
 -g_{(\pi K^*);\Lambda N} f_{NN\pi} G_{N\Lambda K^*}\frac{1}{2M_N}\left[
 \left(\vphantom{\frac{A}{A}} \bm{\sigma}_1\cdot{\bf k}\bm{\sigma}_2\cdot{\bf k}_2
 +\bm{\sigma}_1\cdot{\bf k}_2\bm{\sigma}_2\cdot{\bf k}\right)\right]    
 \frac{1}{\omega_1\omega_2(\omega_1+\omega_2)}\cdot{\cal P}_f , \\
&& \widetilde{V}_{\sigma+T}^{(1)}({\bf q},{\bf k}) = 
 -g_{(\pi K^*);\Lambda N} f_{\Lambda\Sigma\pi} G_{N\Sigma K^*} 
 \frac{1}{M_\Lambda+M_\Sigma}
 \left(\vphantom{\frac{A}{A}} \bm{\sigma}_1\cdot{\bf k}\bm{\sigma}_2\cdot{\bf k}_2
 +\bm{\sigma}_1\cdot{\bf k}_2\bm{\sigma}_2\cdot{\bf k}\right)    
 \frac{1}{\omega_1\omega_2(\omega_1+\omega_2)}\cdot{\cal P}_f.    
\end{eqnarray}
\end{subequations}
The configuration space potentials are:
\begin{subequations}
\label{appeq:MPE.8b}  
\begin{eqnarray}
 V^{(1)}_{\sigma+T}(r) &=& 
 +2g_{(\pi K^*);\Lambda N} f_{NN\pi} G_{N\Lambda K^*}
 \frac{m_\pi}{2M_N} \left(\vphantom{\frac{A}{A}} G_{B,\sigma}^{(1)}(r)   
\bm{\sigma}_1\cdot\bm{\sigma}_2 + G_{B,T}^{(1)}(r)\ S_{12}\right]\cdot{\cal P}_f\ , \\
 V^{(1)}_{\sigma+T}(r) &=& 
 +2g_{(\pi K^*);\Lambda N} f_{\Lambda\Sigma\pi} G_{N\Sigma K^*} 
 \frac{m_\pi}{M_\Lambda+M_\Sigma} \left(\vphantom{\frac{A}{A}} G_{B,\sigma}^{(1)}(r) 
\bm{\sigma}_1\cdot\bm{\sigma}_2 + G_{B,T}^{(1)}(r)\ S_{12}\right]\cdot{\cal P}_f.    
\end{eqnarray}
\end{subequations}
Above, ${\cal P}_f$ is the flavor-exchange operator, discussed in \cite{NRS77,MRS89}.
In addition, we have to multiply these potentials with the isoscalar
factors appearing in the Hamiltonian (\ref{eq:MPE.2}). For example for $K-\rho$ and
$K-\phi$ pairs this factor is $+\sqrt{3}/4$ respectively $-1/4$, etc.
%---------------------------------------------------------------------------------

%-----------------------------------------------------------------------------
\section{Exchange Potentials}                    
\label{app:H}     
In this section we follow our multi-channel description formalism in the 
treatment of the exchange potentials \cite{Rij04c}.\\

\begin{figure}[hbt]
%{\includegraphics[250,525][360,705]{ynfig/spinexch1.ps}}
{\includegraphics[250,525][360,705]{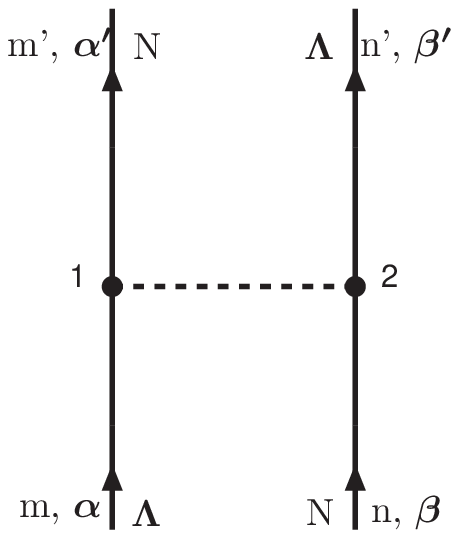}}
  \caption{Particle- and spin-exchange for $\Lambda N$.}             
  \label{fig.spinexch}   
% \end{center}
  \end{figure}

\noindent In the case of the anti-symmetric spin-orbit the exchange potential requires   
some attention, because their special features. 
The potentials in configuration space are described in Pauli-spinor space
as follows
\begin{equation}
 V = V_C + V_\sigma \bm{\sigma}_1\cdot\bm{\sigma}_2 + V_T\ S_{12} 
   + V_{SLS}\ {\bf L}\cdot{\bf S}_+ + V_{ALS}\ {\bf L}\cdot{\bf S}_-
   + V_{Q}\ Q_{12}\ .
\label{appH.1}\end{equation}
Here, the definition of the matrix elements of the spin operators are
defined as follows
\begin{eqnarray}
&& \left(\chi^\dagger_{m'}(\Lambda)\chi^\dagger_{n'}(N)|
 \bm{\sigma}_1\cdot\bm{\sigma}_2|
 \chi^\dagger_{m}(\Lambda)\chi^\dagger_{n}(N)\right) \equiv 
 \left(\chi^\dagger_{m'}(\Lambda)|\bm{\sigma}_1|\chi^\dagger_{m}(\Lambda)\right)
 \cdot
 \left(\chi^\dagger_{n'}(N)|\bm{\sigma}_1|\chi^\dagger_{n}(N)\right)\ ,
\label{appH.2}\end{eqnarray}
and similarly for the SU(2) and SU(3) operator matrix elements.
In Fig.~\ref{fig.spinexch} the labels $(m,n,m',n')$ refer to the spin, and the
labels $(\alpha,\beta,\alpha',\beta')$ refer to unitary spin, like SU(2) or SU(3).
The momenta on line 1 are ${\bf p}$ and ${\bf p}'$ for respectively the initial
and the final state. 
Likewise, the momenta on line 2 are $-{\bf p}$ and $-{\bf p}'$ for respectively 
the initial and the final state. 
                                                                                
%--------------------------------------------------------------------------
\twocolumngrid

%-----------------------------------------------------------------
%-----------------------------------------------------------------------------
In graph Fig.~\ref{fig.spinexch} we encounter the matrix elements
\begin{eqnarray}
&& (\bm{\sigma}_1)_{m',m} = 
 \left(\chi^\dagger_{m'}(N)| \bm{\sigma}_1| \chi^\dagger_{m}(\Lambda)\right),
\nonumber\\ &&
   (\bm{\sigma}_2)_{n',n} = 
 \left(\chi^\dagger_{n'}(\Lambda)| \bm{\sigma}_2| \chi^\dagger_{n}(N)\right)  
\label{appH.3}\end{eqnarray}

%-----------------------------------------------------------------------------
\subsection{Spin-Exchange Potentials}                    
\label{app:H.a}     
In order to project the exchange potentials on the forms in (\ref{appH.1}) we
have to rewrite these matrix elements in terms of those occurring in 
(\ref{appH.2}). This can be done using the spin-exchange operator $P_\sigma$:
\begin{equation}
 P_\sigma = \frac{1}{2}\left( 1 + \bm{\sigma}_1\cdot\bm{\sigma}_2\right)\ .
\label{appH.4}\end{equation}
Properties of this operator are 
\begin{eqnarray*}
&& P_\sigma\dagger = P_\sigma\ \ ,\ \ P_\sigma^2 = 1\ , \\
&& P_\sigma\ \chi_{1,m} \chi_{2,n} = \chi_{1,n} \chi_{2,m}\ , \\
&& P_\sigma\ \sigma_{1,k}\ P_\sigma = \sigma_{2,k}\ , \\ 
&& P_\sigma\ \sigma_{2,k}\ P_\sigma  =  \sigma_{1,k}\ .   
\end{eqnarray*}
Similar properties hold for the flavor-exchange operator $P_f$, but then for the
SU(2) isospin operators $\tau_k$, or the SU(3) octet operators $\lambda_k$.\\
In the following we make only explicit the spin labels, but similar
operations apply to the SU(2) or SU(3) labels.\\

\begin{widetext}
\noindent Using this spin-exchange operator, we find that
\begin{eqnarray}
&& \left(\chi^\dagger_{1,m'}(N)\chi^\dagger_{2,n'}(\Lambda)|
 \bm{\sigma}_1\otimes 1_2 - 1_1\otimes\bm{\sigma}_2|
 \chi^\dagger_{1,m}(\Lambda)\chi^\dagger_{2,n}(N)\right) = \nonumber\\
&& \left(\chi^\dagger_{2,n'}(N)\chi^\dagger_{1,m'}(\Lambda)|
 P_\sigma^\dagger\left(\vphantom{\frac{A}{A}}
 \bm{\sigma}_1\otimes 1_2 - 1_1\otimes\bm{\sigma}_2\right) P_\sigma\ P_\sigma|
 \chi^\dagger_{1,m}(\Lambda)\chi^\dagger_{2,n}(N)\right) = \nonumber\\
&& -\left(\chi^\dagger_{1,m'}(\Lambda)\chi^\dagger_{1,n'}(N)|
 \left(
 \bm{\sigma}_1\otimes 1_2 - 1_1\otimes\bm{\sigma}_2\right)\ P_\sigma|
 \chi^\dagger_{1,m}(\Lambda)\chi^\dagger_{2,n}(N)\right)\ .              
\label{appH.6}\end{eqnarray}
Above, we added the subscripts 1 and 2 to indicate explicitly the baryon line 
that is involved.

%-----------------------------------------------------------------------------
\subsection{Spin- and Strangeness-Exchange Potentials}                    
\label{app:H.b}     
In addition to the spin-exchange, we also have the flavor-exchange operator
$P_f$ active here. So, in total we have to apply $-P_\sigma\ P_f = P_x$, i.e.
the space-exchange operator. This latter relation follows from the anti-symimetry
of the two-baryon states, which implies that only states with $P_f P_\sigma P_x =-1$
are physical. All this implies\\

\noindent 1. For the ALS-potential derived in K-exchange etc. one has in 
(\ref{appH.1}), considering both spin- and flavor-exchange, the operator
\begin{equation}    
 {\rm ALS}\ \Rightarrow \frac{1}{2}
 \left(\bm{\sigma}_1-\bm{\sigma}_2\right)\cdot{\bf L}\ P_x
\label{appH.7}\end{equation}
\noindent 2. For the SLS-potential derived in K-exchange etc. one has in 
(\ref{appH.1}), considering both spin- and flavor-exchange, the operator
$P_f P_\sigma$, but since 
\begin{eqnarray*}
 \bm{\sigma}_1\cdot\bm{\sigma}_2\ \sigma_{1,k} &=& \sigma_{2,k} +i\epsilon_{klm}\
 \sigma_{1,l}\sigma_{2,m}\ , \nonumber\\
 \bm{\sigma}_1\cdot\bm{\sigma}_2\ \sigma_{2,k} &=& \sigma_{1,k} +i\epsilon_{klm}\
 \sigma_{2,l}\sigma_{1,m}\ , 
\end{eqnarray*}
one derives easily that
\begin{eqnarray}
 P_\sigma \left(\bm{\sigma}_1+\bm{\sigma}_2\right)\cdot{\bf L} &=&                   
          \left(\bm{\sigma}_1+\bm{\sigma}_2\right)\cdot{\bf L}\ ,                    
\label{appH.8}\end{eqnarray}
and therefore, similarly to (\ref{appH.6}) we have, with the inclusion of the 
flavor labels,
\begin{eqnarray}
&& \left(\chi^\dagger_{1,m'\alpha'}(N)\chi^\dagger_{2,n'\beta'}(\Lambda)|
 \bm{\sigma}_1\otimes 1_2 + 1_1\otimes\bm{\sigma}_2|
 \chi^\dagger_{1,m \alpha}(\Lambda)\chi^\dagger_{2,n \beta}(N)\right) = \nonumber\\
&& \left(\chi^\dagger_{2,n' \beta'}(N)\chi^\dagger_{1,m' \alpha'}(\Lambda)|
 P_f^\dagger P_\sigma^\dagger\left(\vphantom{\frac{A}{A}}
 \bm{\sigma}_1\otimes 1_2 + 1_1\otimes\bm{\sigma}_2\right)|
 \chi^\dagger_{1,m \alpha}(\Lambda)\chi^\dagger_{2,n \beta}(N)\right) = \nonumber\\
&& \left(\chi^\dagger_{1,m' \alpha'}(\Lambda)\chi^\dagger_{1,n' \beta'}(N)|
 \left(
 \bm{\sigma}_1\otimes 1_2 + 1_1\otimes\bm{\sigma}_2\right)\ P_f|
 \chi^\dagger_{1,m \alpha}(\Lambda)\chi^\dagger_{2,n \beta}(N)\right)\ .              
\label{appH.9}\end{eqnarray}
So, for the SLS-potential derived in K-exchange etc. one has in 
(\ref{appH.1}), considering both spin- and flavor-exchange, the operator
\begin{equation}    
 {\rm SLS}\ \Rightarrow \frac{1}{2}
 \left(\bm{\sigma}_1+\bm{\sigma}_2\right)\cdot{\bf L}\ P_f
\label{appH.10}\end{equation}
This treatment for the SLS-potential also applies to the central-, spin-spin-,
tensor-, and quadratic-spin-orbit potentials as well, of course.\\

\noindent {\it We conclude this section by noticing that we have found, 
using our multi-channel
set-up the same prescriptions for the treatment of the flavor-exchange potentials as in 
\cite{NRS77}. For the treatment of the ALS-potential for $S=\pm 1$-exchange, our
prescription here is more clear. For example in the case of the coupled $^1P_1-^3P_1$
system our prescription is unambiguous, and given by the $P_x$-operator, which is 
the same for both partial-waves coupled in this case.}

%--------------------------------------------------------------------
  \begin{figure}[hbt]
  \begin{center}
% \resizebox{7.0cm}{!}       
  \resizebox{10.0cm}{!}       
% {\includegraphics[200,600][420,780]{ynfig/fig.kexch.ps}}
  {\includegraphics[200,600][420,780]{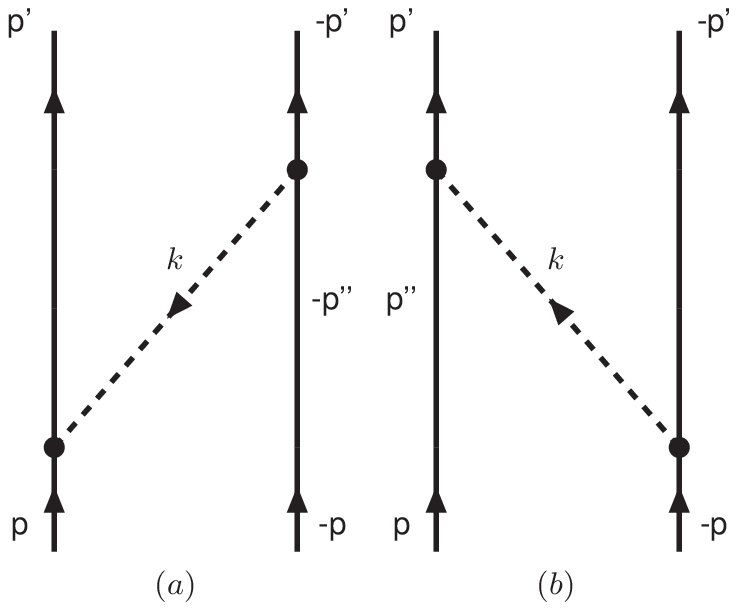}}
 \caption{K- and K$^*$-exchange time-ordered graphs (a) and (b).}         
  \label{fig.kexch}
  \end{center}
  \end{figure}
%-----------------------------------------------------------------
%-----------------------------------------------------------------
\section{Derivation BDI ALS-potentials for strange-meson-exchanges} 
The contributions to the $P_8$-spinor invariant, see \cite{SNRV71},
\begin{equation}
 P_8 = 2
\left(\vphantom{\frac{A}{A}}\bm{\sigma}_1\cdot{\bf q}\bm{\sigma}_2\cdot{\bf k} 
-\bm{\sigma}_1\cdot{\bf k}\bm{\sigma}_2\cdot{\bf q}\right)\ ,    
\label{appG.0}\end{equation}
for $(K, K^*)$-exchange were given by Brown, Downs, and Iddings \cite{BDI70}.
Here we derive these for $(K, K^*, K_1, K_2)$, and in particularly for
the pseudoscalar K within the ps-pv theory.

\subsection{K-exchange ALS-potential (PS-PV Theory)}          
\label{app:G.ps}     
We derive the K-exchange potential using the PV-theory, and show that we
get the BDI-answer for the anti-symmetric spin-orbit potential (ALS).
%---------------------------------------------------------------------------------
For graph $(a)$ we get from the vertices the matrix element
\begin{subequations}\label{appG.1}
\begin{eqnarray}
 (a)&:& -\frac{f_P^2}{m_\pi^2}\left[ \bm{\sigma}_1\cdot{\bf k} +                       
 \frac{2\omega}{M_\Lambda+M_N}\bm{\sigma}_1\cdot{\bf q}\right]
 \left[-\bm{\sigma}_2\cdot{\bf k} +                       
 \frac{2\omega}{M_\Lambda+M_N}\bm{\sigma}_2\cdot{\bf q}\right]
\cdot\frac{1}{2\omega}\frac{-1}{\omega-a} 
\nonumber\\ &=& 
  -\frac{f_P^2}{m_\pi^2}\left[ \bm{\sigma}_1\cdot{\bf k}\bm{\sigma}_2\cdot{\bf k}           
 -\frac{2\omega}{M_\Lambda+M_N}\left( \bm{\sigma}_1\cdot{\bf k}\bm{\sigma}_2\cdot{\bf q}
 -\bm{\sigma}_1\cdot{\bf q}\bm{\sigma}_2\cdot{\bf k}\right)\right]\cdot
 \frac{1}{2\omega(\omega-a)}\ , \\          
 (b)&:& -\frac{f_P^2}{m_\pi^2}\left[ \bm{\sigma}_1\cdot{\bf k} -                       
 \frac{2\omega}{M_\Lambda+M_N}\bm{\sigma}_1\cdot{\bf q}\right]
 \left[-\bm{\sigma}_2\cdot{\bf k} -                       
 \frac{2\omega}{M_\Lambda+M_N}\bm{\sigma}_2\cdot{\bf q}\right]
\cdot\frac{1}{2\omega}\frac{-1}{\omega+a} \nonumber\\
 &=& -\frac{f_P^2}{m_\pi^2}\left[ \bm{\sigma}_1\cdot{\bf k}\bm{\sigma}_2\cdot{\bf k}           
 +\frac{2\omega}{M_\Lambda+M_N}\left( \bm{\sigma}_1\cdot{\bf k}\bm{\sigma}_2\cdot{\bf q}
 -\bm{\sigma}_1\cdot{\bf q}\bm{\sigma}_2\cdot{\bf k}\right)\right]\cdot
\frac{1}{2\omega(\omega+a)}\ ,             
\end{eqnarray}
\end{subequations}
where $a=M_\Lambda-M_N$. 
Summing these contributions gives
\begin{eqnarray}
&& \widetilde{V}_K({\bf q},{\bf k}) =
 -\frac{f_P^2}{m_\pi^2}\left[\vphantom{\frac{A}{A}}
\frac{1}{2\omega}\left\{\frac{1}{\omega-a}+\frac{1}{\omega+a}\right\}\
\bm{\sigma}_1\cdot{\bf k}\bm{\sigma}_2\cdot{\bf k} \right. \nonumber\\
&& \left. 
+\frac{1}{M_\Lambda+M_N}\left\{\frac{1}{\omega-a}-\frac{1}{\omega+a}\right\}\
 \left( \bm{\sigma}_1\cdot{\bf k}\bm{\sigma}_2\cdot{\bf q}
 -\bm{\sigma}_1\cdot{\bf q}\bm{\sigma}_2\cdot{\bf k}\right)\right]\ {\cal P}_f\nonumber\\
&=&             
 -\frac{f_P^2}{m_\pi^2}\left[\vphantom{\frac{A}{A}}
 \bm{\sigma}_1\cdot{\bf k}\bm{\sigma}_2\cdot{\bf k} 
 -2\frac{M_\Lambda-M_N}{M_\Lambda+M_N}
\left(\vphantom{\frac{A}{A}}\bm{\sigma}_1\cdot{\bf k}\bm{\sigma}_2\cdot{\bf q} 
-\bm{\sigma}_1\cdot{\bf q}\bm{\sigma}_2\cdot{\bf k}\right)
\right]\ {\cal P}_f\cdot \frac{1}{\omega^2-a^2}        
\label{appG.2}\end{eqnarray}

We notice that this result corresponds with the answer in the PS-PS theory.
All this in the approximation 
$(M_\Lambda+M_N)^{-1}=(1/M_\Lambda + 1/M_N)/4$.
Now, using the definitions in \cite{SNRV71,MRS89} we have
\begin{eqnarray*}
 P_8 &=& 2
\left(\vphantom{\frac{A}{A}}\bm{\sigma}_1\cdot{\bf q}\bm{\sigma}_2\cdot{\bf k} 
-\bm{\sigma}_1\cdot{\bf k}\bm{\sigma}_2\cdot{\bf q}\right)\ , \\
 P_6 &=& (i/2)\left(\bm{\sigma}_1-\bm{\sigma}_2\right)\cdot{\bf n}\ ,\ \
 {\bf n}={\bf p}\times{\bf p}' = {\bf q}\times{\bf k}\ ,
\end{eqnarray*}
with the relation \cite{BDI70}
$ P_8 = -\left(1+\bm{\sigma}_1\cdot\bm{\sigma}_2\right)\ P_6 
 = 2 {\cal P}_x {\cal P}_f\ P_6$. 
This leads to the following expression 
\begin{eqnarray}
\widetilde{V}_K({\bf q},{\bf k}) &=&
 -\frac{f_P^2}{m_\pi^2}\left[\vphantom{\frac{A}{A}}
 \bm{\sigma}_1\cdot{\bf k}\bm{\sigma}_2\cdot{\bf k} 
 +2\frac{M_\Lambda-M_N}{M_\Lambda+M_N}\cdot
 (i/2)\left(\bm{\sigma}_1-\bm{\sigma}_2\right)\cdot{\bf n}\ {\cal P}_x {\cal P}_f
\right]\ {\cal P}_f\cdot \frac{1}{\omega^2-a^2}        
\label{appG.5}\end{eqnarray}
%---------------------------------------------------------------------------------
\subsection{K$^*$-exchange ALS-potential }          
\label{app:G.pv}     
Upon inspection, we find that the only contribution to the $P_8$-invariant is
given by
\begin{eqnarray}
\widetilde{V}_{K^*}({\bf q},{\bf k}) &\approx& \frac{1}{4}\frac{G_{13} G_{24}}
{\omega^2-a^2}        
\bm{\sigma}_1\cdot\left(\frac{\bf p}{M_N}-\frac{\bf p'}{M_\Lambda}\right)
\bm{\sigma}_2\cdot\left(\frac{\bf p}{M_\Lambda}-\frac{\bf p'}{M_N}\right)
{\cal P}_f \nonumber\\ 
 &=& \frac{1}{4}\frac{G_{13} G_{24}}{\omega^2-a^2}\left[ \bm{\sigma}_1\cdot
 \left\{\left(\frac{1}{M_N}-\frac{1}{M_\Lambda}\right){\bf q} -\frac{1}{2}
 \left(\frac{1}{M_N}+\frac{1}{M_\Lambda}\right){\bf k}\right\}\cdot 
\right.\nonumber\\ && \left.
  \bm{\sigma}_2\cdot
 \left\{\left(\frac{1}{M_\Lambda}-\frac{1}{M_M}\right){\bf q} -\frac{1}{2}
 \left(\frac{1}{M_\Lambda}+\frac{1}{M_N}\right){\bf k}\right\}      
\right]{\cal P}_f  \nonumber\\
 &=& \frac{1}{4}\frac{G_{13} G_{24}}{\omega^2-a^2}\left[ 
\frac{1}{4}\left(\frac{1}{M_N}+\frac{1}{M_\Lambda}\right)^2
\bm{\sigma}_1\cdot{\bf k} \bm{\sigma}_2\cdot{\bf k}
-\left(\frac{1}{M_N}-\frac{1}{M_\Lambda}\right)^2
\bm{\sigma}_1\cdot{\bf q} \bm{\sigma}_2\cdot{\bf q} \right.\nonumber\\
&& \left.
-\frac{1}{2}\left(\frac{1}{M_N^2}-\frac{1}{M_\Lambda^2}\right)
\left(\vphantom{\frac{A}{A}}\bm{\sigma}_1\cdot{\bf q}\bm{\sigma}_2\cdot{\bf k} 
-\bm{\sigma}_1\cdot{\bf k}\bm{\sigma}_2\cdot{\bf q}\right)\right]\ {\cal P}_f,       
\label{appG.12}\end{eqnarray}
which gives the anti-symmetric spin-orbit potential
\begin{eqnarray}
\widetilde{V}_{K^*}({\bf q},{\bf k}) &=& \frac{1}{4}\frac{G_{13} G_{24}}{\omega^2-a^2} 
\left(\frac{1}{M_N^2}-\frac{1}{M_\Lambda^2}\right)
 (i/2)\left(\bm{\sigma}_1-\bm{\sigma}_2\right)\cdot{\bf n}\ {\cal P}_x\ .
\label{appG.13}\end{eqnarray}
\end{widetext}

\noindent Finally, we mention the relation with another sometimes used form
for the antisymmetric spin-orbit. Namely, we have
$  \bm{\sigma}_1\cdot\bm{\sigma}_2\left(\bm{\sigma}_1\times\bm{\sigma}_2\right) =
 -2i (\bm{\sigma}_1-\bm{\sigma}_2) - \bm{\sigma}_1\times\bm{\sigma}_2$,  
so that 
\begin{equation}
 (\bm{\sigma}_1-\bm{\sigma}_2) = i P_\sigma\ (\bm{\sigma}_1\times\bm{\sigma}_2)\ .
\label{appG.14}\end{equation}

%--------------------------------------------------------------------
%\begin{flushleft}
%\rule{16cm}{0.5mm}
%\end{flushleft}
%--------------------------------------------------------------------

%----------------------------------------------------------------------------------------
\acknowledgments
 We wish to thank E. Hiyama, K. Itonaga, T. Motoba, and H.-J. Schulze for many stimulating 
discussions.                                
                                                                                
%--------------------------------------------------------------------------
%\onecolumngrid   
%\end{widetext}

%\clearpage
%\newpage

%\end{widetext}  

\begin{thebibliography}{99}
\bibitem{NRY12a} M.M.\ Nagels, Th.A.\ Rijken, and Y.\ Yamamoto,
 {\it Extended-soft-core Baryon-Baryon Model ESC08,     
    I. Nucleon-Nucleon Scattering }, arXiv:nucl-th/1408.4825 (2014)
\bibitem{NRY12b} M.M.\ Nagels, Th.A.\ Rijken, and Y.\ Yamamoto,
 {\it Extended-soft-core Baryon-Baryon Model ESC08,     
    II. Hyperon-Nucleon Interactions }, this paper preprint 2014.
\bibitem{NRY12c} M.M.\ Nagels, Th.A.\ Rijken, and Y.\ Yamamoto,
 {\it Extended-soft-core Baryon-Baryon Model ESC08,     
    III. S=-2\ Hyperon-hyperon/nucleon Interactions}, preprint 2014.
\bibitem{PTP185}
Th.A.\ Rijken, M.M.\ Nagels, and Y.\ Yamamoto, Progr.\ Theor.\ Phys.\ {\bf 185}, 14 (2010);
Y.\ Yamamoto, T.\ Motoba, and Th.A.\ Rijken, ibid 72; 
E.\ Hiyama, M.\ Kamimura, Y.\ Yamamoto, T.\ Motoba, and Th.A.\ Rijken, ibid 106.
%------------------------------------------------------------------------
\bibitem{Rij04a} Th.A.\ Rijken, Phys.\ Rev.\ {\bf C73}, 044007 (2006).           
\bibitem{Rij04b} Th.A.\ Rijken and Y.\ Yamamoto, 
 Phys.\ Rev.\ {\bf C73}, 044008 (2006).   
\bibitem{Rij04c}  Th.A.\ Rijken and Y.\ Yamamoto, 
                {\it Extended-soft-core baryon-baryon model 
                III, hyperon-hyperon/nucleon interactions},
                arXiv:nucl-th/060807 (2006)
%------------------------------------------------------------------------
\bibitem{Mic69} L.\ Micu, Nucl.\ Phys.\ {\bf B10} (1969) 521;          
                R.\ Carlitz and M.\ Kislinger, 
                Phys.\ Rev.\ D {\bf 2} (1970) 336.
\bibitem{LeY73} A.\ Le\ Yaouanc, L.\ Oliver, O.\ P\'{e}ne, and J.-C.
                Raynal, Phys.\ Rev.\ D {\bf 8} (1973) 2223;
                Phys.\ Rev.\ D {\bf 11} (1975) 1272.
\bibitem{Isg85} N.\ Isgur and J.\ Paton, Phys.\ Rev.\ {\bf D31}, 2910 (1985);
                R.\ Kokoski and N.\ Isgur, Phys.\ Rev.\ {\bf D35}, 907 (1987).

\bibitem{NRS78} M.M.\ Nagels, T.A.\ Rijken, and J.J.\ de Swart,
         Phys.\ Rev.\ D {\bf 17} (1978) 768.
\bibitem{MRS89} P.M.M.\ Maessen, Th.A.\ Rijken, and J.J.\ de Swart,
         Phys.\ Rev.\ C {\bf 40} (1989) 2226.
\bibitem{RSY99} Th.A.\ Rijken, V.G.J.\ Stoks, and Y.\ Yamamoto, 
         Phys.\ Rev.\ C {\bf 59}, 21, (1999).
%---------------------------------------------------------------------------
\bibitem{Hiy00}
E.\ Hiyama, M.\ Kamimura, T.\ Motoba, T.\ Yamada, and Y.\ Yamamoto,
Phys.\ Rev.\ Lett.\  85 (2000) 270.
\bibitem{HT06} O.\ Hashimoto and H.\ Tamura, Progr.\ Part.\ Phys.\ {\bf 57} (2006) 564.

\bibitem{BDI70} J.T.\ Brown, B.W.\ Downs, and C.K.\ Iddings, 
                Ann.\ Phys.\ (N.Y.) {\bf 60}, 148 (1970).
\bibitem{ALS07}
For the OBE-potentials we have included the Brown-Downs-Iddings
anti-symmetric spin-orbit potentials from pseudo-scalar, vector-, and scalar-
meson exchange \cite{BDI70}. Also we derived new anti-symmetric spin-orbit
contributions from MPE. Since we do not fit P-waves for YN, these play no role 
in the construction of the ERSC07-model. Therefore, these potentials will be 
published elsewhere.
\bibitem{note.diffr}
In principle, the off-mass-shell J=0 contribution from the tensor-meson nonet $A_2, K_2$ etc. 
is included with the diffractive soft-core potentials, see e.g. \cite{NRS78,MRS89}.
Although the couplings are zero in ESC08 models, we include these potentials in the text
for completeness.
\bibitem{Odd03} For a review see: C.\ Ewerz, {\it The Odderon in Quantum Chromodynamics},
                hep-ph/0306137. 
\bibitem{Oka00} M.\ Oka, K.\ Shimizu, and K.\ Yazaki, Progr.\ Theor.\ Phys., Suppl.
{\bf 137}, p. 1 (2000).
\bibitem{Fuj07} Y.\ Fujiwara, Y.\ Suzuki, and C.\ Nakamoto, 
 Progr.\ in Part.\ and Nuclear Physics, {\bf 58} (2007) 439.
\bibitem{Tam65} R.\ Tamagaki and H.\ Tanaka, 
  Progr.\ Theor.\ Physics, {\bf 34}, 191 (1965); 
 R.\ Tamagaki, Suppl. Progr.\ Theor.\ Phys., Extra Number, p.242, 1968;
 R.\ Tamagaki, Progr.\ Theor.\ Phys.\ {\bf 39}, 91 (1968).

\bibitem{Dab99} J.\ Dabrowski, 
Phys.\ Rev.\ C 60 (1999) 025205.
\bibitem{Nou02}  H.\ Noumi et al.,
Phys.\ Rev.\ Lett.\ 89 (2002) 072301.

\bibitem{NRS77}  M.M.\ Nagels, T.A.\ Rijken, and J.J.\ deSwart, 
Phys.\ Rev.\ D 15 (1977) 2547.
\bibitem{SNRV71} J.J.\ de Swart, M.M.\ Nagels, T.A.\ Rijken, and P.A.\ Verhoeven,
         Springer Tracts in Modern Physics, {\bf 60}, 138 (1971).
\bibitem{Nag73} M.M.\ Nagels, T.A.\ Rijken, and J.J.\ de Swart,
         Ann.\ Phys.\ (N.Y.) {\bf 79}, 338 (1973).
\bibitem{Dal64} R.H.\ Dalitz and F.\ von Hippel,
         Phys.\ Lett.\ {\bf 10}, 153 (1964).
\bibitem{IZ80} 
C.\ Itzykson and J-B.\ Zuber, {\it Quantum Field Theory}, section 2-3-2,
McGraw-Hill Inc. 1980.
\bibitem{Rij91} Th.A.\ Rijken, Ann.\ Phys.\ (N.Y.), {\bf 208}, 253 (1991).
\bibitem{RS96ab} Th.A.\ Rijken and V.G.J.\ Stoks,
                Phys.\ Rev.\ C 54 (1996) 2869; 
                ibid.\ C 54 (1996) 2869.

\bibitem{HMYR08} E.\ Hiyama, T.\ Motoba, Y.\ Yamamoto, and Th.A.\ Rijken, in preparation.
%---------------------------------------------------------------------
\bibitem{Sto93}
 V.G.J.\ Stoks, R.A.M.\ Klomp, M.C.M.\ Rentmeester, and J.J.\ de Swart,
 Phys.\ Rev.\ C 48 (1993) 792.
\bibitem{Klo93} R.A.M.\ Klomp, private communication (unpublished).
%---------------------------------------------------------------------
\bibitem{Kanda05}
 J.K.\ Ahn et al.,  Nucl. Phys. A 761 (2005) 41. 
\bibitem{Kadyk71} 
 J.A. \ Kadyk, G. \ Alexander, J.H. \ Chan, P. \ Gaposchkin and G.H. \ Trilling,
  Nucl. Phys. B27 (1971) 13.
\bibitem{Kondo00}
 Y.\ Kondo et al.,  Nucl. Phys. A 676 (2000) 371.  
\bibitem{Tak01}  T.\ Takahashi et al.,
Phys.\ Rev.\ Lett.\ 87 (2001) 212502.
\bibitem{Hiy02}
E.\ Hiyama, M.\ Kamimura, T.\ Motoba, T.\ Yamada, and Y.\ Yamamoto,
Phys.\ Rev.\   C66 (2002) 024007.
\bibitem{Khaus00}
 P.\ Khaustov et al., Phys.\ Rev.\ C 61 (2000) 054603. 
%---------------------------------------------------------------------
\bibitem{Stoks93a}
 V.G.J.\ Stoks, R.\ Timmermans, and J.J.\ de Swart, Phys. Rev. C 47 (1993) 512.
\bibitem{Stoks94}
 V.G.J.\ Stoks, R.A.M.\ Klomp, C.P.F.\ Terheggen, and J.J.\ de Swart, 
 Phys. Rev. C 49 (1994) 2950.
\bibitem{Bry72} R.A.\ Bryan and A.\ Gersten, Phys.\ Rev.\ D {\bf 6} (1972) 341.
%---------------------------------------------------------------------
\bibitem{SR97} V.G.J.\ Stoks and Th.A.\ Rijken,
                Nucl.\ Phys.\ {\bf A 613} (1997) 311. 
\bibitem{Ale68} G.\ Alexander, U.\ Karshon, A.\ Shapira,
         G.\ Yekutieli, R.\ Engelmann, H.\ Filthuth, and W.\ Lughofer,
         Phys.\ Rev.\ {\bf 173}, 1452 (1968).
\bibitem{Sec68} B.\ Sechi-Zorn, B.\ Kehoe, J.\ Twitty, and
         R.A.\ Burnstein, Phys.\ Rev.\ {\bf 175}, 1735 (1968).
\bibitem{Eis71} F.\ Eisele, H.\ Filthuth, W.\ F\"olisch, V.\ Hepp,
         E.\ Leitner, and G.\ Zech,
         Phys.\ Lett.\ {\bf 37B}, 204 (1971).
\bibitem{Clo93} F.E. \ Close, and R.G. \ Roberts,
         Phys.\ Lett.\ {\bf B 316}, 165 (1993).
\bibitem{Eng66} R.\ Engelmann, H.\ Filthuth, V.\ Hepp, and E.\ Kluge,
         Phys.\ Lett.\ {\bf 21}, 587 (1966).
\bibitem{Hep68} V.\ Hepp and M.\ Schleich,
         Z.\ Phys.\ {\bf 214}, 71 (1968).
\bibitem{Ste70} D.\ Stephen, Ph.D.\ thesis, University of Massachusetts, 1970.
         Z.\ Phys.\ {\bf 214}, 71 (1968).
%-----------------------------------------------------------------------------
\bibitem{Inou12}
 T.\ Inoue et al, Nucl.\ Phys.\ {\bf A 881} (2012) 28.
%---------------------------------------------------------------------
%NEW BEGIN   YAMAMOTO references.
%---------------------------------------------------------------------
\bibitem{Yam85} 
Y.\ Yamamoto and H.Band\=o,
Prog. Theor. Phys. Suppl. {\bf No.81} (1985), 9.

\bibitem{Yam94}
Y.\ Yamamoto, T.Motoba, H.Himeno, K.Ikeda and S.Nagata,
Prog. Theor. Phys. Suppl. {\bf No.117} (1994), 361.

\bibitem{Yam10}
Y.\ Yamamoto, T.Motoba, Th.A. Rijken,
Prog. Theor. Phys. Suppl. {\bf No.185} (2010), 72. 

\bibitem{YFYR13}
Y.\ Yamamoto, T.Furumoto, N.Yasutake and Th.A. Rijken,
Phys.\ Rev.\ {\bf C 88} (2013), 022801(R).         

\bibitem{YFYR14}
Y.\ Yamamoto, T.Furumoto, N.Yasutake and Th.A. Rijken,
Phys.\ Rev.\ {\bf C 90} (2014), 045805.         
\bibitem{Harada05}
T.\  Harada and Y.\  Hirabayashi,
Nucl.\ Phys.\ A759, 143 (2005).

\bibitem{Kohno06}
M.\ Kohno, Y.\ Fujiwara, Y.\ Watanabe, K.\ Ogata and M.\ Kawai
Phys. Rev. {\bf C74} (2006), 064613.

\bibitem{Sch76} 
R.R.\ Scheerbaum,
Nucl. Phys. {\bf A257} (1976), 77.

\bibitem{TamHashi}
O.\ Hashimoto and H.\ Tamura
Prog. Part. Nucl. Phys. {\bf 57} (2006), 564.
\bibitem{E885}
T.\  Fukuda {\it et al.}, Phys. Rev.{\bf C58} (1998), 1306.
\\
P.\ Khaustov {\it et al.}, Phys. Rev. {\bf C61} (2000), 054603.

\bibitem{Gogami}
T.\ Gogami, PhD-thesis, Tohoku University 2014.
%---------------------------------------------------------------------
\bibitem{GAMil89}
G.A.\ Miller, Phys.\ Rev.\ {\bf C} (1989) 1563.

\bibitem{HYP06}
Th.A.\ Rijken and Y.\ Yamamoto, Proceedings of {\it The IX International Conference
on Hypernuclear and Strange Particle Physics}, edited by J.\ Pochodzalla and Th.\ Walcher,
October 10-14, 2006, p. 279. ISBN-10 3-540-76365-1 Springer Berlin Heidelberg New York.

\end{thebibliography}
\end{document}